\documentstyle[a4,11pt,epsfig]{article}
\newcommand \beq{\begin{eqnarray}}
\newcommand \eeq{\end{eqnarray}}
\newcommand \li{\par\noindent} 
\def\leftrightarrowfill{$ \mathord\leftarrow \mkern-6mu \cleaders
\hbox{$\mkern-2mu \mathord- \mkern-2mu$}\hfill \mkern-6mu \mathord\rightarrow$}
\def\overleftrightarrow#1{ \vbox{\ialign{##\crcr \leftrightarrowfill\crcr
\noalign{\kern-1pt\nointerlineskip}
$\hfil\displaystyle{#1}\hfil$\crcr}}}
\def\sqr#1#2{{\vcenter{\vbox{\hrule height.#2pt \hbox {\vrule width.#2pt
height#1pt \kern#1pt \vrule width.#2pt} \hrule height.#2pt }}}}

\def\tim#1{\mbox{\tiny{\sl{#1}}}}

\begin{document}
\title{{\sf Dispersion relations of mesons in symmetric 
nuclear matter}}
\author{\sf L. Mornas \\ 
\\
{\small{\it Universidad de Oviedo, Departamento de F\'{\i}sica,
E-33007 Oviedo, Spain}} }
\maketitle
\par\noindent {\small PACS: 21.65+f, 24.10.Jv, 21.60.Jz}
\par\noindent {\small keywords: in medium properties of mesons, 
                relativistic RPA}

\begin{abstract} 
We calculate dispersion relations and propagators for the $\sigma$, 
$\omega$, $\pi$, $\rho$, $\delta$, $\eta$ and $a_1$ mesons in
relativistic, dense, hot, symmetric nuclear matter. In addition 
to the usual mixing of the $\sigma-\omega$ system, we obtain 
mixing of  the $\delta$ with the longitudinal $\rho$ mode and of the 
tranverse $\rho$ with the transverse $a_1$ mode. Finally, the 
component of the $a_1$ polarization along the transferred momentum 
modifies the in-medium pion propagator in a way similar to the Migdal 
contact interaction, but with the opposite sign. The spurious
pion condensate as well as the additional contribution from the
$a_1$ meson are removed by a contact term. We compare two ways
of implementing contact term subtraction.
\end{abstract}


\section{Introduction}

This works investigates some points concerning the dispersion relations
and effective masses of mesons in dense, hot, symmetric nuclear 
matter. The dispersion relations will be calculated in the random 
phase approximation (RPA), by performing a linear response analysis 
\cite{DA85}--\cite{MGP01} around the homogeneous Hartree ground state.
This paper reproduces previous results and extends them in two aspects. 

Equations are presented for the six mesons of the Bonn 
model \cite{BonnPot}, {\it i.e.} $\sigma$, $\omega$, $\pi$,
$\rho$, $\delta$ and $\eta$. 
In addition, the contribution of NN loops to the dispersion relation 
of the $a_1$ is investigated, since this meson can be of interest 
from various points of view. One is the construction of chiral 
Lagrangians, since the $a_1$ is the chiral partner of the $\rho$ 
meson \cite{W67-WZ67-GG69} -- \cite{SR97}. Another is the issue of 
the dilepton production in relativistic heavy ion collisions.
As a matter of fact, the $a_1$ \cite{dileptonsa1} couples to 
the $\rho$ meson and may modify its spectral properties. 
The $a_1$ may also be used to describe part of the nucleon-nucleon
interaction which is mediated by correlated $\pi$-$\rho$ exchange 
in the S-wave channel \cite{DBS84,JHS96}. The inclusion of
correlated $\pi$-$\rho$  exchange is known to improve the
Bonn potential \cite{M89}; however one should keep in mind
that the structure of correlated $\pi$-$\rho$ exchange is 
more complex than what can be described by  exchange of particles 
with a sharp mass \cite{JHS96}.
Finally, the $a_1$ meson has sometimes been quoted \cite{S99,ADMSM96}
to provide a means to improve the behaviour of the differential 
nucleon-nucleon cross section for backwards scattering angle. 
This is related to the behavior of the
tensor part of the NN potential near $\vec r=0$. 

As a first part of this work, a discussion of the $a_1$ meson 
dispersion relation and mixing  effects with the $\rho$ and 
$\pi$ mesons is presented in section \S 2. It will be seen that 
the transverse mode of the $a_1$ mixes with that of the $\rho$. 
Since the axial current is not conserved, we will have a part 
of the $a_1$ polarization proportional to $q^\mu q^\nu$, where 
$q^\mu$ is the transferred quadrimomentum. This part mixes 
with the pion and renormalizes its polarization in a way which 
is formally similar to that produced by a contact interaction 
of the Landau-Migdal type. However, it has the opposite sign, 
so that the $a_1$ would enhance the unobserved pion zero sound 
mode. Since experimental evidence shows that pion condensation 
does not occur at saturation density in symmetric matter, this 
spurious mode should be removed. The standard procedure to 
implement this is to introduce a contact interaction.

We are thus lead in a second part to make some considerations 
on the relativistic generalization of the contact term. 
We will compare in section 
\S 3 the results obtained using the Ansatz of Horowitz {\it et al.} 
\cite{HP94,KPH95}, who modify the pion propagator in an {\it ad-hoc} 
way, to a more standard implementation simply consisting of adding 
one or more contact pieces to the Lagrangian density.  It will be 
seen that both procedures may lead to different results, when the 
former is not used with some caution. In particular, this
may explain the discrepancy recently obtained in calculating
the RPA corrections to neutrino-nucleon scattering in proto neutron
star matter \cite{RPLP99,YT00,MP01}. We advocate for the latter 
procedure.

Finally, numerical results are presented in section \S 4.
In symmetric matter, the dispersion relation factors in 
several subsystems. First there is the $\sigma$-$\omega$ sector,
with a transverse $\omega$ mode and a longitudinal mixed
$\sigma$-$\omega$ mode. Since this mode has already been 
studied extensively in the litterature, no numerical results 
are shown here. Then we have a $\delta$-$\rho$-$\pi$-$a1$
sector, which is composed of three parts: a longitudinal
mixed $\delta$-$\rho$ mode, a transverse mixed $\rho$-$a_1$ 
mode and a mixed $\pi$-$a_1$ mode. Dispersion relations
are calculated for these modes and the behaviour of the 
effective masses of the mesons is investigated.
We mention the problem of vacuum fluctuations and of the
choice of a renormalization procedure. As already noticed 
in the case of the $\rho$ meson \cite{MGP01}, different
results are obtained depending of the choice of the renormalization
conditions. At the time of writing, we must regard this as an
unresolved issue.

Among the possible applications of the present results, one can 
quote the calculation of screened nucleon-nucleon potentials
\cite{DP91,Angeles} or cross sections \cite{DM98a,DM98b}  and 
their application to nuclear dynamics 
\cite{MK95}, dilepton production in relativistic heavy ion 
collisions \cite{dileptonsa1}, the calculation of the electronic 
response function \cite{Wehrberger} or of the compressional 
modes of nuclei \cite{DF90,Piek01,ringvangiai01}.
Another field of application concerns the physics of supernovae 
explosions and protoneuton star cooling. In conditions of high 
density and temperature reached in the early phases of neutron 
star formation, it has recently been argued that RPA corrections 
to the neutrino-nucleon scattering cross section could be 
responsible for a sizeable modification of the neutrino mean free
path \cite{RPLP99,YT00,MP01}.
The results obtained in this work may be applied in the limiting case 
of pure neutron matter, by appropriately modifying the degeneracy
factor. In astrophysical applications nevertheless, one has 
to deal with an asymmetric environment. The dispersion relation of 
neutral mesons in asymmetric nuclear matter is presented in a 
companion paper \cite{M01b}. The case of propagation of charged 
mesons in asymmetric matter is currently under consideration.

\section{Meson exchange model}

\subsection{Discussion about the mesons included in this work}

The $\sigma$, $\omega$, $\rho$ and $\pi$ mesons are standard
mesons which are included in the description of the NN interaction
in hadronic models, whose prototype is the Quantum Hadrodynamics
(QHD) developped by Walecka and coworkers \cite{SW86-Se97}.
A vast amount of literature has been devoted to study their 
dispersion relations \cite{DA85}--\cite{MGP01},
\cite{C77}--\cite{TDMG00}.
This section presents a discussion on the relevance of
extending the approach to study more mesons.

The $\delta$ and $\eta$ mesons belong to the set of mesons
exchanged in the Bonn potential \cite{BonnPot}. However, 
they have been less studied than the previous ones. 
One reason is that the authors of \cite{BonnPot} found that 
they only bring a small adjustment to the form of the NN potential.
In mean field studies of nuclear matter, the $\delta$ is usually
not considered on the argument that the exchange of $\rho$ mesons 
is enough to reproduce the asymmetry energy at the mean field level
(see however \cite{KKS98,SST97}). In asymmetric matter, on the 
other hand, the delta meson could be important, since it 
carries isospin.
Interest has been shown for the $\delta$ meson in the context 
of Dirac-Brueckner calculations of the equation of state in 
asymmetric matter. When developping equivalent mean field 
theories with density dependent couplings, in order to 
reproduce the results of the full many-body calculation,
de Jong and Lenske \cite{dJL98} or Shen {\it et al.} \cite{SST97}
have found that it was necessary to introduce the $\delta$ meson 
with a significant coupling strength, {\it e.g.} 
$g_\delta^2(n_{\rm sat})/(4 \pi)=4.61$ at saturation density.

The $\delta$ meson mixes with the $\rho$ meson, just as does 
the $\sigma$ with the $\omega$ in the isoscalar sector. 
It is well known that $\sigma$-$\omega$ mixing is very strong 
(see {\it e.e} \cite{DP91,CL95}), so that one may ask whether 
the same occurs in the case of $\delta$-$\rho$ mixing. 
In relation with the debated case of the interpretation 
of dilepton measurements in heavy ion collisions, the effect 
of $\delta$-$\rho$ mixing has been studied by Teoredescu 
{\it et al.} \cite{TDMG00}.

The $a_1$ is very massive ($m_a$=1260 MeV), and is for this
reason usually not included in meson exchange models, which
take 1 GeV as a reasonable cutoff energy scale. There are
however theoretical reasons to consider this meson. 
As mentioned in the introduction, there is a need
of keeping this meson from symmetry arguments since the 
$a_1$ is the chiral partner of the $\rho$ 
\cite{W67-WZ67-GG69} -- \cite{SR97}.

For the purpose of reproducing the dilepton measurements
of CERES and HELIOS, an enormous amount of work has been spent 
on calculating the $\rho$ spectral function, as this meson yields 
the dominant contribution to dilepton production in the vector 
dominance model. The controversy arised, as to whether the dilepton 
enhancement at lower invariant mass was due to a reduction of the 
$\rho$ meson mass in the medium and a signal of chiral restoration 
in the Brown-Rho scaling picture \cite{BR91} (B/R scenario), 
or could it be explained by purely hadronic models as an increased 
width and shifted strength due $\rho\pi\pi$ coupling \cite{RCW97}
(R/W scenario). It was suggested \cite{dileptonsa1,KRBR00} that 
one may expect a substantial contribution to the width from 
$\rho\pi a_1$ loops. 

An other ground to study the $a_1$ meson is to determine
whether it could simulate short range corrections in the pion
exchange potential \cite{S99,ADMSM96}. As a matter of fact, Fourier 
transforming the expression obtained for the pion exchange 
contribution to the transition matrix yields a $\delta(r)$ piece 
in the NN potential in coordinate space. The standard procedure 
is to introduce the so-called Landau-Migdal parameter $g'$ as 
a contact interaction to remove this piece. This term is necessary 
to achieve a good description of the proton-neutron cross section 
at large scattering angles $\theta \simeq 180^o$ \cite{GL94,DAM01}.
The parameter $g'$ then also enters in the expression of the pion
polarization and further has the nice property of removing spurious
zero-sound modes from the pion dispersion relation, which would
be responsible for the onset of pion condensation at unrealistically 
low densities \cite{MSTV90}. 

In a relativistic formalism, it is less clear how to implement this 
term in a simple fashion. For example, Horowitz {\it et al.}
\cite{HP94,KPH95} have suggested an Ansatz making a replacement in 
the pion propagator. An other possibility would be to add to the 
Lagrangian a contact term with pseudovector coupling. This is the 
method used by Sch\"afer {\it el al.} \cite{SDEM94}, who add to their 
Lagrangian a term $g_a (\overline\psi \gamma_5 \gamma^\mu \tau \psi)
(\overline\psi \gamma_5 \gamma_\mu \tau \psi)$ in order to fit the 
NN cross section at large scattering angles.
Equivalently Engel {\it et al.} \cite{ADMSM96} suggest to add 
to the Lagrangian a piece $g_a \overline\psi \gamma_5 \gamma^\mu 
\psi a_\mu$  and take the limit of infinitely heavy $a^\mu$ field.
In section \ref{Horowitz-vs-Schaefer}, we compare the effects
of the Ans\"atze of Horowitz {\it et al.} or of Sch\"afer
{\it et al} on the one hand, and of true $a_1$ exchange on the other 
hand, on the dispersion relation of the $\pi$ and $\rho$ mesons.
It will be seen that the mixing of the $a_1$ meson with the pion 
acts as a Landau-Migdal term with the ``wrong'' sign. 

\subsection{Outline of derivation}

We will take the following Lagrangian density
\beq
{\cal L} &=& {\cal L}_{NN} + {\cal L}_{N\Phi}+ {\cal L}_{\Phi\Phi} 
+ {\cal L}_{\Phi_1\Phi_2}  + {\cal L}_{\rm CT} \label{TheLagrangian} \\
{\cal L}_{NN} &=& \overline\psi \biggl[ {i \over 2} \gamma. \overleftrightarrow 
\partial -m \biggr] \psi \qquad ; \qquad 
{\cal L}_{N\Phi} =  \overline\psi \Phi(x) \psi \\
& & {\rm with} \quad \Phi(x)=
g_\sigma\, \sigma + g_\delta\, \vec \delta . \vec \tau - g_\omega\, \gamma^\mu 
\omega_\mu -{f_\pi \over m_\pi} \gamma_5 \gamma^\mu \partial_\mu \vec \pi . 
\vec \tau -i g_\eta\, \gamma_5 \eta  \nonumber \\
& & \qquad \qquad \quad  -g_\rho\, \gamma^\mu \vec \rho_\mu . \vec \tau 
+ i {f_\rho \over 2 m} \sigma^{\mu\nu} \partial_\nu \vec \rho_\mu .\vec \tau 
-g_a\, \gamma_5 \gamma^\mu \vec a_\mu . \vec \tau 
\label{interaction} \\
{\cal L}_{\Phi\Phi} &=& {1 \over 2} \partial^\mu \sigma \partial_\mu \sigma 
-{1 \over 2} m_\sigma^2 \sigma^2  +{1 \over 2} \partial^\mu \vec \delta 
\partial_\mu \vec \delta -{1 \over 2} m_\delta^2 \vec \delta . \vec \delta 
+{1 \over 2} \partial^\mu \vec\pi \partial_\mu \vec\pi -{1 \over 2} m_\pi^2 
\vec\pi . \vec \pi +{1 \over 2} \partial^\mu \eta \partial_\mu \eta 
-{1 \over 2} m_\eta^2 \eta^2 \nonumber \\ 
& & - {1 \over 4} F^{\mu\nu} F_{\mu\nu} +{1 \over 2} m_\omega^2 \omega_\mu 
\omega^\mu - {1 \over 4} \vec R^{\mu\nu} \vec R_{\mu\nu} +{1 \over 2} m_\rho^2 
\vec \rho_\mu \vec \rho^\mu - {1 \over 4} \vec {\cal A}^{\mu\nu} \vec 
{\cal A}_{\mu\nu} +{1 \over 2} m_a^2 \vec a_\mu \vec a^\mu \nonumber \\
{\cal L}_{\Phi_1\Phi_2} & \ni & -{1 \over 3} b\, m\, \sigma^3 -{1 \over 4} c\, 
\sigma^4 - {1 \over 2} g_{\sigma\pi\pi}\, m_\sigma\, \sigma \pi^2 
+g_{\rho\pi\pi}\, \vec \rho^\mu . 
(\partial_\mu \vec \pi \times \vec \pi )+{1 \over 2} g_{\rho\pi\pi}^2\, 
(\vec \rho^\mu \times \vec \pi) .(\vec \rho_\mu \times \vec \pi)  \nonumber \\
& & +g_{a\rho\pi}\, \vec \pi . (\vec \rho_\mu \times \vec a^\mu) 
+ g'_{a\rho\pi}\, (\vec R^{\mu\nu} \times \vec {\cal A}_{\mu\nu} ).\vec \pi
\eeq

The first two lines are the Lagrangian for the nucleon and its interaction
with the meson fields. We chose the pseudovector coupling for the pion
since its phenomenology is better, namely it permits to reproduce
the pion-nucleon scattering. The pseudoscalar coupling will also be
considered in section \S \ref{pion}.

The following two lines contain the kinetic terms for the mesons. The
last two lines contain some meson-meson interactions which are important 
from the point of view of phenomenology. In particular, we may have 
$\sigma\pi\pi$, $\rho\pi\pi$ and $\rho\pi a_1$ couplings, which contribute
to the spectral width of the $\sigma$, $\rho$ and $a_1$ mesons. The
sigma self couplings  $b\, m\, \sigma^3 +  c\, \sigma^4/4$ improve
the description of the effective nucleon mass and incompressibility
modulus.

One could consider more complicated forms of the Lagrangian, which would 
be obtained by enforcing chiral symmetry \cite{SW92,KR94}, 
\cite{FRS94} -- \cite{BFMNLQS01}. Accordingly, more meson-meson coupling
terms would appear in ${\cal L}_{\Phi_1\Phi_2}$. Since our purpose 
here was mainly to investigate NN loops, we will not consider 
such terms in this work. In fact, as will be seen in the following, 
the derivation method  based on linear response analysis, which we
follow here, discards meson-meson correlations at an early
stage of the calculation, so that terms involving meson couplings 
only appear at the mean field level. Since we deal in this paper
with symmetric matter, most of these terms actually disappear,
since they would involve expectation values of the pion, rho or
$a_1$ fields, all of which vanish in this approximation.
Only terms involving the $\sigma$ field will survive.
Meson loops should of course appear in a more realistic treatment.

The term ${\cal L}_{\rm CT}$ represents a counterterm Lagrangian in order
to handle the vacuum divergences; its form is given in Appendix A.3.

The dispersion relations are derived from a linear response 
analysis in the Wigner operator formalism. By applying a perturbation
to the relativistic Hartree equilibrium, one obtains homogeneous
equations for the perturbed meson fields $\phi_1(q)$ in the form 
$D(q) \phi_1(q) =0$, where  $\phi_1 (q)$ is a vector formed by 
the components of the perturbations to the various fields $\sigma$, 
$\omega$, $\pi$, $\rho$,
$\delta$, $\eta$ and $a_1$, and $D(q)$ is a matrix which contains
the dispersion relations. The condition that these homogeneous 
equations admit non trivial solutions is that the determinant of 
$D(q)$ vanishes. As $D(q)$ is in general non diagonal, the
determinant will only partially factorize. A standard
example is the case of the $\sigma$-$\omega$ subsystem, where
the $\omega$ transverse mode factorizes out, while the longitudinal
mode of the $\omega$ mixes with the $\sigma$ mode \cite{DP91}
It was shown in this work that the results obtained coincide with 
the one-loop approximation from Green's function formalism.

We define the Wigner operator
\beq
F_{\alpha\beta}(x,p)=\int {d^4\, R \over (2 \pi)^4}\, e^{-i p.R}\ 
\overline\psi_\beta(x+{R\over 2}) \otimes \psi_\alpha(x-{R \over 2})
\eeq
It obeys the kinetic equation (as well as a conjugate equation)
\beq
\left[ i {\gamma.\partial \over 2} + (\gamma.p -m) \right] F (x,p) =
- \int \, {d^4\, R \over (2 \pi^4)}\, d^4\, \xi \ e^{-i (p-\xi).R }
\ \Phi(x-{R \over 2}) F(x,\xi) 
\label{kinetic}
\eeq
with $\Phi(x)$ defined as in Eq. (\ref{interaction})
At this level, $F(x,p)$ as well as the fields $\sigma(x)$, $\omega(x)$,
$\pi(x)$, $\delta(x)$ ...{\it etc} contained in $\Phi$ are operators.
In order to obtain the RPA approximation, two basic assumptions are 
made:
\li --- Correlations are neglected. When taking the statistical 
average of Eq. (\ref{kinetic}), it amounts to replacing in the
right hand side the average of the product $<\Phi F>$ by the product
of averages $<\Phi><F>$. It is at this point that we neglect
the contribution of meson loops to the dispersion relations.
They could be restored by releasing this assumption, or be
reintroduced by hand at the end of the calculation. We will
not consider them in this work.
We will omit the $<>$ denoting statistical averages henceforward 
in order to simplify the notations.
\li --- It is assumed that there exists an uniform, unpolarized 
equilibrium given by the relativistic Hartree approximation, 
and that one may perform an expansion around this equilibrium 
(remember the notations $F$, $\sigma$ ... represent from now on 
the statistical averages $<F>$, $<\sigma>$ ...)
\beq
  && F(x,p) = F_H(p) + F_1(x,p) \quad , \quad \sigma(x) = \sigma_H 
+\sigma_1(x) \quad , \quad \vec \delta(x) = \vec \delta_H 
+ \vec \delta_1(x) \nonumber \\
  && \omega^\mu(x) = \omega_H^\mu + \omega_1^\mu(x) \qquad \quad \ \ 
, \quad \vec\rho^\mu(x) = \vec\rho_H^\mu + \vec\rho_1^\mu(x)  
\label{perturb} \\
 && \vec\pi(x) = \vec\pi_1 (x) \qquad \qquad \qquad \quad , \quad 
\eta(x) = \eta_1(x) \qquad \quad \ \ , \quad
\vec a^\mu(x)= \vec a_1^\mu(x) \nonumber
\eeq
In eqs. (\ref{perturb}) the pion, eta and $a_1$ contributions vanish
due to parity arguments in unpolarized matter. Moreover, in symmetric 
matter the chemical potentials of the proton and neutron coincide,
so that  $\rho_H$ and $\delta_H$ will vanish as well, since they are
proportional to the difference between the proton and neutron
density, and between the proton and neutron effective masses 
respectively. (We do not take into account the tiny difference due 
to the electromagnetic interaction.)
  
With these approximations, we obtain after linearizing and Fourier transforming 
Eq. (\ref{kinetic}) the first order perturbation to the nucleon Wigner function
\beq
F_1(q,p)= G\left( p-{q \over 2}\right)\, \Phi(q)\, F_H\left( p+{q \over 2}
\right) + F_H \left( p-{q \over 2} \right)\, \Phi(q)\, 
G\left( p+{q \over 2}\right)
\label{wign1}
\eeq
with
\beq
\Phi(q) &=& -g_\sigma \sigma_1(q) + g_\omega \gamma^\mu \omega_{1\, \mu}(q)
-g_\delta \vec\delta_1(q).\vec\tau  + i\, {f_\pi \over m_\pi}\, \gamma_5 
\gamma^\mu q_\mu \vec\pi_1(q).\vec\tau \nonumber \\
& & + g_\rho \gamma^\mu \vec\rho_{1\, \mu}(q).\vec\tau +{f_\rho \over 2 m} 
\sigma^{\mu\nu} q_\nu \vec\rho_{1\, \nu}(q).\vec\tau +g_a \gamma_5 \gamma^\mu 
\vec a_{1\, \mu}(q).\vec\tau 
\label{Sq}\\
G(p) &=& {\gamma.P +M \over P^2 -M^2 \pm i \epsilon}
 \\
F_H(p) &=& S(p) \varphi(p) \nonumber \\
& & {\rm with}\quad S(p)= \gamma.P+M \\
& & {\rm and}\quad \varphi(p)= {d \over (2 \pi)^3}\, \delta(P^2 -M^2) 
\left[ \theta(p_0) n(p) + \theta(-p_0) \overline n(p) 
- \theta(-p_0)  \right] \nonumber
\eeq
In symmetric nuclear matter, $d=2$ is the isospin degeneracy. $n(p)$
and $\overline n(p)$ are the Fermi-Dirac distribution functions for the
(quasi)-particles and antiparticles respectively. $M=m-g_\sigma \sigma_H$ is
the effective mass and $P^\mu=p^\mu -g_\omega \omega_H^\mu$ is
the effective momentum.
Finally, we insert this solution in the linearized equations of the mesons. 
For example, we have for the $a_1$ meson
\beq
\left[ q^\mu q^\nu +(-q^2+m_a^2) g^{\mu\nu} \right] \vec a_{1\, \nu}(q)=
-g_{a\rho\pi} \vec \pi_1(q)\times \vec \rho_H^\mu + 
g_a \int d^4\, p Tr[ \gamma_5 \gamma^\mu \vec \tau F_1(q,p) ]
\eeq
Since we assume in this work that we are in symmetric
matter, the Hartree component of the $\rho$ field vanishes.
Inserting (\ref{wign1}) in this expression, we can recast 
it in the form
\beq
\left[ -q^\mu q^\nu +(q^2-m_a^2) g^{\mu\nu} \right] \vec a_{1\, \nu}(q)
=  \Pi_{a \pi}^\mu(q)\ \vec \pi_1(q) + \Pi_{a \rho}^{\mu\nu}(q)\ 
\vec \rho_{1\, \nu}(q) +\Pi_{aa}^{\mu\nu}(q)\ \vec a_{1\, \nu}(q)
\eeq
where, for example
\beq
\Pi_{a \rho}^{\mu\nu} = \int d^4 p\, {\rm Tr} \left[  g_a \gamma_5 
\gamma^\mu \tau_i\,  S(p-{q \over 2}) \left( g_\rho \gamma^\nu 
+ {f_\rho \over 2 m} \sigma^{\nu\lambda} q_\lambda \right)  \tau_j\,  
S(p+{q \over 2}) \right]  \left\{ {\varphi (p+{q \over 2}) 
- \varphi (p-{q \over 2}) \over  2 p.q -i \epsilon} \right\}
\eeq 
Terms such as $\Pi_{a \omega}$ vanish in symmetric matter since the trace
over isospin will involve the difference of proton and neutron distribution
functions.
\smallskip
The dispersion relations may be summarized in matricial form $D(q) \phi(q)=0$
In symmetric matter the dispersion relation decouples in three blocks:
($\sigma$-$\omega$), ($\eta$) and ($\rho$-$\delta$-$a_1$-$\pi$):

\beq
D(q)= \left( \matrix{ D_{\sigma\sigma} & D_{\sigma\omega}^\nu & 0 & 0 & 0 & 0 & 0 \cr
    D_{\sigma\omega}^\mu & D_{\omega\omega}^{\mu\nu} & 0 & 0 & 0 & 0 & 0 \cr
    0 & 0 & D_{\eta\eta} &  0 &  0 &  0 &  0 \cr
    0 & 0 & 0 & D_{\delta\delta} & D_{\delta\rho}^\nu &  0 &  0 \cr
    0 & 0 & 0 & D_{\delta\rho}^\alpha & D_{\rho\rho}^{\mu\nu} 
                                              & D_{a \rho}^{\mu\nu} &  0 \cr
    0 & 0 & 0 & 0 & D_{a \rho}^\mu & D_{aa}^{\mu\nu} & D_{a \pi}^\nu \cr
    0 & 0 & 0 & 0 & 0 & -D_{a \pi}^\mu & D_{\pi\pi} \cr} \right)
\qquad \phi(q) = \left( \matrix{ \sigma_1 \cr
\omega_{1\, \nu} \cr \eta_1 \cr \delta_1 \cr \rho_{1\, \nu} \cr
a_{1\, \nu} \cr \pi_1 } \right)
\eeq
with
\beq
D_{\sigma\sigma} &=& q^2 - m_\sigma^2 + \Pi_{\sigma\sigma} \quad ; \quad
D_{\omega\omega}^{\mu\nu} = q^\mu q^\nu + (m_\omega^2 - q^2) g^{\mu\nu}
+ \Pi_{\omega\omega}^{\mu\nu} \quad ; \quad
D_{\sigma\omega}^\mu = \Pi_{\sigma\omega}^\mu \nonumber \\
D_{\eta\eta} &=& q^2 -m_\eta^2 + \Pi_{\eta\eta} \nonumber \\
D_{\delta\delta} &=& q^2 - m_\delta^2 + \Pi_{\delta\delta} \quad ; \quad
D_{\rho\rho}^{\mu\nu} = q^\mu q^\nu + (m_\rho^2 - q^2) g^{\mu\nu}
+ \Pi_{\rho\rho}^{\mu\nu} \quad ; \quad
D_{\delta\rho}^\mu = \Pi_{\delta\rho}^\mu \nonumber \\
D_{aa}^{\mu\nu} &=& q^\mu q^\nu + (m_a^2 - q^2) g^{\mu\nu}
+ \Pi_{a}^{\mu\nu} \quad ; \quad D_{a\rho}^{\mu\nu} = \Pi_{a\rho}^{\mu\nu} 
\nonumber \\
D_{\pi\pi} &=& q^2 - m_\pi^2 + \Pi_{\pi\pi} \qquad ; \qquad 
D_{a \pi}^\mu = \Pi_{a \pi}^\mu \nonumber
\eeq
In the preceding equations, the meson self couplings were not 
taken into account. If they are present, they appear as mean 
field modifications. The dispersion relations in this case are
given by making the following replacements for the $\sigma$ 
and $\pi$ masses
\beq
m_\sigma^2 \rightarrow M_\sigma^2= m_\sigma^2 +2 b m \sigma_H
+3 c \sigma_H^2 \quad , \quad m_\pi^2 \rightarrow 
M_\pi^2= m_\pi^2 + g_{\sigma\pi\pi}\, m_\sigma\, \sigma_H
\nonumber 
\eeq
We will decompose the polarizations on the usual orthogonal set
of tensors and vectors as follows
\beq
\Pi_{\omega\omega}^{\mu\nu} &=& - \Pi_{\omega\, L} L^{\mu\nu} 
- \Pi_{\omega\, T} T^{\mu\nu} \qquad ; \qquad \Pi_{\sigma \omega}^{\mu} 
= \Pi_{\sigma\omega} \eta^\mu \nonumber \\
\Pi_{\rho\rho}^{\mu\nu} &=& - \Pi_{\rho\, L} L^{\mu\nu} - \Pi_{\rho\, T} 
T^{\mu\nu} \qquad ; \qquad \Pi_{\delta\rho}^{\mu} = \Pi_{\delta\rho} 
\eta^\mu \nonumber \\
\Pi_{aa}^{\mu\nu} &=& -\Pi_{a\, L} L^{\mu\nu} 
- \Pi_{a\, T} T^{\mu\nu} -\Pi_{a Q} Q^{\mu\nu}  \\
\Pi_{a\rho}^{\mu\nu} &=& i\ \Pi_{a \rho} 
\epsilon^{\mu\nu\alpha\beta} q_\alpha \eta_\beta \qquad ; \qquad
\Pi_{a \pi}^\mu = i\ \Pi_{a \pi} q^\mu \nonumber
\eeq
with
\beq
\eta^\mu &=& u^\mu - {q.u\over q^2}\, q^\mu \nonumber \\
L^{\mu\nu} &=& {\eta^\mu \eta^\nu \over \eta^2} \qquad ; \qquad 
T^{\mu\nu} = g^{\mu\nu} -  {\eta^\mu \eta^\nu \over \eta^2} 
- {q^\mu q^\nu \over q^2}  \\
Q^{\mu\nu} & =&  {q^\mu q^\nu \over q^2}  \qquad ; \qquad 
E^{\mu\nu} = \epsilon^{\mu\nu\alpha\beta} q_\alpha \eta_\beta
\nonumber
\eeq
The explicit expression of the polarizations can be found in the Appendix.

We obtain the dispersion relations by equating the determinant of the 
matrix $D$ to zero 
\beq
{\rm Det}(D)&=& m_\rho^2 m_\omega^2  \left[ (q^2 -m_\omega^2 +\Pi_{\omega L})
(q^2 -m_\sigma^2 + \Pi_{\sigma}) + \eta^2 \Pi_{\sigma\omega} \right] 
\times\left[ q^2 - m_\omega^2 + \Pi_{\omega T} \right] ^2\times 
\nonumber \\
& & \times \left[ q^2 -m_a^2 +\Pi_{a L} \right] \times 
(m_a^2 -\Pi_{a Q}) \times \left[ q^2 -m_\pi^2 + \Pi_\pi - 
{q^2 \Pi_{a \pi}^2 \over m_a^2 -\Pi_{a Q} } \right] \times  
\nonumber \\
& & \times \left[ (q^2 -m_\rho^2 + \Pi_{\rho T}) (q^2 -m_a^2 
+ \Pi_{a T}) + \eta^2 q^2 \Pi_{a \rho}^2 \right]^2 \times 
\nonumber \\
& & \times \left[ (q^2 - m_\rho^2 + \Pi_{\rho L})(q^2 - m_\delta^2 
+ \Pi_\delta) + \eta^2 \Pi_{\delta\rho^2}) \right] \times 
\left[ q^2 -m_\eta^2 + \Pi_\eta \right] =0 
\eeq

\subsection{Propagators for the 
$\sigma$-$\omega$-$\pi$-$\rho$-$\delta$-$\eta$-$a_1$ model}

The propagators $G$ are obtained by inverting the dispersion 
matrix $D$. 

The $\sigma-\omega$ sector has already been studied thoroughfully 
(see {\it e.g.} \cite{DP91,GDP94,LH89,KS88,CPS92,JP94,CL95}). 
We obtain as usual
\beq
G_{\sigma} &=& {1 \over q^2 -m_\sigma^2 + \Pi_\sigma} 
\label{Gsig} \\
G_{\omega}^{\mu\nu} &=& -G_{\omega T} T^{\mu\nu} -G_{\omega L} 
L^{\mu\nu} -G_{\omega Q} Q^{\mu\nu} \\
& & {\rm with}\quad G_{\omega L} ={ q^2 -m_\sigma^2 + \Pi_\sigma \over
(q^2 - m_\omega^2 + \Pi_{\omega L})(q^2 -m_\sigma^2 + \Pi_\sigma) +
\eta^2 \Pi_{\sigma\omega}^2} \label{GvL} \\
& & \phantom{\rm with\quad } G_{\omega T} ={1 \over q^2 -m_\omega^2 + 
\Pi_{\omega T}}  \label{GvT} \\
& & \phantom{\rm with\quad } G_{\omega Q} ={-1 \over m_\omega^2} \\
G_{\sigma\omega}^\mu &=& {\Pi_{\sigma\omega} \eta^\mu \over
(q^2 - m_\omega^2 + \Pi_{\omega L})(q^2 -m_\sigma^2 + \Pi_\sigma) +
\eta^2 \Pi_{\sigma\omega}^2} \label{Gsv}
\eeq
The $\sigma$ meson mixes with the longitudinal part of the $\omega$ meson.
The mixing effect is known to be strong. A zero sound mode may appear
in the longitudinal mode, with a strength which depends on the values of the
cutoff parameters in the nucleon form factors \cite{DP91,CL95}. 

The $\eta$ meson decouples from the other ones in symmetric matter,
therefore the calculation of its propagator is trivial. We have:
\beq
G_{\eta} &=& {1 \over q^2 -m_\eta^2 + \Pi_\eta} \label{Geta} 
\eeq
If we had not taken into account the $a_1$ meson, the $\rho$-$\delta$ 
system would decouple and reproduce the pattern of the $\sigma$-$\omega$
in the isovector sector. Now the $\rho$ also couples to the
$a_1$ through the polarization $\Pi_{a \rho}^\mu$. Nevertheless,
the calculation shows that the dispersion relation of the 
($\rho$-$\delta$-$a_1$-$\pi$) subsystem factorizes in three terms: 
a mode where the $\delta$ meson mixes with the longitudinal part of the
$\rho$, a mode where the transverse contribution from the $a_1$ 
mixes with the transverse part of the $\rho$, and a mode where the pion 
dispersion relation is modified by a term coming from the part of
the $a_1$ meson which is parallel to $q^\mu q^\nu$. We have
\beq
G_{\delta} &=& {q^2 -m_\rho^2 +\Pi_{\rho L} \over 
(q^2 -m_\rho^2 +\Pi_{\rho L})(q^2 -m_\delta^2 + \Pi_\delta)
+ \eta^2 \Pi_{\rho\delta}^2} \label{Gd} \\
G_{\rho}^{\mu\nu} &=& -G_{\rho T} T^{\mu\nu} -G_{\rho L} L^{\mu\nu}
-G_{\rho Q} Q^{\mu\nu} \\
& & {\rm with}\quad G_{\rho L} ={ q^2 -m_\delta^2 + \Pi_\delta \over
(q^2 - m_\rho^2 + \Pi_{\rho L})(q^2 -m_\delta^2 + \Pi_\delta) +
\eta^2 \Pi_{\rho\delta}^2} \label{GrL} \\
& & \phantom{\rm with\quad} G_{\rho T} ={(q^2 -m_a^2+ \Pi_{a T}) \over
(q^2 -m_a^2+ \Pi_{a T})(q^2 -m_\rho^2 +\Pi_{\rho T}) 
+\eta^2 q^2 \Pi_{\rho a}^2} 
\label{GrT} \\
& & \phantom{\rm with\quad} G_{\rho Q} ={-1 \over m_\rho^2} \\
G_{\delta\rho}^\mu &=& {\Pi_{\delta\rho} \eta^\mu \over
(q^2 -m_\rho^2 +\Pi_{\rho L})(q^2 -m_\delta^2 + \Pi_\delta)
+ \eta^2 \Pi_{\rho\delta}^2}
\label{Gdr} \\
G_a^{\mu\nu} &=& -G_{a T} T^{\mu\nu} -G_{a L} L^{\mu\nu}
-G_{a Q} Q^{\mu\nu} \\
& & {\rm with\ } G_{a L} ={ 1 \over q^2 -m_a^2 + \Pi_{a L}} 
\label{GaL} \\
& & \phantom{\rm with\ } G_{a T} ={(q^2 -m_\rho^2+ \Pi_{\rho T}) \over
(q^2 -m_\rho^2+ \Pi_{\rho T})(q^2 -m_a^2 + \Pi_{a T}) 
+ \eta^2 q^2 \Pi_{\rho a}^2}
\label{GaT}  \\
& & \phantom{\rm with\ } G_{a Q} ={-(q^2 -m_\pi^2+ \Pi_{\pi}) \over
(m_a^2- \Pi_{a Q}) (q^2 -m_\pi^2+ \Pi_{\pi}) - q^2 \Pi_{a \pi}^2} 
\label{GaQ} \\
G_{\pi} &=& {1 \over q^2 -m_\pi^2 +\widetilde \Pi_\pi} \quad ; \quad
\widetilde \Pi_\pi= \Pi_\pi - { q^2 \Pi_{a \pi}^2 \over
m_a^2 - \Pi_{a Q}}  \label{Gpi} \\
G_{a\pi}^\mu &=& -i G_{a \pi} q^\mu \quad ; \quad
G_{a \pi} = {-\Pi_{a \pi} \over
(q^2 -m_\pi^2 +\Pi_\pi)(m_a^2-\Pi_{a Q}) -q^2 \Pi_{a \pi}^2}
\label{Gapi} \\
G_{a\rho}^{\mu\nu} &=& {- i\, \Pi_{a \rho} \epsilon^{\mu\nu\rho\lambda} 
q_\rho \eta_\lambda \over (q^2 -m_\rho^2+ \Pi_{\rho T})
(q^2 -m_a^2 + \Pi_{a T}) + \eta^2 q^2 \Pi_{\rho a}^2}
\label{Garho}
\eeq

The mixing with the $a_1$ meson acts as a modification to the
pion polarization. Using the relations existing between $\Pi_\pi$,
$\Pi_{a \pi}$ and $\Pi_{a Q}$ (see Appendix A), it can be recast into
the form
\beq
\widetilde \Pi_\pi = {q^2 \Pi_\pi \over q^2 + \displaystyle{g_a^2 
\over m_a^2}\displaystyle{m_\pi^2 \over f_\pi^2} \Pi_\pi }
\eeq
It therefore comes out in a form similar as would a Landau-Migdal interaction. 
We would obtain exactly the Landau-Migdal form in the relativistic case
with the replacement 
\beq
{g_a^2  \over m_a^2} \rightarrow\, -\, g' {f_\pi^2 \over m_\pi^2}
\label{a-vs-LM}
\eeq
We must note however the sign in the previous equation. One would need a 
fictitious $a_1$ field with an imaginary coupling so that equation 
(\ref{a-vs-LM}) may be fulfilled.
Contrary to some expectations raised in the literature \cite{S99}, 
the $a_1$ meson cannot be used to improve the short range behaviour of the
pion-nucleon interaction; instead it contributes an additional term
which must be compensated by {\it e.g.} a residual contact interaction. 
 
Also when calculating the potential generated by the exchange of $\pi$ 
and $a_1$ mesons in vacuum, one obtains after taking the semiclassical 
limit $k/m \ll 1$ and also $k/m_a \ll 1$
\beq
V_{\pi + a}^{\tim{NR}} = - \left( {f_\pi \over m_\pi} \right)^2
{1 \over 3} \left[ \vec\sigma_1.\vec\sigma_2 -{m_\pi^2 \over k^2 
+ m_\pi^2}  \vec\sigma_1.\vec\sigma_2 + {S_{12} \over k^2 + m_\pi^2}
\right] + {g_a^2 \over m_a^2} \left[ -  \vec\sigma_1.\vec\sigma_2
+ {\cal O}(k^2) \right]
\eeq
and the term responsible for the $\delta(\vec r)$ singularity
$ \vec\sigma_1.\vec\sigma_2$ after Fourier transforming to
position space would be removed by the same choice 
$(g_a^2/ m_a^2) \rightarrow\, -\, (1/3) (f_\pi^2 / m_\pi^2)$.

\section{Short range behavior}
\label{Horowitz-vs-Schaefer}

As is well known, and also will be seen in section \S \ref{pion}, the 
dispersion relations derived in the preceding paragraph would lead to 
an excessive softening of the pion mode and to pion condensation,
which is not observed experimentally. This is due to the fact that we
have not yet taken into account the effect of short range corrections
at this level of approximation. 
A related shortcoming of the pion exchange model is that
it gives a vanishing differential cross section at scattering
angle $\theta=180^o$ in the neutron-proton exchange reaction,
in glaring contradiction with the experiment. Both features are due 
to the fact that the contribution to the NN potential arising
from pion exchange $V_\pi(r)$ contain a piece $\delta(\vec r)$ 
singular at the origin. This is an artefact of the model due
to the assumption that the particles are pointlike. Such
singular pieces in fact appear for all kinds of meson exchange.
Whereas it is enough for the $\sigma$, $\omega$ terms entering the
definition of the central potential to smooth this divergence by 
folding it with  a form factor of the type
\beq 
g_\alpha^2 \rightarrow g_\alpha^2 \left( {\Lambda_\alpha^2 - 
m_\alpha^2 \over \Lambda_\alpha^2 - q^2} \right) \quad
\alpha \in \{ \sigma, \omega, \pi, \rho, ... \},
\nonumber
\eeq
it is necessary in the case of the pion to remove the $\delta(\vec r)$
function in order to arrive at a correct description of the NN cross
cross section and spin transfer observables \cite{GL94}.

Short range corrections of the desired type would arise in a many body 
calculation from higher order correlations at the $\pi NN$ vertex
\cite{Dickoff}. In a simpler approach, the usual practice is to 
Fourier transform the potential to coordinate space and there 
remove the delta function by hand. This is equivalent to adding 
to the potential the Migdal contact term $g' \sigma_1.\sigma_2 
\tau_1.\tau_2$. The delta substraction procedure for the pion alone 
amounts to taking $g'=0.333$. A further contribution comes from the rho.
Phenomenology favors higher values $g'=0.5 - 0.9$. A recent analysis
\cite{SST99} extracted $g'=0.6$ from data on the Gamow Teller resonance.
This procedure works well in the non-relativistic limit \cite{Oset} 
but has the disadvantage of not beeing covariant.

In the following we examine two procedures which have been suggested
in the litterature to introduce the Landau-Migdal short range correction 
in a covariant way. 

\subsection{Propagator replacement Ansatz}

Horowitz {\it et al} \cite{HP94,KPH95} suggest to replace the vertex 
and propagator of the pion as follows:
\beq
  \left\{ \matrix{\Gamma_\pi & =&\gamma_5 \gamma^\mu q_\mu \cr
                  G^0_\pi & =& \displaystyle{1 \over q^2 -m_\pi^2 } \cr } \right\}
\rightarrow 
  \left\{ \matrix{\Gamma_\pi^\mu & =&\gamma_5 \gamma^\mu \cr
            G_\pi^{0\, \mu\nu} & =& \displaystyle{q^\mu q^\nu \over q^2 -m_\pi^2 } 
                  -g' g^{\mu\nu} \cr } \right\}
\eeq
In the medium, the pi and rho propagators obey the coupled Dyson equation 
$ G = G^0 + G^0\, \Pi\, G $ or equivalently, $[G]^{-1} = [G_0]^{-1} - \Pi$, 
or explicitely
\beq
[G]^{-1} = \left( \matrix{q^\mu q^\nu + (m_\rho^2 - q^2) g^{\mu\nu} 
+\Pi_{\rho\rho}^{\mu\nu} & \Pi_{\rho \pi}^{\mu\nu} \cr 
\Pi_{\pi\rho}^{\mu\nu} & \displaystyle{q^\mu q^\nu
\over g' [ q^2 -g' (q^2 -m_\pi^2)]} -\displaystyle{g^{\mu\nu} \over g'} 
+\Pi_{\pi\pi}^{\mu\nu} \cr} \right)
\label{dispHor}
\eeq
The ``pion'' polarization which appears in this expression is given by
\beq
\Pi_{\pi\pi}^{\mu\nu} &=& -\left({f_\pi \over m_\pi}\right)^2 
\int d^4 p {\rm Tr}\left[\gamma_5 
\gamma^\mu S(p-{q \over 2})\gamma_5 \gamma^\nu S(p+{q \over 2})\right] 
\left\{ {\varphi (p-{q \over 2}) - \varphi (p+{q \over 2}) \over  
2 p.q -i \epsilon} \right\} \\
&=& -\Pi_{\pi T} T^{\mu\nu} -\Pi_{\pi L} L^{\mu\nu} -\Pi_{\pi Q} Q^{\mu\nu} 
\eeq
and the usual pseudovector pion polarization is recovered by 
contracting with $q_\mu q_\nu$
\beq
\Pi^{PV} = -\Pi_\pi^{\mu\nu} q_\mu q_\nu = - q^2 \Pi_{\pi Q}
\eeq
This ``pion'' mixes with the $\rho$ meson, due to the polarization
\beq
\Pi_{\pi\rho}^{\mu\nu} &=& i \int d^4 p {\rm Tr} \left[ \left({f_\pi \over m_\pi}
\gamma_5 \gamma^\mu \right)S(p-{q \over 2}) \left( g_\rho \gamma^\nu 
+ {f_{\rho} \over 2 m} \sigma^{\nu\lambda} q_\lambda \right) S(p+{q \over 2})
\right] 
\left\{ {\varphi (p-{q \over 2}) - \varphi (p+{q \over 2}) \over  
2 p.q -i \epsilon} \right\} \nonumber \\
&=& \Pi_{\pi\rho}\, \epsilon^{\mu\nu\rho\lambda} q_\rho \eta_\lambda 
\eeq
The full dispersion relation in the $\pi$-$\rho$ sector is obtained by taking 
the determinant of  (\ref{dispHor}).
We have
\beq
{\rm Det}(G^{-1}) &=& m_\rho^2 \left[ ( {1 \over g'} +\Pi_{\pi T}) 
(q^2-m_\rho^2+\Pi_{\rho T}) -\eta^2 q^2 \Pi_{\rho \pi}^2 \right]^2 \times
\left[ q^2 -m_\rho^2 + \Pi_{\rho L} \right] \times \nonumber \\
& & \times \left[ q^2 - m_\pi^2 -{q^2 \Pi_{\pi Q} \over  1 + g' \Pi_{\pi Q}} 
\right] \times (1 +g' \Pi_{\pi L}) \times {(1 +g' \Pi_{\pi Q}) \over 
g' [ q^2 -g' (m_\pi^2 -q^2) ]}
\label{detdispHor}
\eeq
By inversion of (\ref{dispHor}), we obtain the propagators
\beq
G_\rho^{(H)\mu\nu} &=& -G_{\rho L}^{(H)} L^{\mu\nu} - G_{\rho T}^{(H)} T^{\mu\nu}
- G_{\rho Q}^{(H)} Q^{\mu\nu} \nonumber \\
  & &  G_{\rho L}^{(H)} = {1 \over q^2 -m_\rho^2 + \Pi_{\rho L}} \\
  & &  G_{\rho T}^{(H)} = {1 \over q^2 -m_\rho^2 + \Pi_{\rho T} -\eta^2 q^2
             g' \displaystyle{\Pi_{\rho\pi}^2 \over 1 +g' \Pi_{\pi T}}} 
 \label{GrTHor} \\
  & &  G_{\rho Q}^{(H)} = {-1 \over m_\rho^2}\\ 
G_{\rho\pi}^{(H)\mu} &=&  G_{\rho\pi}\, \epsilon^{\mu\nu\rho\lambda} 
q_\rho \eta_\lambda \nonumber \\
  & & G_{\rho \pi}^{(H)}= {-\Pi_{\rho\pi} \over (q^2-m_\rho^2+\Pi_{\rho T}) 
( \displaystyle{1 \over g'} +\Pi_{\pi T}) -\eta^2 q^2 \Pi_{\rho \pi}^2} \\
G_\pi^{(H)\mu\nu} &=& -G_{\pi L}^{(H)} L^{\mu\nu} - G_{\pi T}^{(H)} T^{\mu\nu}
- G_{\pi Q}^{(H)} Q^{\mu\nu} \nonumber \\
  & &  G_{\pi L}^{(H)} = {g' \over 1 + g' \Pi_{\pi L}} \\ 
  & &  G_{\pi T}^{(H)} = { (q^2 -m_\rho^2 + \Pi_{\rho T}) \over
(q^2 -m_\rho^2 + \Pi_{\rho T}) ( \displaystyle{1 \over g'} +  \Pi_{\pi T} )
- \eta^2 q^2  \Pi_{\rho \pi}^2 } \\
  & &  G_{\pi Q}^{(H)}=  { [ g' (m_\pi^2 -q^2) +q^2 ] \over
   (m_\pi^2 -q^2) (1+g' \Pi_{\pi Q}) +q^2 \Pi_{\pi Q}}
\eeq 

Using the relations between the polarizations $\Pi_{a \rho}$, $\Pi_{\pi\rho}$,
$\Pi_{a L}$, $\Pi_{\pi L}$, it can be checked that one can go from the
equations for the $\rho$-$a_1$-$\pi$ system to those resulting from the
Ansatz of Horowitz by taking the limit $q^2/m_a^2 \rightarrow 0$
and replacing $(g_a/m_a)^2 (m_\pi/f_\pi)^2$ by $-g'$, so that we see again
that the mixing with the $a_1$ meson acts as a Landau-Migdal term with
the opposite sign.

\subsection{Contact term}
\label{contact}

The procedure consisting of adding a contact term to the Lagrangian 
is straightforward
\beq
{\cal L} \ni \overline\psi \left[ {i \over 2} \overleftrightarrow
     \partial -m -{f_\pi \over m_\pi} \gamma_5 \gamma^\mu \partial_\mu
     \vec\pi.\vec\tau \right] \psi - g_A (\overline\psi \gamma_5
     \gamma_\mu \vec\tau\psi). (\overline\psi \gamma_5 \gamma^\mu\vec\tau\psi)
\eeq
The new term does not directly modify the field equations for
the pi meson. Rather, it contributes an additional pseudovector
term to the evolution equation for the Wigner function.
Following the derivation method used above, after neglecting all
correlations, we obtain at first order
\beq
&& \left[ \gamma^\mu (p_\mu - {q_\mu \over 2}) -m  \right] F_1(x,p)
\ni \left[ i {f_\pi \over m_\pi} \gamma_5 \gamma^\mu q_\mu \vec\tau.\vec\pi_{1}(q) 
+ 2 g_A \gamma_5 \gamma^\mu \vec\tau.\vec P_1^\mu (q) \right] F_H(p + {q \over 2})
\\
&& \left[ -q^2 + m_\pi^2 \right] \vec\pi_1 = i {f_\pi \over m_\pi} 
\vec P_1^\mu q_\mu
\eeq
with
\beq
\vec P_1^\mu(q) = \int d^4 k\ {\rm Tr}\left[ \gamma_5 \gamma^\mu \vec\tau F_1(q,k) 
\right]
\eeq 
Taking the trace of Eq. (\ref{wign1}), where $\Phi(q)$ should now
be replaced by $\Phi(q) +2 g_A \gamma_5 \gamma_\mu \vec\tau \vec P_1^\mu(q)$, 
with $\gamma_5 \gamma^\mu$ and integrating, yields a self consistent equation 
for $P_1^\mu$
\beq
\vec P_1^\mu(q) = i {f_\pi \over m_\pi}\, \Pi^{\mu\nu}_{\tim{AA}} q_\nu\, 
\vec\pi_1(q) + 2 g_A\, \Pi^{\mu\nu}_{\tim{AA}}\, \vec P_{1\, \nu}(q)
\eeq
The polarization 
\beq
\Pi^{\mu\nu}_{\tim{AA}} &=& \int d^4 p\ {\rm Tr}\left[\gamma_5 
\gamma^\mu \tau G(p-{q\over 2})  \gamma_5\gamma^\nu \tau F_H(p + {q \over 2}) 
+ (G \leftrightarrow F_H) \right] \nonumber \\
&=& -\Pi^Q_{\tim{AA}} Q^{\mu\nu} -\Pi^T_{\tim{AA}} T^{\mu\nu} 
-\Pi^L_{\tim{AA}} L^{\mu\nu}
\nonumber
\eeq
has already been met before.
The equation is easily solved as 
\beq
\vec P_1^\nu q_\nu = i \left({f_\pi \over m_\pi}\right) {q^2 \Pi^{Q}_{\tim{AA}} 
\over 1 -2 g_A  \Pi^{Q}_{\tim{AA}}}\ \vec\pi_1
\eeq
Replacing this solution in the first order term of the field equation 
for the pion, we obtain the dispersion relation of the pion modified 
by a contact term
\beq
\left[ q^2 - m_\pi^2 - \left( {f_\pi \over m_\pi}\right)^2
{q^2  \Pi^{Q}_{\mbox{\tiny AA}} \over 1 -2 g_A  \Pi^{Q}_{\tim{AA}}}
\right] =0
\eeq
With $ -(f_\pi/m_\pi)^2 q^2 \Pi^{Q}_{\tim{AA}}=\Pi_{\tim{PV}}$, and
defining $2 g_A (m_\pi/ f_\pi)^2 = g'$, we recover the standard expression 
for the pion pseudovector polarization modified by the Landau-Migdal 
$g'$ parameter.
\beq
\widetilde\Pi_{\pi} = - \left( {f_\pi \over m_\pi}\right)^2
{q^2  \Pi^{Q}_{\mbox{\tiny AA}} \over 1 -2 g_A  \Pi^{Q}_{\tim{AA}}}
= {\Pi_{\tim{PV}} \over q^2 - g' \Pi_{\tim{PV}}}
\label{PiPtilde}
\eeq

Apart from redefining the pion polarization, the $g_A$ contact term
also mixes with the transverse part of the $\rho$ meson. The result
is given in Eq. (\ref{PiRtilde}). Let us nevertheless study before 
a more general choice of the contact interaction. 
The rho meson exchange potential also gives rise to a delta
function in the coordinate representation of the potential
(see eq. (\ref{deltapot})). This delta can be eliminated independently 
from the one appearing in the pion potential if we add another contact 
term  for the $\rho$ meson
\beq
{\cal L} \ni \overline\psi \left[ {i \over 2} \overleftrightarrow
       \partial -m - g_\rho \gamma^\mu \vec\rho_\mu.\vec\tau \right] 
       - {1 \over 4} R_{\mu\nu} R^{\mu\nu} + {1 \over 2} m_\rho^2 
       \rho_\mu \rho^\mu - g_R (\overline\psi\gamma_\mu \vec\tau\psi). 
        (\overline\psi\gamma^\mu \vec\tau\psi) 
\eeq
When solving the equation for the Wigner function, there is an 
additional contribution to the isovector density.
At first order of the linearization procedure, we have 
\beq
&& \left[ \gamma^\mu (p_\mu - {q_\mu \over 2}) -m - \gamma^\mu 
(g_\rho \rho_{H\, \mu} + 2 g_R I_{H\, \mu})\tau_3 \right] F_1(x,p) 
\nonumber \\
&& \qquad \qquad = \left[ g_\rho \gamma^\mu \vec\tau \vec\rho_{1\, \mu} (q) 
+ 2 g_R\gamma^\mu\vec\tau.\vec R_{1\, \mu}(q) \right] F_H (p+{q \over 2}) + ... 
\eeq
with
\beq
\vec R_{1}^\mu(q) =\int d^4 k\ {\rm Tr}\left[\gamma^\mu \vec\tau F_1(q,k) \right]
\eeq
In symmetric matter, the mean value of the rho Hartree field and 
the contribution to the isospin current 
$I_H^\mu=\int d^4 p\ Tr[\gamma^\mu \tau F_H]$ vanish.
$\vec R_1^\mu(q)$ obeys the self-consistent equation
\beq
\vec R_1^\mu (q)= - g_\rho\, \Pi_{\tim{VV}}^{\mu\nu}\, \vec\rho_{1\, \nu} 
-2\, g_R\,  \Pi_{\tim{VV}}^{\mu\nu}\, \vec R_{1\, \nu} 
\eeq
where $\Pi_{\tim{VV}}^{\mu\nu}$ is the vector contribution to the
$\rho$ polarization $\Pi_{\rho}^{\mu\nu} = g_\rho^2  \Pi_{\tim{VV}}^{\mu\nu}
+ 2 g_\rho f_\rho \Pi_{\tim{VT}}^{\mu\nu} + f_\rho^2 \Pi_{\tim{TT}}^{\mu\nu}$. 
It is solved by projecting this relation on $T^{\mu\nu}$, $L^{\mu\nu}$ 
and $Q^{\mu\nu}$.
In the general case that pion, rho and $a_1$ fields are present,
and we moreover introduce both the contact terms $g_A (\overline\psi \gamma_5
\gamma_\mu \tau \psi)^2$ and $g_R (\overline\psi \gamma^\mu \tau \psi)^2$,
the various conponents obey coupled equations. As before, the longitudinal 
modes decouple, whereas the transverse mode of the $a_1$ is mixed with
the $\rho$ and the pion mode is modified by the part of the $a_1$ polarization
which is parallel to $q^\mu q^\nu$. We arrive at the dispersion relations

\beq
&& \left[ q^2 - m_\pi^2 - \left( {f_\pi \over m_\pi} \right)^2  
{ \Pi_{\tim{AA}}^Q q^2 \over 1 - \left(2 g_A + (g_a^2/m_a^2)\right) 
\Pi_{\tim{AA}}^Q} \right] \vec\pi_1(q) =0  \\ 
&& (q^2 - m_\rho^2 + \widetilde\Pi_{\rho\, T}) T^{\mu\nu} \vec\rho_{1\, \nu}
+ \widetilde \Pi_{\rho a\, T} T^{\mu\nu} \vec a_{1\, \nu} =0  \\
&& (q^2 - m_a^2 + \widetilde\Pi_{a\, T}) T^{\mu\nu} \rho_{1\, \nu}
+ \widetilde \Pi_{a \rho\, T} T^{\mu\nu} \vec\rho_{1\, \nu} =0  \\
&& (q^2 - m_\rho^2 + \widetilde \Pi_{\rho\, L} ) L^{\mu\nu} \vec\rho_{1\, \nu} 
=0 \\
&&  (q^2 - m_\rho^2 + \widetilde \Pi_{a\, L} ) L^{\mu\nu} \vec a_{1\, \nu} =0 
\label{piarhogAgR}
\eeq

The explicit form of the polarizations corrected by the contact terms
$g_A$, $g_R$ is given in Appendix B. When $g_R$=0, the polarizations
of the $\rho$ reduce to
\beq
\widetilde\Pi_{\rho T} &=& \Pi_{\rho T} + {2 g_A (\Pi_{\tim{AR}})^2 
\over 1-2 g_A \Pi_{\tim{AA}}^T} \nonumber \\
\widetilde\Pi_{\rho L} &=& \Pi_{\rho L}
\label{PiRtilde}
\eeq

The equations (\ref{PiPtilde},\ref{PiRtilde}) are in fact the same as 
obtained from Horowitz propagator replacement Ansatz (see the third factor 
in eq. (\ref{detdispHor}) and eq. (\ref{GrTHor})). One should notice
however that the later Ansatz, apart from producing the desired 
result for the pi and transverse rho propagators, also introduces
additional spurious terms in the dispersion relation (\ref{detdispHor})
arising from the unphysical part of the new ``pion'' propagator 
orthogonal to $q^\mu q^\nu$ like {\it e.g.} $ 1 +g' \Pi_{\pi T}$ 
and $ 1 +g' \Pi_{\pi L}$.  

\section{Dispersion relations: numerical results}

In this section we study the dispersion relations which result from
the models described in the previous sections. We will concentrate 
on the $\delta$-$\rho$-$\pi$-$a_1$ sector, since the 
$\sigma$-$\omega$ branches have been studied extensively 
elsewhere.  

\subsection{Model parameters}
\label{sectparam}

When not explicitely stated otherwise, the parameters 
are chosen to be those of the Bonn potential \cite{BonnPot}. 
They are summarized in the following table. 

\begin{center}
\begin{tabular}{|c|c|c|c|}
\hline
meson $\alpha$ & $m_\alpha$  [MeV] &  $\displaystyle{g_\alpha^2 \over 4 \pi}$ &
$\Lambda_\alpha$ \\
\hline
\hline
$\sigma$ & 550. & 8.2797 & 2000. \\
\hline
$\omega$ & 782.6 & 20 & 1500. \\
\hline
$\pi$ & 138. & 14.6 & 1300. \\
\hline
$\rho$ & 769. & 0.81 & 2000. \\
  & & ( $f_\rho/g_\rho$=6.1) & ($n_\rho$ =2) \\
\hline
$\delta$ & 983. & 1.1075 & 2000. \\
\hline
$\eta$ & 548.8 & 5 & 1500. \\
\hline
\end{tabular}
\end{center}

The coupling of the pion is taken in this Bonn model to be 
$g_\pi^2/(4 \pi)=14.6$. One usually prefers now a somewhat lower
value $g_\pi^2/(4 \pi) \simeq 13.67$ so that $f_\pi =m_\pi g_\pi/(2 m)
\simeq 0.965$. An other issue is the value of the cutoff, which
is taken to be 1300 MeV in the Bonn model, whereas most calculations
and experimental determinations of this parameter would give
$\Lambda_\pi \simeq 800$ MeV.

Data for the coupling of the $a_1$ meson to the nucleon is scarce. 
A simple realization of the chiral symmetry is $g_a =g_\rho$. From 
early chiral models by Weinberg and Wess \& Zumino \cite{W67-WZ67-GG69} 
one obtains the relation $g_a=m_a (f_\pi/m_\pi)$. With $m_a$=1260 MeV 
and $f_\pi \simeq 0.965$ as determined from NN scattering data, this 
yields $g_a \simeq 8.79$. More recent implementations of the chiral 
model \cite{KR94,SR97} point out the difficulty in adjusting all
known data on the width and mass of the $a_1$ and quote values
ranging from 3.8 to 18 \cite{SR97}. Another value used in the 
literature is $g_a= 6.44$ \cite{DBS84}. Reference \cite{ATG00}
also provides us with a rather low value for the cutoff 
$\Lambda_{a NN}= 809$ MeV. Here we will work with $g_a = 8.8$ and 
a higher value of the cutoff $\Lambda_{a NN}= 2000$ MeV.

\subsection{Longitudinal $\delta$-$\rho$ mode} 

The dispersion relations were calculated with the parameters of 
Machleidt's Bonn potential \cite{BonnPot} using three alternative 
renormalization schemes. As discussed {\it e.g.} in \cite{MGP01}, 
where the dispersion relation of the rho was studied without mixing,
the polarizations contain a diverging contribution from
vacuum fluctuations which has to be renormalized. One may
define several subtraction schemes, which, unfortunately, 
lead to very different behaviors of the mesons effective masses.
In \cite{MGP01} two classes of renormalization procedures
were identified
\smallskip
\li {\bf -- } a first class which renormalizes divergences of the form
$M^2/\epsilon$ by subtracting a general counterterm
$ A_0 + A_1 \sigma + A_2 \sigma^2$, and adjust the constants
so that the polarization $\Pi$ and its derivatives $ (\partial \Pi 
/ \partial \sigma)$, $(\partial^2 \Pi / \partial \sigma^2)$ 
vanish at some point. In this way, one tries to minimize the 
effect of introducing new couplings with the sigma field in the 
counterterm Lagrangian, which are not present in the original 
physically motivated Lagrangian. In references \cite{DP91,DPS89}, 
the renormalization point is chosen to be the mass shell 
$q^2=m_\alpha^2$ for the polarization as well as its derivatives. 
In \cite{KS88} Kurasawa and Suzuki subtract the polarization at 
$q^2 = m_\alpha^2$ but the derivatives at $q^2=0$. 
Using this procedure, the $\omega$ meson mass decreases with 
density, however the $\rho$ mass increases due to the tensor
coupling $f_\rho$.
\smallskip
\li {\bf -- }  a second class which preserves the structure of 
contributions $M^2/\epsilon = (m -g_\sigma \sigma)^2/\epsilon$
by subtracting a countertern in the form  $ A (m -g_\sigma \sigma)^2$. 
$A$ is determined by setting $A=0$ at some point (at $q^2=m_\alpha^2$)
Since we have only one parameter $A$, it is not possible to minimize 
the new couplings to the $\sigma$ field introduced in the counterterm 
Lagrangian.
Using this procedure, the $\rho$ meson mass decreases with density.
\smallskip

The reader is referred to \cite{MGP01} and Appendix A.2 for further details.
Once a renormalization procedure is chosen, it should be used for all mesons.

The first renormalization scheme (scheme A) used in this paper belongs 
to the first class for all mesons. For the $\rho$ meson, it is the 
``scheme 2'' described in the Appendix C of \cite{MGP01}. For the 
$\delta$ meson, it is obtained by replacing everywhere $m_\sigma$ 
and $g_\sigma$ by $m_\delta$ and $g_\delta$ in the expression for 
the vacuum polarization of the $\sigma$ given in Ref. \cite{DPS89}

The second renormalization scheme (scheme B) belongs to the second 
class for all mesons. For the $\rho$ meson, it is the ``scheme 3''
described in the Appendix C of \cite{MGP01}.

A third renormalization scheme (scheme C) is used for comparison 
with calculations published in the litterature \cite{TDMG00}. For
the rho meson, it uses as in \cite{TDMG00} the procedure of 
Shiomi and Hatsuda \cite{SH94}, which according to the 
classification given above, belongs to the second class. 
Moreover, the polarization obtained in this paper subtract 
the vacuum polarization at all $k$, so that it vanishes identically 
in the vacuum. In contrast, the vacuum polarization of schemes 
A and B does not vanish away  from the mass shell. For the 
$\delta$ meson, it uses as in \cite{TDMG00} the procedure of
 Kurasawa and Suzuki (KS) (thus belonging 
to the first class defined above) and then subtract the vacuum 
polarization at all $k$, $\Pi_{vac}(M,k) = \Pi_{KS}(M,k)
 - \Pi_{KS}(m,k)$.

Several parameter sets were tried besides those given in the 
table of \S \ref{sectparam}. In all cases, we find normal mode 
branches for the $\delta$ and longitudinal $\rho$ (see Fig. 1). 
For the $\rho$, there are moreover two heavy meson branches. 
Contrary to the very similar $\sigma$-$\omega$ system, no zero 
sound mode was found.

Effective masses can be defined as usual as the solution(s) 
$m_i^*= \omega$ of the dispersion relation $D(\omega,\vec k)=0$
at $\vec k=0$. With the renormalization scheme A, the $\rho$ 
meson mass increases with density and the $\delta$ meson mass 
also slightly increases (see Fig. 2). With renormalization 
scheme B, the $\rho$ meson mass first decreases, reaches a 
minimum at $\sim 1.8\ \rho_0$ and then slowly increases, whereas 
the delta meson mass increases reaches a maximum at 
$\sim 0.8\ \rho_0$ and then decreases. 
The $\rho$ and $\delta$ masses are almost constant with temperature 
(see Fig. 3) with both $m_\rho^*$  and $m_\delta$ slightly decreasing.
This figure was plotted using renormalization scheme A.
In scheme C we reproduced the results of \cite{TDMG00}, with both
the rho and delta masses decreasing as a function of density.

This illustrates again the difficulties met with the standard
renormalization techniques, which preclude a reliable prediction
of the behavior of the effective meson masses in the medium.
One cannot simply drop the contribution of the vacuum term by 
performing a normal ordering, since the contribution left out 
depends on the density through the effective mass of the nucleon. 
Moreover, pathologies appear in the dispersion relation (kinks, 
no clean normal modes) if one attempts to do so 
\cite{DPS89,MGP01,LH89}. 
One could in principle apply a subtraction procedure at the 
one-loop order, however the various schemes used in the litterature 
\cite{DP91,KS88,SH94} lead to widely different results. This is 
related to the fact that we are dealing with an effective theory 
which should enforce the scalings and symmetries of the underlying 
more fundamental theory, whereas the approximation made to the 
full many-body theory (here, RPA) blurs these concepts.
There is some hope that one could solve the problem by
applying ``naturalness'' and symmetry arguments \cite{Furnstahl,LP01}. 
At the time of writing however, we must consider this a 
still unsolved problem.

\vskip 0.2cm

One sometimes defines a mixing angle \cite{DM97} by
\beq
\theta_{\delta\rho L} = {1 \over 2} \arctan \left[ { 2 \sqrt{|\eta^2 |} 
\Pi^{\rho\delta} \over m_\delta^2 - m_\rho^2 - \Pi_{\rho L} 
+ \Pi_\delta} \right]
\eeq
which is obtained from the diagonalization of the mass matrix
for the mixed dispersion relation of two mesons $A$ and $B$
in the timelike region $q^2 >0$, $\eta^2 < 0$:
\beq
{\cal M} &=& \left[ \matrix{ m_A^2 - \Pi_A & - \sqrt{|\eta^2|} \Pi_{AB} \cr
- \sqrt{|\eta^2|} \Pi_{AB} & m_B^2 - \Pi_B \cr} \right]  \\
\noalign{\medskip}
{\rm det}\ \bigl[ q^2 I - {\cal M} \bigr] = 
&& (q^2-m_A^2 +  \Pi_A) (q^2-m_B^2 +  \Pi_B) + \eta^2  \Pi_{AB}^2
\nonumber 
\eeq

The mixing angle is calculated at the solutions of the dispersion 
relation, there is therefore one for each branch. 
The mixing angle is represented for renormalization schemes A and B 
in Fig. 4 as a function of density at vanishing temperature for a momentum 
$k=300$ MeV. Schemes B and C yield values of $\theta_{\delta\rho L}$ of the 
order of a few degrees. The higher values found in scheme A results in fact 
no so much from the strength of the mixing, but rather from the fact that 
the difference of the effective masses of the $\delta$ and $\rho$ happens 
to be smaller in this scheme.

The behavior with $k$ is non monotonous. It is represented in Fig. 5 in
renormalization scheme B at $T=0$ and $n_B=3\ n_{\rm sat}$.
We also calculated the effect of finite temperature. The mixing angle 
decreases with $T$ in all renormalization schemes. This might reduce the 
effectiveness of $\delta$-$\rho$ mixing as a mechanism invoked by
the authors of \cite{TDMG00} for dilepton production.

\subsection{Transverse $a_1$-$\rho$ mode and longitudinal $a_1$}

Here we investigate the dispersion relation of the $a_1$ meson, 
and in what measure does the mixing with the $a_1$
meson modify the transverse $\rho$ dispersion relation.

The dispersion relation for the longitudinal $a_1$ mode is 
represented on Fig. 6 with renormalization schemes A and B.
Only normal branches appear.

The $a_1$-$\rho$ mixing vanishes at $\vec k$=0. It therefore does not 
affect the effective masses, and $m_a^*$ coincides as calculated
from the transverse or longitudinal dispersion relations. 
The $a_1$ effective mass is plotted on Fig. 7. With renormalization 
scheme A, it decreases slightly with increasing density. When using
renormalization scheme B, a non monotonous behavior is obtained for 
the $a_1$ mass, which first presents a steep increase at low density, 
and then decreases again, in contradiction with the behavior expected 
from QCD sum rules \cite{HKL93,P95,DEI90}. In the case of the $a_1$ 
meson therefore, the renormalization scheme A seems in better agreement 
with the results expected from other QCD-based models, whereas we saw 
that for the $\rho$ meson, scheme B would have seemed preferable. 
A third renormalization scheme was also tried (let us call it scheme C), 
using the same procedure as for the $\delta$ in \cite{TDMG00},
(that is, subtracting the vacuum at all $k$ from the expression
obtained from the procedure of Kurasawa and Suzuki, see previous 
section and Appendix A.2). In this scheme, the effective mass of the 
$a_1$ decreases. The behavior of $m_a^*$ as a function of temperature
was also investigated; for all renormalization schemes it is almost
constant as a function of temperature.

At finite $k$, the $a_1$-$\rho$ mixing sets on. It does not
appreciably affect the position of the normal branches. Nevertheless,
it amplifies somewhat the (spurious) zero sound branch which may appear 
in the transverse part of the dispersion relation at high density.
This mode corresponds to the $\rho$ meson and is due to the 
high value of the tensor coupling $f_\rho/g_\rho=6.1$ of the Bonn 
potential. It is weaker and appears at a higher density with the 
vector dominance value $f_\rho/g_\rho=3.7$  and disappears completely 
if $f_\rho$=0. Note however that a high value of $f_\rho$ is also 
supported by QCD sum rule calculations \cite{Zhu}. This mode is 
present in both schemes a high enough density, but is stronger in 
scheme A. Such a mode would lead to divergences, or at least an 
enhancement of Friedel oscillations of the $\rho$ contribution 
to the screened nucleon-nucleon potential \cite{MGP01}. 
It could  be eliminated if a strong cutoff is applied to the $\rho$
meson, of the order of 1200 - 1300 MeV. An other possibility
is to use the contact term introduced in section \S \ref{contact}, 
which we will need anyway for the pion. It was checked that 
the spurious zero sound branch appearing in the $\rho$ dispersion
relation is removed at all densities by the contact term
used in next section.

The strength of the $a_1$-$\rho$ mixing can be estimated by 
calculating the mixing angle, as explained in the previous section.
In Fig. 8 it is shown at $T=0$ and $k$= 300 MeV.

\subsection{Pion mode}
\label{pion}

If the dispersion relation is calculated with pure pseudovector coupling,
and neither mixing with the $a_1$ meson nor Ans\"atze for short range
corrections (\`a la Horowitz or Landau-Migdal contact term) are taken
into account, a zero sound mode is found already at saturation density
in renormalization scheme A. The short range corrections must be added
since such a mode is not observed. Before we pass to examine this point,
let us first make a few more observations. 

Contrary to the non relativistic case, the zero sound mode is 
found to disappear 
again at higher density. Such a behavior was already noticed by Dawson 
and Piekarewicz \cite{DawPiekcond}. The zero-sound problem is
in part due to the high value of the cutoff $\Lambda_\pi$ of the Bonn
potential model, whereas several studies favour a lower value 
$\Lambda_\pi= 800$ MeV. If we choose $g_\pi^2/(4 \pi)$ and 
$\Lambda_\pi=800$ MeV, there is only a tiny zero sound branch 
around $k=250$ MeV, between $n_B = 0.72\ n_{\rm sat}$ and $n_B=1.2\ 
n_{\rm sat}$.

In renormalization scheme B, no zero sound mode is found when 
pure pseudovector coupling of the pion is used without mixing 
with the $a_1$. However this renormalization scheme appears less
favorable when we consider the behavior of the effective pion
mass. Whereas the pion mass stays approximately constant with
scheme A, it strongly decreases with density when scheme B is used.

In both cases, it is possible to adjust the effective mass at 
saturation density as measured in recent experiments \cite{mefpi}
by adding a small admixture of  pseudoscalar coupling
( ${\cal L} \ni \overline\psi\left( i \alpha g_\pi \gamma_5 \vec\pi . 
\vec \tau\right.  $ 
$ \left. + (1-\alpha) (f_\pi/m_\pi) \gamma_5 \gamma^\mu \partial_\mu 
\vec\pi \vec \tau \right) \psi + {1 \over 2} g_{\sigma\pi\pi} m_\sigma 
\sigma \pi^2$ \cite{DMP01}. 
Such an admixture has been considered from the theoretical 
\cite{NFSLQMP98} as well as phenomenological \cite{GLMBG94,Gross}
point of views. Gross {\it et al.} \cite{Gross} could allow a value 
as high as 25 \% of PS admixture whereas Goudsmit {\it et al.} 
\cite{GLMBG94} found an optimal value of 3.5 \%.
 
The $\sigma \pi^2$ term must be introduced in order
to compensate the contribution of the pseudoscalar coupling to
the pion-nucleon scattering. The $\pi-N$ scattering length 
can be reproduced 
when choosing $g_{\sigma\pi\pi}= \alpha^2 (g_\pi^2/g_\sigma) m_\sigma$.
With this value, the pion mass can be adjusted to the value extracted
from measurements of the pionic atom 
\cite{mefpi} $m_\pi(n_{\rm sat}) = 1.1 \pm 0.03$,
with $\alpha=0.0735$ in renormalization scheme A and $\alpha=0.118$
in renormalization scheme B.

As was to be expected, the zero sound branches are amplified when
the mixing with the $a_1$ meson is taken into account, since it 
modifies the pion dispersion relation formally as would a Landau-Migdal
contact term but with the opposite sign. Then it is present in 
renormalization scheme B as well. 
The zero sound branch can be eliminated as before by adding a contact term.
The relevant formula when a PS admixture and a $\sigma \pi^2$ term are
also present is 
\beq
&& q^2 -m_\pi^2 - g_{\sigma\pi\pi} m_\sigma \sigma + \widetilde\Pi_\pi=0
\nonumber \\
&& \widetilde \Pi_{\pi} = \left[ \left( {f_\pi \over m_\pi} \right)^2 
\Pi_{\tim{PV}} + g_\pi^2 \Pi_{\tim{PS}} +2 g_\pi\, {f_\pi \over m_\pi} \,
\Pi_{\rm mix} \right] - { \left( \displaystyle{ g_a^2 \over m_a^2} +2
g_A \right) \left( g_\pi \Pi_{\pi a}^{\tim{PS}\, \mu} q_\mu + 
\displaystyle{f_\pi \over m_\pi} \Pi_{\pi a}^{\tim{PV}\, \mu} q_\mu
\right)^2 \over q^2 +  \left( \displaystyle{ g_a^2 \over m_a^2} -2
g_A \right) \Pi_{\tim{PV}}} \nonumber \\
&& \Pi_{\rm mix} = {1 \over 2 M} \Pi_{\tim{PV}}\ ,\quad \Pi_{\tim{PS}}
= \left({1 \over 2 M}\right)^2 \Pi_{\tim{PV}} 
-8 \int d^4\, p\ \varphi(p)
\nonumber \\
&& \Pi_{\pi a}^{\tim{PV}\, \mu} q_\mu = \Pi_{\tim{PV}} \ , \quad
\Pi_{\pi a}^{\tim{PS}\, \mu} q_\mu =  {1 \over 2 M}
\Pi_{\tim{PV}}
\eeq
(For clarity we draw the coupling constants out of the definition 
of the polarizations in this equation.)  
Mixing with the $a_1$ can somewhat increase the effective pion mass.
This effect is less important after including the 
pseudoscalar admixture.

The Horowitz Ansatz was also studied. The part of the dispersion relation
which corresponds to the pion mode has the required form of a pseudovector
coupling with a short range correction of the Landau-Migdal contact type. 
This removes the zero sound branch which appeared there for a pure
pseudovector coupling. However, this Ansatz does also affect the 
rho meson mode through mixing with the fictitious "pion" introduced
there. It is seen in dispersion relation Eq. (\ref{dispHor}) that there 
are also factors $1 + g' \Pi_{\pi T}$, $1 + g' \Pi_{\pi L}$ .
It was found that these terms give spurious branches in the
spacelike region whatever renormalization scheme was used. The most 
favorable case was that of scheme A, which removes the branch due to 
$1 + g' \Pi_{\pi L}$  entirely and part of that due $1 + g' \Pi_{\pi T}$  
at low density. These branches are an artefact
of the unphysical components of the auxiliary pseudovectorial
``pion'' and should be removed carefully by hand at the end of the 
calculation. We therefore prefer using the contact Lagrangian
of section \S \ref{contact}, since this is free from such
problems from the onset, moreover it permits easily to allow
for a PS admixture, whereas Horowitz Ansatz can only be
implemented for a pure pseudovector coupling.

\section{Conclusions}

This work gathers and extends results on the dispersion retations of 
mesons in relativistic nuclear matter at high density and finite
temperature, as obtained in the RPA approximation to the 
quantum hadrodynamics. Besides the ``standard'' mesons $\sigma$,
$\omega$, $\pi$, $\rho$, results for additional mesons $\delta$,
$a_1$ are given. Both mix with the $\rho$ meson, the $\delta$ with 
the longitudinal mode and the $a_1$ with the transversal mode.
They therefore may represent an additional source of modification 
of the dilepton production at finite density and temperature.
Also, new interest arises in the $\delta$ meson in the context of the 
description of asymmetric matter in density dependent mean field 
theories \cite{dJL98}.

Various ways of introducing short range corrections in order to 
eliminate unobserved zero sound modes at saturation density were examined.
It is seen that the $a_1$ meson mixes with the pion so that it acts
exactly as would a Landau-Migdal contact term, but with the "wrong" sign.
The Ansatz by Horowitz comes with the right sign, however it introduces
spurious branches in the transverse channel which badly affect the 
dispersion relation in the spacelike region. A simple standard 
contact interaction in the Lagrangian does the job best.

It was shown that the results are affected by the way the renormalization 
is performed in order to regulate the high momenta divergences. Without 
any renormalization, there appear kink structures in the effective meson 
masses as a function of density, no matter what a strong cutoff is applied. 
The expected branches are recovered when applying a renormalization 
procedure. Among the possible subtraction procedures, two schemes A and B 
were introduced and their predictions compared. Such important results 
as the behavior of the effective meson masses or the presence of zero 
sound modes differ widely whether one or the other scheme is used.
In order to be consistent, a same scheme should be used for all mesons. 
Schemes of the A class have been used in previous litterature to study 
$\sigma$, $\omega$ and $\pi$ mesons. Scheme A yields results more in 
agreement with other models for the behavior of the $\pi$, $\sigma$
and $a_1$ masses, but would predict an increasing $\rho$ meson mass.
On the other hand, schemes of the B class which give a decreasing 
rho meson mass, as favored by theoretical models and experimental data,
would predict strongly increasing $\sigma$ and $a_1$ masses, and
a rapidly dropping pion mass. This last result stands in strong 
disagreement with the expected behavior of a Goldstone boson and
the recent experimental determination of \cite{mefpi}.

A renormalization procedure which would respect the scalings and
symmetries \cite{Furnstahl} of the underlying more fundamental 
theory of which the effective Lagrangian means to be a low energy 
approximation  is clearly needed.

\vskip 1cm

\noindent{\Large{\bf Acknowledgements}}

The author gratefully acknowledges enlightning discussions with 
J. Diaz Alonso. This work was supported in part by the spanish grant 
n$^{\underline{\rm o}}$ MCT-00-BFM-0357.

\section*{Appendix A: Polarizations}

There are contributions from the meson self interaction terms and from
the particle hole insertions

The (retarded) particle hole insertions are given by
\beq
\Pi^{AB}(p) = \int d^4 p\ {\rm Tr} \left[ \Gamma^A S(p-{k\over 2})
\Gamma^B S(p+{k \over 2}) \right]  \left\{ {\varphi (p+{q \over 2}) 
- \varphi (p-{q \over 2}) \over  2 p.q -i\, \epsilon\ {\rm sign}( p_0)} 
\right\}
\eeq
where 
\beq 
S(p)&=&\gamma.P+M, \nonumber \\
\varphi(p) &=& {d \over (2 \pi)^3}\, \delta(P^2 -M^2) 
\left[ \theta(p_0) n(p) + \theta(-p_0) \overline n(p) 
- \theta(-p_0)  \right], \nonumber
\eeq
$d=2$ is the isospin degeneracy,
and the $\Gamma^A$ are the vertices, respectively for the
$\sigma$, $\omega$, $\pi$, $\rho$, $a_1$ couplings to the nucleon
\beq
\Gamma^A \in (g_\sigma, g_\omega \gamma^\mu, i g_\pi \gamma_5,
i (f_\pi/m_\pi) \gamma_5 \gamma^\mu q_\mu,
g_\rho \gamma^\mu +{f_\rho \over 2 M} \sigma^{\mu\nu}q_\nu,
g_{a_1}\gamma_5 \gamma^\mu)
\eeq

As an example, we have
\beq
& & Tr[  \{ g_{a_1} \gamma_5 \gamma^\mu \} G_N(p-{k\over 2})
\{ g_{a_1} \gamma_5 \gamma^\nu \} G_N(p+{k \over 2})] = \nonumber \\
& & \phantom{fantome}
4 g_{a_1}^2 [(-p^2 -M^2 +{k^2 \over 4}) g^{\mu\nu} -{1 \over 2} k^\mu k^\nu +2
p^\mu p^\nu ] \\
& & Tr[ \{ g_\rho \gamma^\mu -{f_\rho \over 2 m} \sigma^{\mu\alpha}k_\alpha \} 
G_N(p-{k\over 2})  \{ g_{a_1} \gamma_5 \gamma^\nu \} G_N(p+{k \over 2})]  =
\nonumber \\
& & \phantom{fantome}
Tr[  \{ g_{a_1} \gamma_5 \gamma^\mu \} G_N(p-{k\over 2})
 \{ g_\rho \gamma^\nu + {f_\rho \over 2 m} \sigma^{\nu\alpha}k_\alpha \} 
G_N(p+{k \over 2})]  = \nonumber \\
& & \phantom{fantome}
 4 i g_{a_1} (g_\rho + f_\rho) \epsilon^{\mu\nu\alpha\beta} k_\alpha
p_\beta
\eeq

In symmetric nuclear matter, the polarization matrix have the components
\beq
\left[ \matrix{ \Pi_{\sigma\sigma} & \Pi_{\sigma\omega}^\nu 
                 & 0 & 0 & 0 & 0 & 0 \cr
                \Pi_{\sigma\omega}^\mu & \Pi_{\omega\omega}^{\mu\nu} 
                 & 0 & 0 & 0 & 0 & 0 \cr
                 0 & 0 & \Pi_{\eta\eta} &  0 &  0 &  0 &  0 \cr
                 0 & 0 & 0 & \Pi_{\delta\delta} & \Pi_{\delta\rho}^\nu 
                 &  0 &  0 \cr
                 0 & 0 & 0 & \Pi_{\delta\rho}^\alpha & 
                 \Pi_{\rho\rho}^{\mu\nu}  & \Pi_{a \rho}^{\mu\nu} 
                 &  0 \cr
                 0 & 0 & 0 & 0 & \Pi_{a \rho}^\mu & \Pi_{aa}^{\mu\nu} 
                 & \Pi_{a \pi}^\nu \cr
                 0 & 0 & 0 & 0 & 0 & - \Pi_{a \pi}^\mu 
                 & \Pi_{\pi\pi} \cr} \right]
\eeq
In asymmetric nuclear matter the mixing polarizations 
$\Pi_{\omega\rho}^{\mu\nu}$, $\Pi_{\sigma\rho}^{\mu}$,
$\Pi_{\sigma\delta}$, $\Pi_{\delta\omega}^{\mu}$, 
$\Pi_{\omega a_1}^{\mu \nu}$ ... would not vanish. 
Asymmetric matter is discussed in a companion paper \cite{M01b}.

The polarizations involving the $a_1$ meson are related to others by 
the following equations

\beq
\Pi_{aa}^{\mu\nu} &=& {g_a^2 \over g_\omega^2} \Pi_{\omega\omega}^{\mu\nu}
+ {g_a^2 \over (f_\pi/m_\pi)^2} {g^{\mu\nu} \over q^2} \Pi_{\pi\pi}^{\rm PV} 
\ \Rightarrow \Pi_{a Q} = - {g_a^2 \over (f_\pi/m_\pi)^2} {1 \over q^2}  
\Pi_{\pi\pi}^{\rm PV} \nonumber \\
\Pi_{a \pi}^\mu &=& i {g_a \over (f_\pi/m_\pi)} {q^\mu \over q^2} 
\Pi_{\pi\pi}^{\rm PV} 
\eeq

These relations are used to simplify the expression of the pion propagator
appearing in Eq. (\ref{Gpi})
\beq
\widetilde \Pi_\pi = \Pi_{\pi\pi} - q^2 {\Pi_{a \pi}^2 \over m_a^2 - \Pi_{a Q}}
= {q^2 \Pi_{\pi\pi} \over q^2 + \displaystyle{g_a^2\over m_a^2} 
\displaystyle{m_\pi^2 \over f_\pi^2} \Pi_{\pi\pi}}
\eeq

\vskip 0.5cm

\subsection*{A.1 - $\underline{\mbox{\sf Polarizations, real matter 
part}}$}   

\vskip 0.5cm

The matter part of the polarizations is given by
\beq
\Pi_\sigma^{\rm (mat)} &=&  - d\, {g_\sigma^2 \over 2 \pi^2} 
\left[ 2\, X_0 +\left( {q^4 \over 4} -M^2 q^2 \right) {\cal I}_0^0 
\right]  \\
\Pi_{\sigma\omega}^{\rm (mat)} &=& - d\, {g_\sigma g_\omega \over 
2 \pi^2}  M  q^2 \left[ {\cal I}_0^1 -{\omega \over k} {\cal I}_1^0 
\right] \eta^\mu  \\
\Pi_\omega^{\mu\nu}&=& -\Pi_{\omega\, L} L^{\mu\nu} -\Pi_{\omega\, T} 
T^{\mu \nu}\ ; \nonumber \\
\Pi_{\omega L}^{\rm (mat)} &=& - d\, {g_\omega^2 q^2 \over 2 \pi^2} 
\left[ {\cal I}_0^2 -{\cal I}_2^0 \right] \\
\Pi_{\omega T}^{\rm (mat)} &=& - d\, {g_\omega^2 \over 4 \pi^2} 
\left[ M^2 q^2 {\cal I}_0^0 +(\omega^2 + k^2)\left( {\cal I}_0^2 
+{\cal I}_2^0 \right) -4\, \omega k {\cal I}_1^1 \right] \\
\Pi_\delta^{\rm (mat)} &=& - d\, {g_\delta^2 \over 2 \pi^2} \left[ 2\, X_0
+\left( {q^4 \over 4} -M^2 q^2 \right) {\cal I}_0^0 \right]  \\
\Pi_{\delta\rho}^{\rm (mat)} &=& - d\, {g_\delta \over 4 \pi^2} \left(
2 M g_\rho + {f_\rho \over 2 m} q^2 \right) q^2 \left[ {\cal I}_0^1
-{\omega \over k} {\cal I}_1^0 \right] \eta^\mu  \\
\Pi_\rho^{\mu\nu}&=& -\Pi_{\rho\, L} L^{\mu\nu} -\Pi_{\rho\, T} 
T^{\mu \nu}\ ; \nonumber \\
\Pi_{\rho L}^{\rm (mat)} &=& - d\, {g_\rho^2 q^2 \over 2 \pi^2} 
\left[ {\cal I}_0^2 -{\cal I}_2^0 \right] - d\, \left( 
{f_\rho \over 2 m} \right) 
g_\rho {M q^4 \over  2 \pi^2} {\cal I}_0^0 \nonumber \\
& & -d\, \left( {f_\rho \over 2 m} 
\right)^2 {q^2 \over 2 \pi^2} \left[ M^2 q^2 {\cal I}_0^0 + 
k^2 {\cal I}_0^2 + \omega^2 {\cal I}_2^0 -2\, \omega k {\cal I}_1^1 
\right] \\
\Pi_{\rho T}^{\rm (mat)} &=& -d\, {g_\rho^2 \over 4 \pi^2} \left[ M^2 q^2 
{\cal I}_0^0 +(\omega^2 + k^2)\left( {\cal I}_0^2 +{\cal I}_2^0 \right) 
-4\, \omega k {\cal I}_1^1 \right] \nonumber \\
& & - d\, \left( {f_\rho \over 2 m} \right)^2
{q^4 \over 4 \pi^2} \left[ M^2 {\cal I}_0^0 + {\cal I}_0^2 -
{\cal I}_2^0 \right] - d\, \left( {f_\rho \over 2 m} \right) g_\rho {M q^4
\over 2 \pi^2} {\cal I}_0^0 \\
\Pi_{\pi\, {\rm\small PV}}^{\rm (mat)} &=& - d\, {q^4 M^2 \over 2 \pi^2}
\left( {f_\pi \over m_\pi} \right)^2 {\cal I}_0^0 \\
\Pi_{a_1}^{\mu\nu}&=& - \Pi_{a\, L} L^{\mu\nu} -\Pi_{a\, T} 
T^{\mu \nu} -\Pi_{a\, Q} Q^{\mu\nu}; \nonumber \\
\Pi_{a\ L}^{\rm (mat)} &=& d\, {g_a^2 q^2 \over 2 \pi^2} 
\left[ M^2 {\cal I}_0^0 -{\cal I}_0^2 +{\cal I}_2^0 \right] \\
\Pi_{a\ T}^{\rm (mat)} &=& d\, {g_a^2 \over 4 \pi^2} \left[ M^2 q^2 
{\cal I}_0^0 -(\omega^2 + k^2)\left( {\cal I}_0^2 +{\cal I}_2^0 \right) 
+4\, \omega k {\cal I}_1^1 \right] \\
\Pi_{a\ Q}^{\rm (mat)} &=& d\, {g_a^2 \over 2 \pi^2} q^2 M^2 {\cal I}_0^0 \\
\Pi_{a\rho}^{\mu\nu} &=& i d\, {g_a \over 4 \pi^2} \left( g_\rho + 2\, M\,
{f_\rho \over 2 m} \right) \epsilon^{\mu\nu\alpha\beta} 
q_\alpha \eta_\beta\ q^2 \left[ {\cal I}_0^1 -{\omega \over k} 
{\cal I}_1^0 \right] 
\eeq

The integrals ${\cal I}_m^n$ which appear in these expressions are
given by
\beq
{\cal I}_m^n(\omega, k) &=& \int_0^\infty dp\ p^2 \left( \sqrt{p^2+M^2}
\right)^{n-1} {\rm I}_m(p,\omega,k)\, \Bigl[ n(p) + (-1)^{m+n}\, \overline n(p) 
\Bigr] \nonumber \\
{\rm I}_m(p,\omega,k) &=& \int_{-1}^{+1} du\, {(p\, u)^m \over 
(w \sqrt(p^2+M^2) -p k u-q^2/2) (w \sqrt(p^2+M^2) -p k u+q^2/2)}
\eeq

\vskip 0.5cm

\subsection*{A.2 - $\underline{\mbox{\sf Polarizations, vacuum part}}$}

\vskip 0.5cm

The vacuum part of the polarizations comes from the $-\theta(-p^0)$
term in the definition of the relativistic Hartree approximation 
to the Wigner function. It is divergent but can be renormalized
by dimensional regularization and subtraction of appropriate counterterms.
The standard procedure is exposed in \cite{DP91,C77}. In the case
of the $\rho$ and $\pi$ mesons, the derivative coupling makes the
model nonrenormalizable in the usual sense. In practice however, 
a renormalization scheme can be defined {\it at a given order} of 
perturbation / loop expansion. One has the choice between several 
renormalization schemes, which were divided in \cite{MGP01} into 
two subclasses: the ``increasing rho mass'' and the ``decreasing 
rho mass''
classes, in reference to the much debated issue of the interpretation
of dilepton production in heavy ion collisions and the Brown-Rho
scaling conjecture.  To the former class belong earlier procedures 
by Chin or Kurasawa and Suzuki \cite{C77,KS88}. In the latter class we find
more recent procedures used by Shiomi and Hatsuda \cite{SH94,TDMG00}.
Both classes have advantages and drawbacks. In any case, a same scheme
should be used for all mesons. 

To the Lagrangian (\ref{TheLagrangian}) we will have to add the following
counterterm Lagrangian in order to perform the dimensional regularization 
and renormalization of diverging vacuum terms
\beq
{\cal L}_{\rm CT} &=& Z_\sigma (\partial_\mu \sigma)^2 
+ A_\sigma \sigma^2 + B_\sigma \sigma^3 + C_\sigma \sigma^4 
+ Z_\omega F^{\mu\nu} F_{\mu\nu} +
\nonumber \\ 
& & + \left( Z_\rho +Y_\rho\, \sigma + X_\rho\, \sigma^2 \right) R^{\mu\nu}
+ W_\rho (\partial^\mu R_{\mu\lambda})(\partial_\nu R^{\nu\lambda}) 
\nonumber \\
& & + (A_\pi + B_\pi\, \sigma + C_\pi\, \sigma^2) \pi^2 \nonumber \\
& & + Z_a {\cal A}^{\mu\nu} {\cal A}_{\mu\nu} + \left( A_a + B_a\, \sigma 
+ C_a\, \sigma^2 \right) a^\mu a_\mu + \left( F + G_\sigma\, \sigma 
+H_\sigma\, \sigma^2 \right) (a^\mu \partial_\mu \pi)
\eeq

\vskip 0.5cm

$\underline{\mbox{\sf Renormalization scheme A}}$   

The first renormalization scheme is obtained by imposing that the
polarization vanish in vacuum ($M=m$) on the mass shell $q^2=m_\alpha^2$,
for each meson $\alpha=\sigma,\, \omega,\, \delta,\, \rho,\, \pi,\, a_1,\,
\eta$. We also require that the derivative with respect to $q^2$ vanishes
on the mass shell for the mesons $\rho$ and $a_1$.
For all mesons except the $\omega$, divergences of the type 
${M^2 \over \epsilon},\, \epsilon \rightarrow 0$ appear, which require
for their cancellation counterterms like $ A_\sigma \sigma^2 
+ B_\sigma \sigma^3 + C_\sigma \sigma^4 $. The additional constants
are fixed by imposing that the derivatives with respect to the $\sigma$
field vanish.
\beq
{\partial \Pi \over \partial \sigma}\bigg|_{\rm shell}=0 \quad , \qquad
{\partial^2 \Pi \over \partial \sigma^2}\bigg|_{\rm shell}=0 
\eeq
The expressions given here are that of reference \cite{GDP94} for
the $\sigma$ and $\omega$. We have:
\beq
\Pi_{\sigma\sigma}^{\rm (vac\, A)} &=& {d \over 2}\, {g_\sigma^2 \over 2 \pi^2} 
\left[ 6 M^2 \log M -q^2 (\log M + \theta) + 4 M^2 \theta - (4 -m_\sigma^2)\, 
\theta_\sigma  \right. \nonumber \\
& & \qquad \ + (q^2 - m_\sigma^2) \left( \theta_\sigma -(4 -m_\sigma^2)\,
\theta_{\sigma q} \right) \nonumber \\
& & \qquad \ +(1-M)\left( 6 -m_\sigma^2 +(4-m_\sigma^2)
\theta_{\sigma m} + 8\, \theta_\sigma \right) \nonumber \\
& & \left. \qquad \ -{1 \over 2} (1-M)^2 \left( 
18 + m_\sigma^2 +(4 -m_\sigma^2)\,\theta_{\sigma mm}+8\, \theta_\sigma 
+16\, \theta_{\sigma m} \right) \right]
\eeq
where we express all quantities in units of the nucleon mass (``$m$=1'') and
with the definitions
\beq
\theta &=& \theta(q^2,M^2) = y \int_0^\infty {dx \over (x^2+y)\sqrt{x^2+1}}
\qquad {\rm with}\ y=1-{q^2 \over 4 M^2} 
\nonumber \\
\theta_\sigma &=& \theta(m_\sigma^2,m^2)\quad , \qquad \theta_{\sigma m } 
= {\partial \theta \over \partial M}\bigg|_{q^2=m_\sigma^2,
M=m}\quad , \qquad \theta_{\sigma mm} = {\partial^2 \theta \over 
\partial M^2}\bigg|_{q^2=m_\sigma^2,M=m} \\
& & \theta_{\sigma q } 
= {\partial \theta \over \partial (q^2)}\bigg|_{q^2=m_\sigma^2,M=m}
\nonumber
\eeq
The expression for $\Pi_\delta^{\rm (vac\, A)}$ is identical to that for
$\Pi_\sigma^{\rm (vac\, A)}$ when replacing $g_\sigma$, $m_\sigma$, $\theta_\sigma$ 
... by $g_\delta$, $m_\delta$, $\theta_\delta$ ... For the $\omega$ meson
the renormalized vacuum polarization is given by
\beq
\Pi_{\omega\omega}^{\mu\nu{\rm (vac\, A)}}&=& {d \over 2}\,  
{g_\omega^2 \over 3 \pi^2} 
\biggl[ \log M + \theta + {2 M^2 \over q^2} (\theta-1) -{1 \over m_\omega^2} 
\left( (2 +m_\omega^2)\, \theta_\omega -2 \right) \biggr] 
\biggl\{ q^2 g^{\mu\nu} -q^\mu q^\nu \biggr\}  
\label{vacV}
\eeq
For the $\rho$ we take the ``scheme 2'' expression of reference \cite{MGP01}.
\beq
\Pi_{\rho\rho}^{\mu\nu{\rm (vac\, A)}} &=& {d \over 2}\, 
\biggl\{ \quad \ \, {g_\rho^2 \over 3 \pi^2} \quad
\biggl[ \log M + \theta + {2 M^2 \over q^2} (\theta-1) -{1 \over m_\rho^2} 
\left( (2 +m_\rho^2)\, \theta_\rho -2 \right) \biggr]  \nonumber \\
& & +{2 \over \pi^2}\! \left( {f_\rho \over 2 m} \right) \!
\biggl[ M \left( \log M + \theta \right) -\theta_\rho -(1-M) \left( 
\theta_\rho +1 +\theta_{\rho m} \right) \biggr] \nonumber \\
& & +\!\!\left(\! {f_\rho \over 2 m} \!\right)^2\!\!\! {1 \over 6 \pi^2} \biggl[
6\, M^2 \log M + 8\, M^2 \theta + q^2 (\log M + \theta) - 
(8 +m_\rho^2)\, \theta_\rho  \nonumber \\
& & \qquad \qquad \quad +(1-M) \left( 6 + m_\rho^2+16\, \theta_\rho 
+ (8 + m_\rho^2)\, \theta_{\rho m} \right) \nonumber \\
& & \qquad \qquad \quad  -{1 \over 2}(1-M)^2 \left( 18 -m_\rho^2 
+16\, \theta_\rho +32\, \theta_{\rho m} + (8 +m_\rho^2) \theta_{\rho mm}  
\right) \nonumber \\
& & \quad \qquad \qquad -(q^2 - m_\rho^2) \left( (8 +m_\rho^2)\, \theta_{\rho q} 
+ \theta_\rho \right) \biggr]  \biggr\} \ 
\biggl\{ q^2 g^{\mu\nu} -q^\mu q^\nu \biggr\} 
\eeq
For the pion with pseudovector coupling, we obtain
\beq
\Pi_{\pi\pi\ \tim{PV}}^{\rm (vac \, A)} &=& -{d \over 2}\, 
{2 \over \pi^2} \left( {f_\pi \over m_\pi} 
\right)^2 q^2 \biggl[ M^2 (\log M + \theta ) -\theta_\pi 
+ (1-M) \left( 1+ 2 \theta_\pi + \theta_{\pi m} \right) \nonumber \\
& & \qquad \qquad \qquad \quad -{1 \over 2}(1-M)^2\left( 3+2\, \theta_\pi  
+4\, \theta_{\pi m} +\theta_{\pi mm} \right) \biggr] \\
\Pi_{a \pi}^{\mu {\rm (vac\, A)}} &=& i {q^\mu \over q^2} {g_a \over (f_\pi/ m_\pi)} 
\Pi_{\pi\pi}^{\rm (vac\, A)} 
\eeq
Since the polarization of the $a_1$ meson can be written as 
$\Pi_a^{\mu\nu}= (g_a/g_\omega)^2 \Pi_\omega^{\mu\nu} +g_a^2 (m_\pi /f_\pi)^2 
\Pi_{\pi}^{\tim{PV}} g^{\mu\nu}/q^2$, it is tempting to apply the renormalization
procedure with this decomposition. In this case, we will have
$\Pi_{aa}^{\mu\nu{\rm (vac)}}= \Pi_{1}\, ( q^2 g^{\mu\nu} -q^\mu q^\nu ) 
+ \Pi_{2}\, g^{\mu\nu}$ where the expression of the $\Pi_{1}$ contribution 
is obtained from that of $\Pi_\omega^{\rm (vac)}$ by replacing 
$g_\omega$, $m_\omega$, $\theta_\omega$ by $g_a$, $m_a$, $\theta_a$.
and that of $\Pi_{2}$ is proportional to $\Pi_{\pi\pi}^{\rm (vac)}$
when replacing $f_\pi/m_\pi$, $m_\pi$, $\theta_\pi$ ... by $g_a$, $m_a$,
$\theta_a$ ...
However, the $\Pi_2$ term is not orthogonal to $q^\mu q^\nu$, whereas
the polarization of the $a_1$ enters in the dispersion relations with the 
decomposition $\Pi_{a\, T} T^{\mu\nu} +\Pi_{a\, L} L^{\mu\nu} 
+\Pi_{a\, Q}\, q^\mu q^\nu/q^2$. It is therefore preferable to
apply the renormalization procedure to $\Pi_{aa}^{\mu\nu{\rm (vac)}}= 
\Pi_{a\, LT}^{\rm (vac)} (g^{\mu\nu} -q^\mu q^\nu/q^2 ) 
+ \Pi_{a\, Q}^{\rm (vac)} q^\mu q^\nu/q^2$. The $\Pi_{a\, LT}^{\rm (vac)}$
then happens to have the same structure as the polarization of the 
$\sigma$ meson. We obtain
\beq
\Pi_{aa}^{\mu\nu{\rm (vac\, A)}} &=& {d \over 2}\, {g_a^2 \over 3 \pi^2} 
\left[ 6 M^2 \log M -q^2 (\log M + \theta) + 4 M^2 \theta - (4 -m_a^2)\, 
\theta_a \right. \nonumber \\
& & \qquad \ + (q^2 - m_a^2) \left( \theta_a -(4 -m_a^2)\,
\theta_{a q} \right) \nonumber \\
& & \qquad \ +(1-M)\left( 6 -m_a^2 +(4-m_a^2)
\theta_{a m} + 8\, \theta_a \right) \nonumber \\
& & \left. \qquad \ -{1 \over 2} (1-M)^2 \left( 
18 + m_a^2 +(4 -m_a^2)\,\theta_{a mm}+8\, \theta_a 
+16\, \theta_{a m} \right) \right] \left\{ g^{\mu\nu} - {q^\mu q^\nu \over q^2}
\right\} \nonumber \\
& & \!\! - {d \over 2}\, 
{2 g_a^2 \over \pi^2} \biggl[\, M^2 (\log M + \theta ) -\theta_a 
+ (1-M) \left( 1+ 2 \theta_a + \theta_{a m} \right) \nonumber \\
& & \qquad \ -{1 \over 2}(1-M)^2\left( 3+2\, \theta_a  +4\, \theta_{a m} 
+\theta_{a mm} \right) \biggr]  {q^\mu q^\nu \over q^2}
\eeq
The vacuum contribution to the mixed polarizations $\Pi_{\sigma\omega}^\mu$,
$\Pi_{\delta\rho}^\mu$ and $\Pi_{a \rho}^{\mu\nu}$ vanishes.

\vskip 0.5cm

$\underline{\mbox{\sf Renormalization scheme B}}$  

In this second renormalization scheme, we notice that there appears in the 
polarizations diverging contributions of the type  $M/\epsilon$, $M^2/\epsilon$, 
$\epsilon \rightarrow 0$. Instead of cancelling this divergences with
counterterms of the form $a + b \sigma + c \sigma^2$, the original structure
is preserved by subtracting only counterterms in the combination 
$ a (m -g_\sigma \sigma)$ or $a (m -g_\sigma \sigma)^2$. The unknown 
constant $a$ is determined by imposing $\Pi|_{\rm shell}=0$. No conditions
are imposed on the derivatives. 

Originally this scheme was motivated by the fact that it yields a decreasing
effective $\rho$ mass \cite{MGP01}. When applying it to the other mesons, we
obtain
\beq
\Pi_{\sigma}^{\rm (vac\, B)} &=& {d \over 2}\, 
{g_\sigma^2 \over 2 \pi^2} \left[ 6 M^2 \log M + 
4 M^2 \theta - q^2 (\log M + \theta) - (4 - m_\sigma^2) \theta_\sigma \right.
\nonumber \\
& & \left. \ + (M^2-1)\left((4-m_\sigma^2) m_\sigma^2 \theta_{\sigma q} 
-4 \theta_\sigma \right) + (q^2 - m_\sigma^2) \left( \theta_\sigma 
- (4 - m_\sigma^2) \theta_{\sigma q} \right) \right] 
\eeq
The polarization of the $\delta$ meson is identical when relacing
$g_\sigma$, $m_\sigma$, $\theta_\sigma$ ... by $g_\delta$, $m_\delta$, 
$\theta_\delta$ ... 

Since no divergences proportional to $M$ or $M^2$ occur in the polarization
of the $\omega$ meson, its expression is left unchanged by this renormalization
scheme and its expression still given by Eq. (\ref{vacV}).
We recall the expression obtained for the $\rho$ meson in scheme B ($\equiv$ 
scheme 3 of \cite{MGP01})
\beq
\Pi_{\rho}^{\rm (vac\, B)} &=&{d \over 2}\, 
\left\{ {g_\rho^2 \over 3 \pi^2} \bigg[ 
\ln M + \theta + {2 M^2 \over q^2}(\theta-1) -{1 \over m_\rho^2}
\left(\, (2 + m_\rho^2)\, \theta_\rho -2  \right)  \bigg] 
\right. \nonumber \\
& & +\left( {f_\rho \over 2 m} \right)^2 {1 \over 6 \pi^2} \bigg[ 
6 M^2 \ln M  + 8 M^2 \theta + k^2 ( \ln M  + \theta) -(8+m_\rho^2)\, 
\theta_\rho   \nonumber \\
& & \qquad \qquad \qquad \quad 
-(q^2 - m_\rho^2) \left(\, (8 + m_\rho^2)\, \theta_{\rho q} 
+ \theta_\rho  \right)
+(M^2-1) \left( m_\rho^2 (8 + m_\rho^2)\, \theta_{\rho q} 
-8  \theta_\rho  \right)  \bigg] \nonumber \\
& &  \left. +{2 \over \pi^2} \left( {f_\rho \over 2 m} \right) g_\rho \bigg[
M ( \ln M  + \theta  -  \theta_\rho )  \bigg]  \right\} \left\{
q^2 g^{\mu\nu} - q^{\mu} q^\nu \right\}
\eeq
For the pion with pseudovectorial coupling, we obtain
\beq
\Pi_{\pi\, {\tim{PV}}}^{\rm (vac\, B)} = -{d \over 2}\, {2 \over \pi^2} 
\left( {f_\pi \over m_\pi} \right)^2 q^2 M^2 \, \left[ \log M + \theta - \theta_\pi
\right]
\eeq
Finally, the polarization of the $a_1$ meson is given in scheme B by
\beq
\Pi_{aa}^{\mu\nu{\rm (vac\, B)}} &=& - \Pi_{a\, LT}^{\rm (vac\, B)} \left( g^{\mu\nu}
-{q^\mu q^\nu \over q^2} \right) - \Pi_{a\, Q}^{\rm (vac\, B)}{q^\mu q^\nu \over q^2}
\nonumber \\
\Pi_{a\, LT}^{\rm (vac\, B)} &=& {d \over 2}\, 
{g_a^2 \over 3 \pi^2} \left[ 6 M^2 \log M + 
4 M^2 \theta - q^2 (\log M + \theta) - (4 - m_a^2) \theta_a \right. \nonumber \\
&& \left. + (M^2-1)\left((4-m_a^2) m_a^2 \theta_{a q} -4 \theta_a \right) 
+ (q^2 - m_a^2) \left( \theta_a - (4 - m_a^2) \theta_{a q} \right) \right] \\
\Pi_{a\, Q}^{\rm (vac\, B)} &=& 2 {d \over 2}\, {g_a^2 \over \pi^2} \, M^2\, 
\left[ \log M + \theta - \theta_a \right] 
\eeq

\vskip 0.5cm

\subsection*{A.3 - $\underline{\mbox{\sf Polarizations, 
imaginary part}}$}   

\vskip 0.5cm

All imaginary parts can be expressed in terms of three integrals
$E_1$, $E_2$, $E_3$. At finite temperature, for the calculation of
the retarded polarizations, we have

\par\noindent {\bf --} for spacelike momentum:
\beq
E_n &=&  \int dy \left[ \left( y+{\omega \over 2} \right)^{n-1} 
\theta(y-y_L) \Bigl\{ n(y) -n(y+\omega) \Bigr\} \right. 
\nonumber \\
          & & \left. \qquad +(-1)^n
\left( y-{\omega \over 2} \right)^{n-1} \theta(y-y_U) 
\left\{ \overline n(y) - \overline n(y-\omega) \right\} \right] 
\eeq 
\par\noindent {\bf --} for timelike momentum with $q^2 < 4 M^2$,
the imaginary parts vanish: $E_n = 0$, 
\par\noindent {\bf --} for timelike momentum with $q^2 > 4 M^2$:
\beq
E_n &=& (-1)^{n-1} \int_M^\infty  dy 
\left( y -{\omega \over 2} \right)^{n-1} 
\left[ \theta(y-y_L) - \theta(y-y_U) \right]  
\left\{ n (\omega -y) + \overline n (y) -1 \right\}
\eeq
with
\beq
&& n(y)=\left[ e^{\beta (y-\mu)} +1 \right]^{-1} \ , \quad
\overline n(y)=\left[ e^{\beta (y+\mu)} +1 \right]^{-1} 
\nonumber \\
&& y_L = \left| {k \sqrt{\Delta} - \omega \over 2} \right|  
\ , \quad y_U = \left| {k \sqrt{\Delta} + \omega \over 2} 
\right|  \ , \quad \Delta= 1 - 4 {M^2 \over q^2} \nonumber
\eeq
In the following equations $d$ is the degeneracy parameter
($d=2$ for symmetric nuclear matter).
\beq
{\cal I}m\ \Pi_\sigma &=& - d {g_\sigma^2 \over 2 \pi k} 
\left( M^2 -{q^2 \over 4} \right) E_1 \nonumber \\
{\cal I}m\ \Pi_{\omega\, L} &=&  d {g_\omega^2 \over 2 \pi k} 
\left[ {q^2 \over 4} E_1 - {q^2 \over k^2} E_3 \right]  
\nonumber \\
{\cal I}m\ \Pi_{\omega\, T} &=& d {g_\omega^2 \over 4 \pi k} 
\left[ \left( M^2 + {q^2 \over 4} \right) E_1 + 
{q^2 \over k^2} E_3 \right] \nonumber \\
{\cal I}m\ \Pi_{\sigma\omega} &=& - d {g_\sigma g_\omega 
\over 2 \pi k^3 }\ q^2 M\ E_2 \nonumber \\
{\cal I}m\ \Pi_{\rho\, L} &=&  d {1\over 2 \pi k} 
\left\{ g_\rho^2 \left[ {q^2 \over 4} E_1 - {q^2 \over k^2} 
E_3 \right]  + g_\rho {f_\rho \over 2 m} \ q^2 M \
E_1 + \left(  {f_\rho \over 2 m} \right)^2 \left[ M^2 q^2\
E_1 +{q^4 \over k^2} E_3 \right] \right\}
\nonumber \\
{\cal I}m\ \Pi_{\rho\, T} &=&  d {1 \over 4 \pi k} \left\{ 
g_\rho^2 \left[ \left( M^2 + {q^2 \over 4} \right) E_1 + 
{q^2 \over k^2} E_3 \right] 
+ 2 g_\rho {f_\rho \over 2 m} \ q^2 M \
E_1  \right. \nonumber \\
& & \left. \qquad + \left(  {f_\rho \over 2 m} \right)^2 \left[
\left( M^2 q^2 + {q^4 \over 4} \right) E_1 
-{q^4 \over k^2} E_3 \right] \right\}
\nonumber \\
{\cal I}m\ \Pi_\pi^{\tim{PV}} &=& - d {M^2 q^2 \over 2 \pi k} 
\left( {f_\pi \over m_\pi} \right)^2 E_1 \nonumber \\
{\cal I}m\ \Pi_\pi^{\tim{PS}} &=& - d {g_\pi^2 q^2 \over 8 \pi k} 
 E_1 \nonumber \\
{\cal I}m\ \Pi_{a\, L} &=&  - d {g_a^2 \over 2 \pi k} 
\left[ \left( M^2 -{q^2 \over 4} \right) E_1 
+ {q^2 \over k^2} E_3 \right]  \nonumber \\
{\cal I}m\ \Pi_{a\, T} &=& -d {g_a^2 \over 4 \pi k} 
\left[ \left( M^2 - {q^2 \over 4} \right) E_1 - 
{q^2 \over k^2} E_3 \right] \nonumber \\
{\cal I}m\ \Pi_{a\, Q} &=& - d {g_a^2 \over 2 \pi k} 
M^2\ E_1 \nonumber \\
{\cal I}m\ \Pi_{a \pi} &=&  d {g_a \over 2 \pi k} 
\left( {f_\pi \over m_\pi} \right)\left( {\alpha \over M} + 
1-\alpha \right) M^2\  E_1 \nonumber \\
{\cal I}m\ \Pi_{a \rho} &=&  d {g_a \left( g_\rho + 2 M (f_\rho/2 m)
\right) \over 4 \pi k^3} q^2 \ E_2 \nonumber \\
{\cal I}m\ \Pi_\eta^{\tim{PS}} &=& - d {g_\eta^2 q^2 \over 8 \pi k} 
 E_1 \nonumber 
\eeq

\section*{Appendix B: Polarizations modified by contact terms}

In meson exchange models of the NN interaction, the potential
is obtained by calculating the transition amplitude
${\cal M} = \sum_{a,b=\sigma,\omega,\pi,\rho,\delta,a_1,\eta}
(\overline{U_3}\, \Gamma_a\, U_1)\ G^{ab}\
(\overline{U_3}\, \Gamma_a\, U_1)$, with $\Gamma^a$ and
$G^{ab}$ beeing the relevant vertices and meson propagators.
After taking the semiclassical limit and multiplying by the
minimal relativity factors, one finally performs a Fourier
transformation to obtain the potential in coordinate space.
Besides the desired Yukawa-like terms with ranges characterized 
by the inverses of the meson masses, one also obtains a singular 
contribution at the origin $\delta(\vec r)$ due to the assumption 
that all particles are pointlike.

In a $\sigma$-$\omega$-$\pi$-$\rho$ model, the singular
contribution is given by \cite{BS69}
\beq
V(r) &\ni& - \left[ {g_\sigma^2 \over 4 m^2} + {g_\omega^2 \over 2 m^2}
+ {g_\rho (g_\rho + f_\rho) \over 2 m^2} \vec\tau_1.\vec\tau_2 \right]
\delta(\vec r) \nonumber \\
&& - \left[ {g_\omega^2 \over 2 m^2} + 
{g_\pi^2 \over 4 m^2} \vec \tau_1.\vec\tau_2 +  {(g_\rho + f_\rho)^2
\over 2 m^2} \vec \tau_1.\vec\tau_2 \right] \delta(\vec r)\, {1 \over 3}
\, \vec\sigma_1.\vec\sigma_2
\label{deltapot}
\eeq

It is not always enough to smooth this singularity by convoluting 
the result with form factors, especially in the case of the pion.
It is then desirable to remove the singular piece by subtracting
one or more contact terms. If this method is chosen, the
dispersion relations of the mesons will be modified as follow.

\vskip 0.5cm
$\underline{\mbox{\sf $\pi$--$a_1$--$\rho$ sector}}$  
\vskip 0.5cm

When a contact interaction is introduced as
\beq
{\cal L} \ni   - g_A (\overline\psi\gamma_5\gamma_\mu \vec\tau\psi). 
        (\overline\psi\gamma_5\gamma^\mu \vec\tau\psi) 
- g_R (\overline\psi\gamma_\mu \vec\tau\psi). 
        (\overline\psi\gamma^\mu \vec\tau\psi) 
\nonumber
\eeq
the dispersion relations of the $\pi$--$a_1$--$\rho$ sector are given
by (\ref{piarhogAgR}). The polarizations which appear in these expressions
are given explicitely in this Appendix.
\beq
\widetilde \Pi_{\rho\, L} &=& {f_\rho \over 2 m} \left( g_\rho \Pi_{\tim{TV}}^L 
+ {f_\rho \over 2 m} \Pi_{\tim{TT}}^L \right) + { \left( g_\rho 
+ 2 g_R {f_\rho \over 2 m} \Pi_{\tim{TV}}^L \right) \left( g_\rho \Pi_{\tim{VV}}^L
+ {f_\rho \over 2 m} \Pi_{\tim{VT}}^L \right) \over \left( 1 -2 g_R 
\Pi_{\tim{VV}}^L \right)} \\
\widetilde \Pi_{\rho\, T} &=& \left(g_\rho + 2 g_R {f_\rho \over 2 m} 
\Pi_{\tim{TV}}^T \right) R_{\rho\, T} + ( 2 g_A   {f_\rho \over 2 m} 
\Pi_{\tim{TA}}^T ) P_{\rho\, T} 
+  {f_\rho \over 2 m} \left( g_\rho \Pi_{\tim{TV}}^T +  {f_\rho \over 2 m} 
\Pi_{\tim{TT}}^T \right) \\
\widetilde \Pi_{a\, L} &=& {g_a^2 \Pi_{\tim{AA}}^L \over 1 -2 g_A 
\Pi_{\tim{AA}}^L } \\
\widetilde \Pi_{a\, T} &=& {f_\rho \over 2 m} g_a \Pi_{\tim{TA}}^T +
\left(g_\rho + 2 g_R {f_\rho \over 2 m} \Pi_{\tim{TV}}^T
\right)  R_{a\, T} + ( 2 g_A   {f_\rho \over 2 m} \Pi_{\tim{TA}}^T ) P_{a\, T} \\
T^{\mu\nu} P_{1\, \nu} &=& P_{\rho\, T} T^{\mu\nu} \rho_{1\, \nu} +
P_{a\, T} T^{\mu\nu} a_{1\, \nu}  \\
T^{\mu\nu} R_{1\, \nu} &=& R_{\rho\, T} T^{\mu\nu} \rho_{1\, \nu} +
R_{a\, T} T^{\mu\nu} a_{1\, \nu}  \nonumber \\
P_{\rho\, T} &=& \left[ \left(1 -2 g_R \Pi^T_{\tim{VV}}\right) \left(g_\rho 
\Pi_{\tim{AV}}^T + {f_\rho \over 2 m} \Pi_{\tim{AT}}^T \right) +2 g_R 
\Pi_{\tim{AV}}^T  \left(g_\rho \Pi_{\tim{VV}}^T + {f_\rho \over 2 m} 
\Pi_{\tim{VT}}^T \right) \right] / {\cal D} \nonumber \\
P_{a\, T} &=& \left[ \left(1 -2 g_R \Pi^T_{\tim{VV}}\right)  
g_a \Pi_{\tim{AA}}^T 
 + 2 g_R g_a \Pi_{\tim{AV}}^T \Pi_{\tim{VV}}^T  \right] / {\cal D} \nonumber \\
R_{\rho\, T} &=& \left[ \left(1 -2 g_A \Pi^T_{\tim{AA}}\right) \left(g_\rho 
\Pi_{\tim{VV}}^T + {f_\rho \over 2 m} \Pi_{\tim{VT}}^T \right) +2 g_A 
\Pi_{\tim{VA}}^T  \left(g_\rho \Pi_{\tim{AV}}^T + {f_\rho \over 2 m} 
\Pi_{\tim{AT}}^T \right) \right] / {\cal D} \nonumber \\
R_{a\, T} &=& \left[ \left(1 -2 g_A \Pi^T_{\tim{AA}}\right)  g_a \Pi_{\tim{VA}}^T 
 + 2 g_A g_a \Pi_{\tim{VA}}^T \Pi_{\tim{AA}}^T  \right] / {\cal D} \nonumber \\
{\cal D} &=& \left(1 -2 g_A  \Pi_{\tim{AA}}^T  \right)\left(1 -2 g_R 
\Pi^T_{\tim{VV}}\right) -4 g_A g_R \Pi_{\tim{AV}}^T \Pi_{\tim{VA}}^T
\eeq
with the definitions
$\Pi_{\rho\rho}^{\mu\nu}= g_\rho^2 \Pi_{\tim{VV}}^{\mu\nu} +2 g_\rho \left( 
\displaystyle{f_\rho \over 2 m} \right) \Pi_{\tim{VT}}^{\mu\nu} 
+\left( \displaystyle{f_\rho \over 2 m} \right)^2 \Pi_{\tim{TT}}^{\mu\nu}$,
$\Pi_{aa}^{\mu\nu}= g_a^2 \Pi_{\tim{AA}}^{\mu\nu}$  and
$\Pi_{a \rho}^{\mu\nu}= g_a g_\rho \Pi_{\tim{AV}}^{\mu\nu}
+g_a \left( \displaystyle{f_\rho \over 2 m} \right) \Pi_{\tim{AT}}^{\mu\nu}$.

\vskip 0.5cm
$\underline{\mbox{\sf $\sigma1$--$\omega$ sector}}$  
\vskip 0.5cm

Contact terms of the form $\delta(\vec r)$ also appear in the
contributions of the $\sigma$ and $\omega$ mesons exchange to the 
NN potential. Although the problem is less acute than in the case
of the spin dependent part of the interaction mediated by the pion,
one could wish for consistency \cite{BLLQMN00} to remove these 
$\delta(\vec r)$  contributions by the introduction of further 
contact terms
\beq
{\cal L} \ni   - g_V (\overline\psi\gamma_\mu\psi). 
        (\overline\psi\gamma^\mu\psi) 
   - g_S (\overline\psi \psi). (\overline\psi \psi) 
\nonumber
\eeq
These terms modify the dispersion relations as follows
\beq
&& [-q^2+m_\sigma^2] \sigma_1 = g_\sigma^2 \widetilde \Pi_\sigma \sigma_1
 + g_\sigma g_\omega \widetilde \Pi_{\sigma\omega}^\mu \omega_{1\, \mu} \\
&& [-q^\mu q^\nu +(q^2 - m_\omega^2) g^{\mu\nu} ] \omega_{1\, \nu}
= g_\sigma g_\omega \widetilde \Pi_{\sigma\omega}^\mu \sigma_1 +
 g_\omega^2 \left( -\widetilde \Pi_{\omega\, T} T^{\mu\nu} - 
\widetilde\Pi_{\omega\, L} L^{\mu\nu} \right) \omega_{1\, \nu}  \\
&& {\rm with}\quad \widetilde\Pi_\sigma = { \left( 1 -2\, g_V\, 
\Pi_{\omega\, L} \right) \Pi_\sigma -2\, g_V\, \Pi_{\sigma\omega}^2\, \eta^2
\over \left( 1-2\, g_V\, \Pi_{\omega\, L} \right)\left( 1 + 2\, g_S\, 
\Pi_\sigma \right) -4\, g_S g_V\, \Pi_{\sigma\omega}^2\, \eta^2 } \\
&& \qquad \quad \widetilde\Pi_{\sigma\omega}^\mu =
 \widetilde\Pi_{\omega\sigma}^\mu = { g_\sigma g_\omega \widetilde 
\Pi_{\sigma\omega}^\mu \over  \left( 1-2\, g_V\, \Pi_{\omega\, L} \right)
\left( 1 + 2\, g_S\, \Pi_\sigma \right) -4\, g_S g_V\, 
\Pi_{\sigma\omega}^2\, \eta^2 }  \\
&& \qquad \quad \widetilde\Pi_{\omega\, T} = { \Pi_{\omega\, T} \over
1 -2 g_V  \Pi_{\omega\, T} }  \\
&& \qquad \quad \widetilde\Pi_{\omega\, L} ={ \left( 1 + 2\, g_S\, 
\Pi_\sigma \right)\Pi_{\omega\, L}+2 g_S \Pi_{\sigma\omega}^2 \eta^2
\over  \left( 1-2\, g_V\, \Pi_{\omega\, L} \right)\left( 1 + 2\, g_S\, 
\Pi_\sigma \right) -4\, g_S g_V\, \Pi_{\sigma\omega}^2\, \eta^2 }
\eeq

\newpage

\begin{figure}[htb]
\mbox{%
\parbox{8.2cm}{\epsfig{file=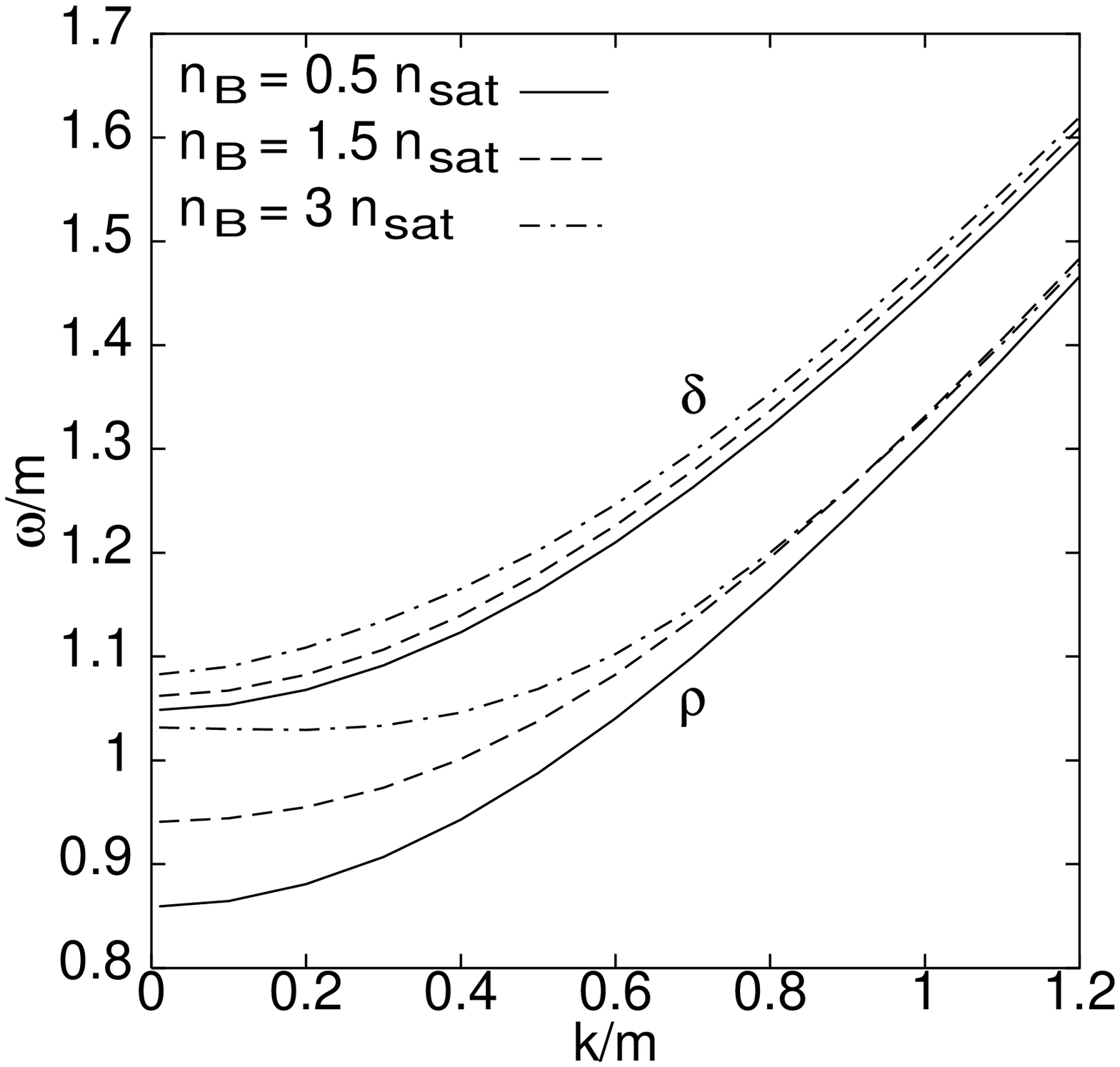,width=8.2cm}}
\parbox{0.2cm}{\phantom{a}}
\parbox{8.2cm}{\epsfig{file=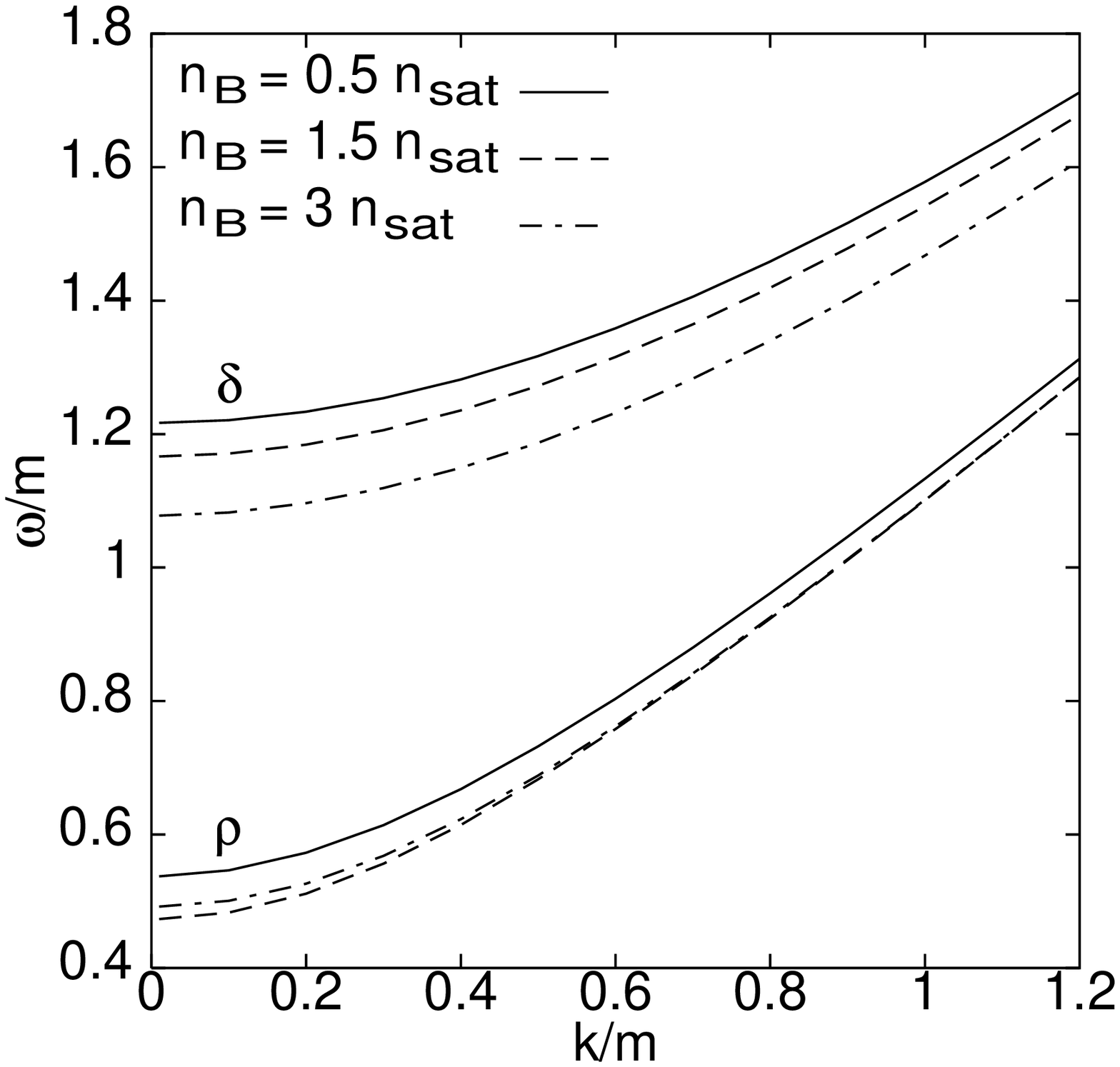,width=8.2cm}}
}
\vskip 0.1cm
\mbox{%
\parbox{16cm}{{\bf Fig. 1} Dispersion relation of the $\delta$-$\rho$
longitudinal mode, as calculated with renormalization scheme A (left panel) 
or B (right panel), at vanishing temperature and three values of the 
density: 0.5, 1.5 and 3 times the saturation density}
}
\end{figure}

\begin{figure}[htb]
\mbox{%
\parbox{8.2cm}{\epsfig{file=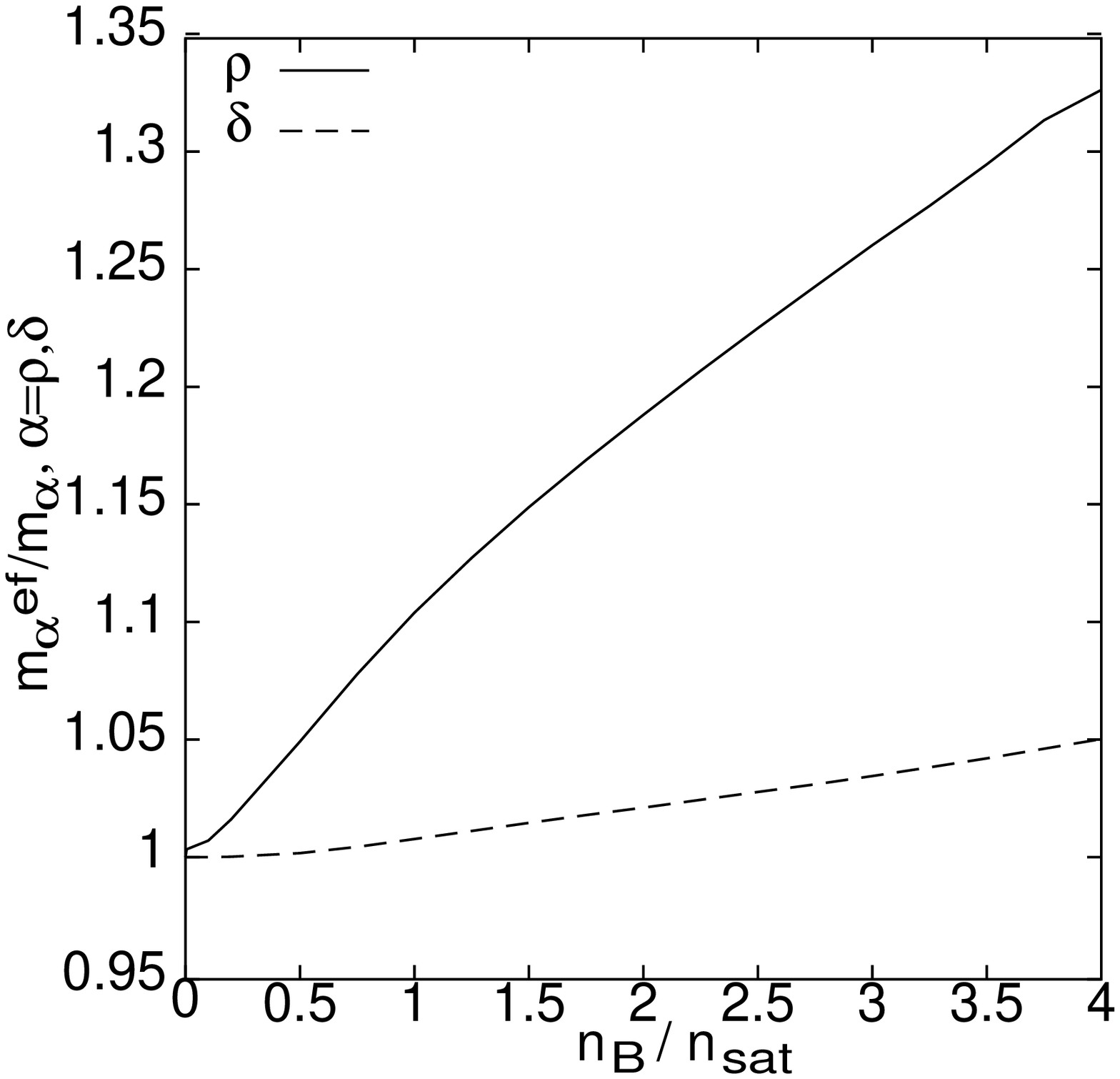,width=8.2cm}}
\parbox{0.1cm}{\phantom{a}}
\parbox{8.2cm}{\epsfig{file=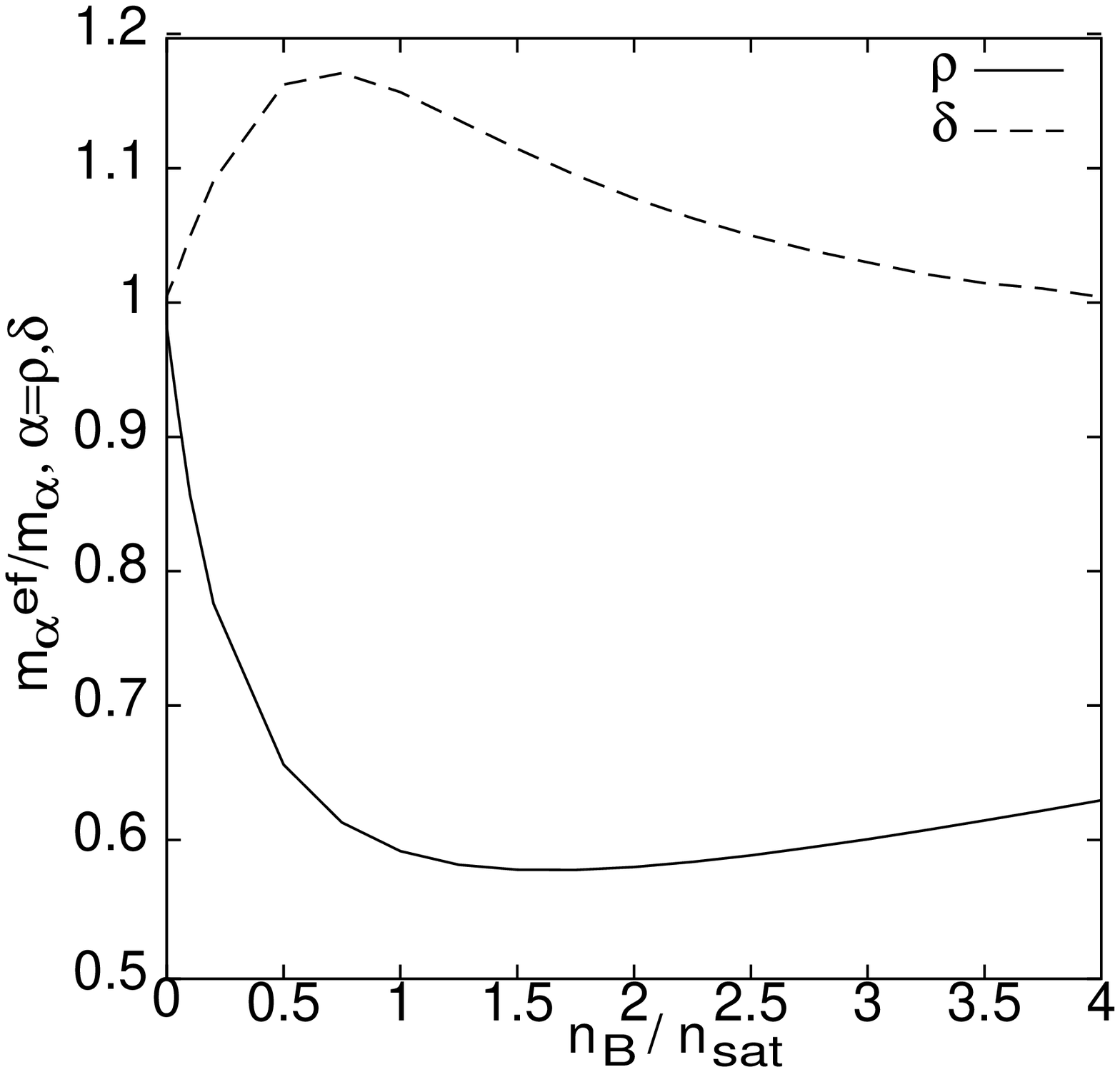,width=8.2cm}}
}
\vskip 0.1cm
\mbox{%
\parbox{8.2cm}{\epsfig{file=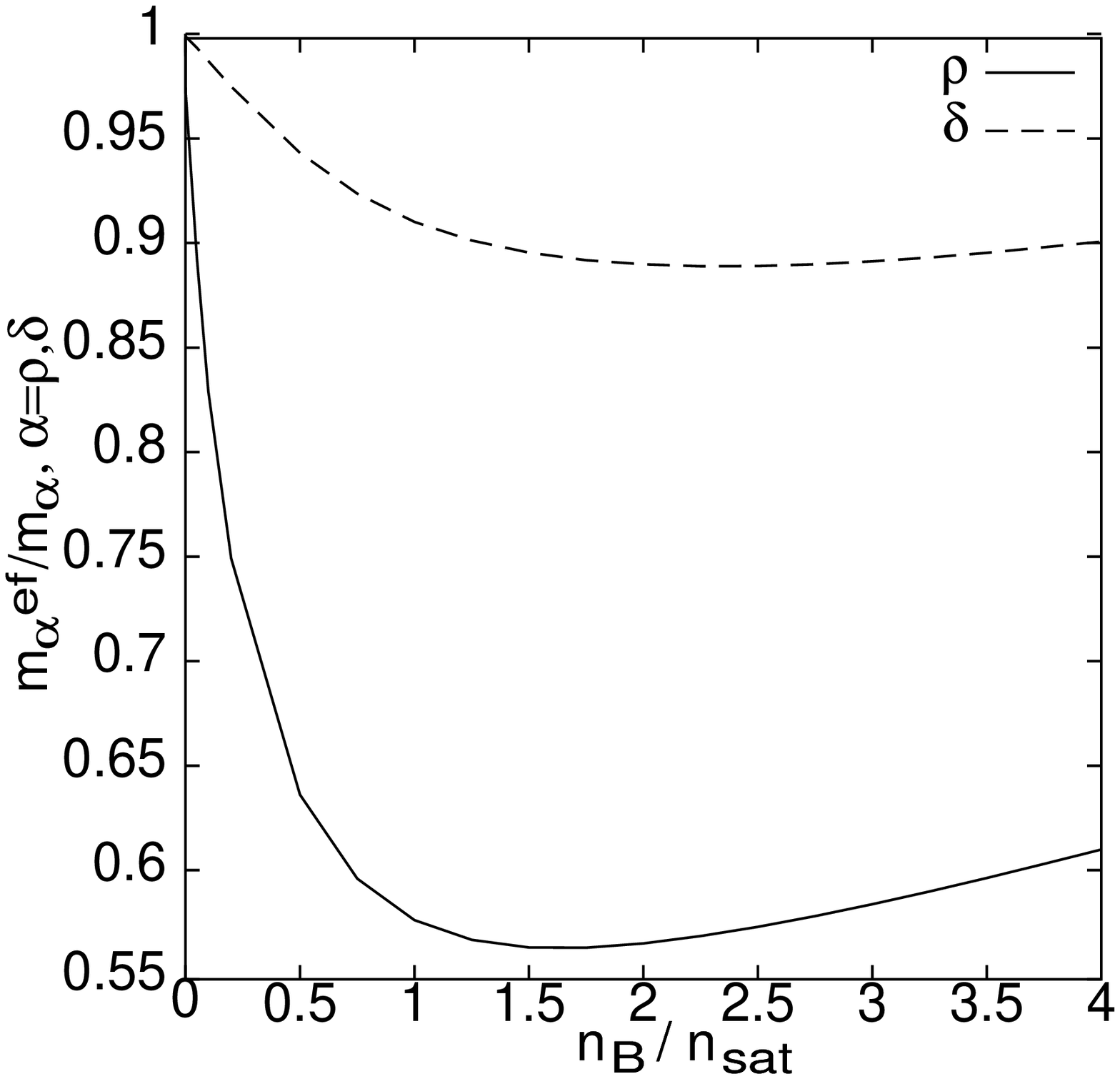,width=8.2cm}}
\parbox{0.1cm}{\phantom{a}}
\parbox{8.2cm}{{\bf Fig. 2} Effective masses of the $\delta$ and $\rho$  mesons as a 
function of density, as calculated with renormalization scheme A (left panel) 
or B (middle panel) or C (right panel), at vanishing temperature.}
}
\end{figure}

\begin{figure}[htb]
\mbox{%
\parbox{8.2cm}{\epsfig{file=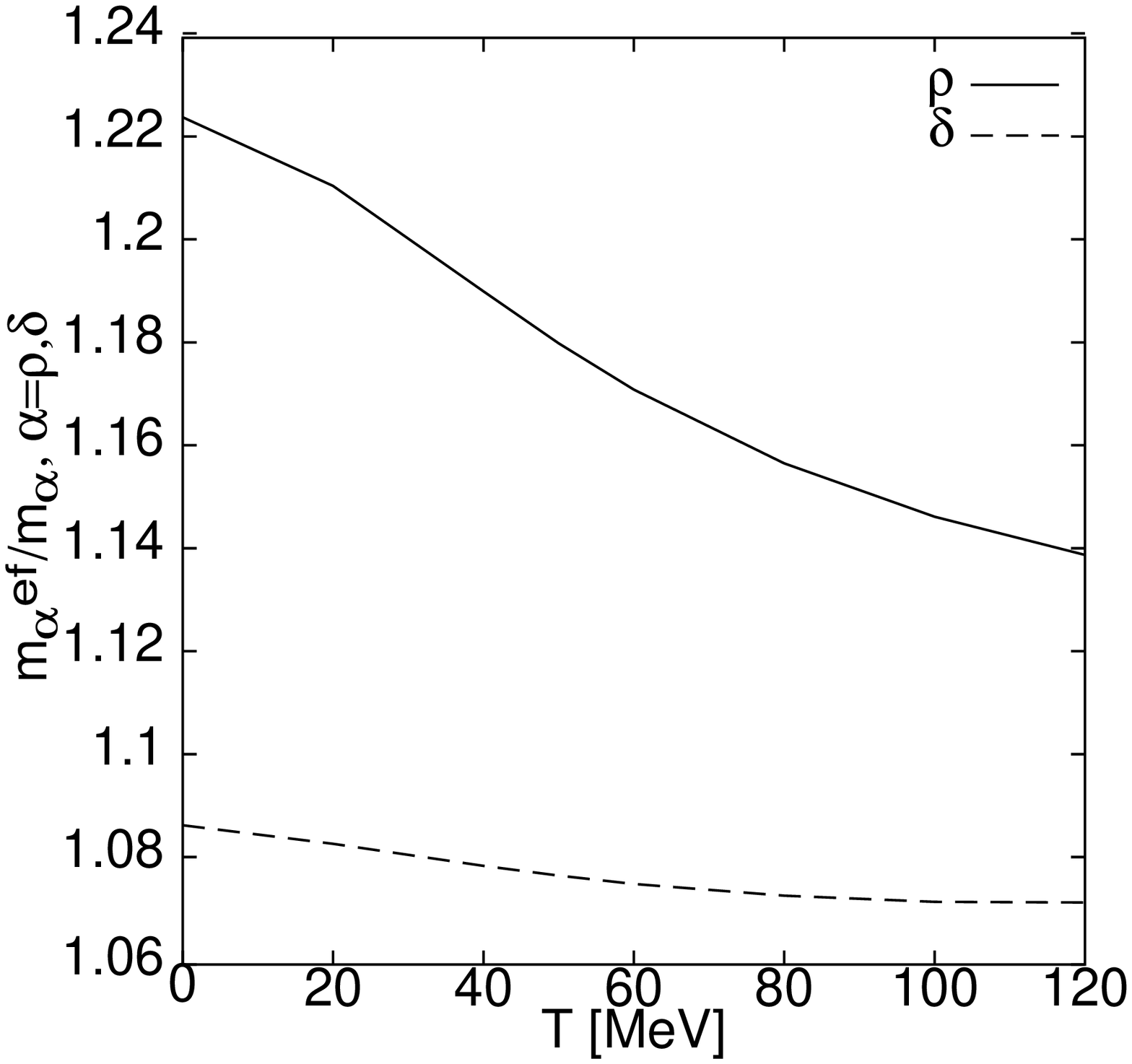,width=8.2cm}}
\parbox{0.1cm}{\phantom{a}}
\parbox{8.2cm}{{\bf Fig. 3} Effective masses of the $\delta$
and $\rho$ as a function of temperature at three times saturation density in renormalization scheme A.}
}
\end{figure}

\begin{figure}[htb]
\mbox{%
\parbox{8.2cm}{\epsfig{file=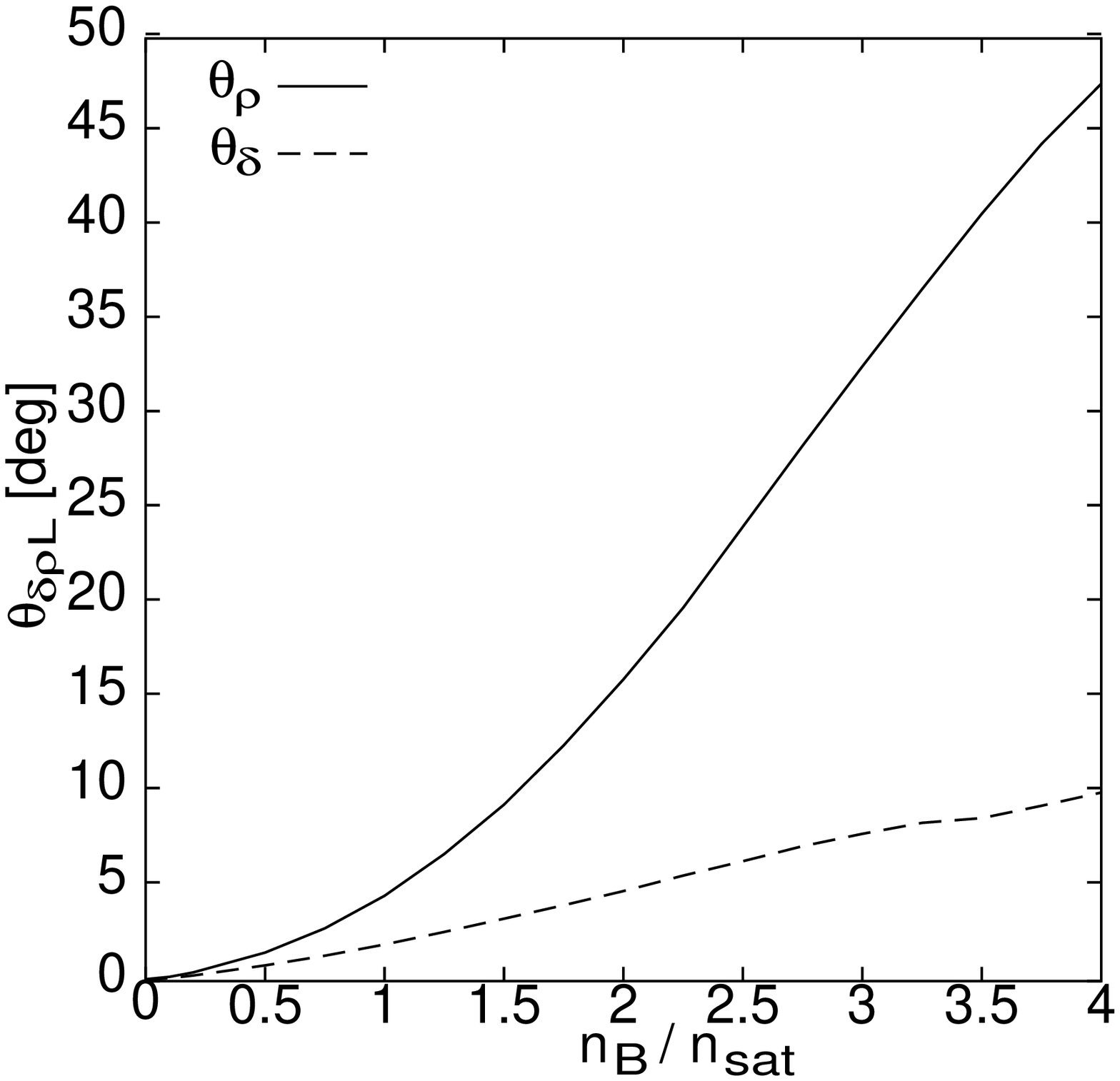,width=8.2cm}}
\parbox{0.1cm}{\phantom{a}}
\parbox{8.2cm}{\epsfig{file=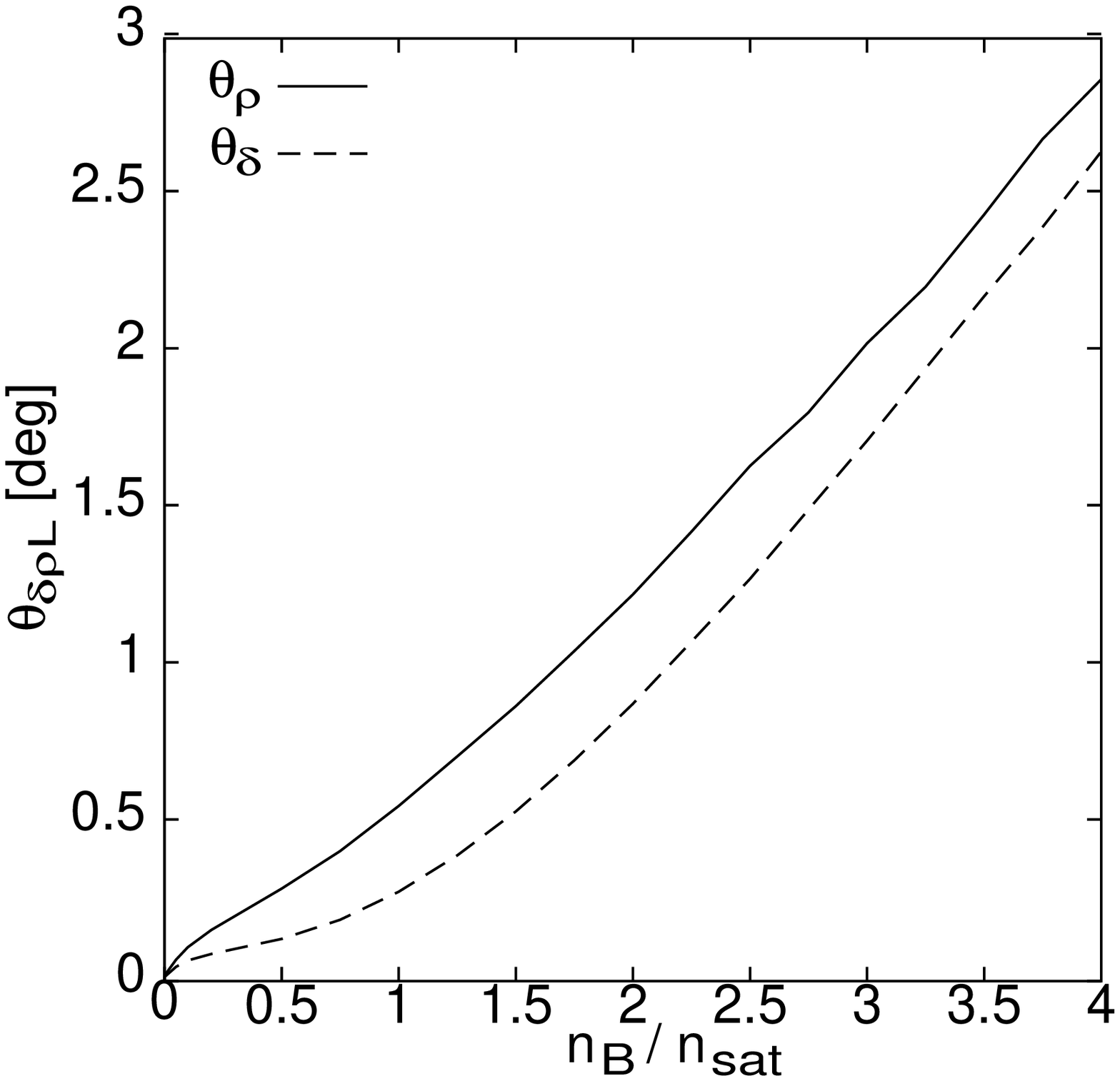,width=8.2cm}}
}
\vskip 0.1cm
\mbox{%
\parbox{16cm}{{\bf Fig. 4} Mixing angle beween the $\delta$ and 
$\rho$ in the longitudinal mode at $T=0$ and $k=300$ Mev, represented as a 
function of density, and calculated with renormalization scheme A (left panel) 
or B (right panel).}
}
\end{figure}

\begin{figure}[htb]
\mbox{%
\parbox{8.2cm}{\epsfig{file=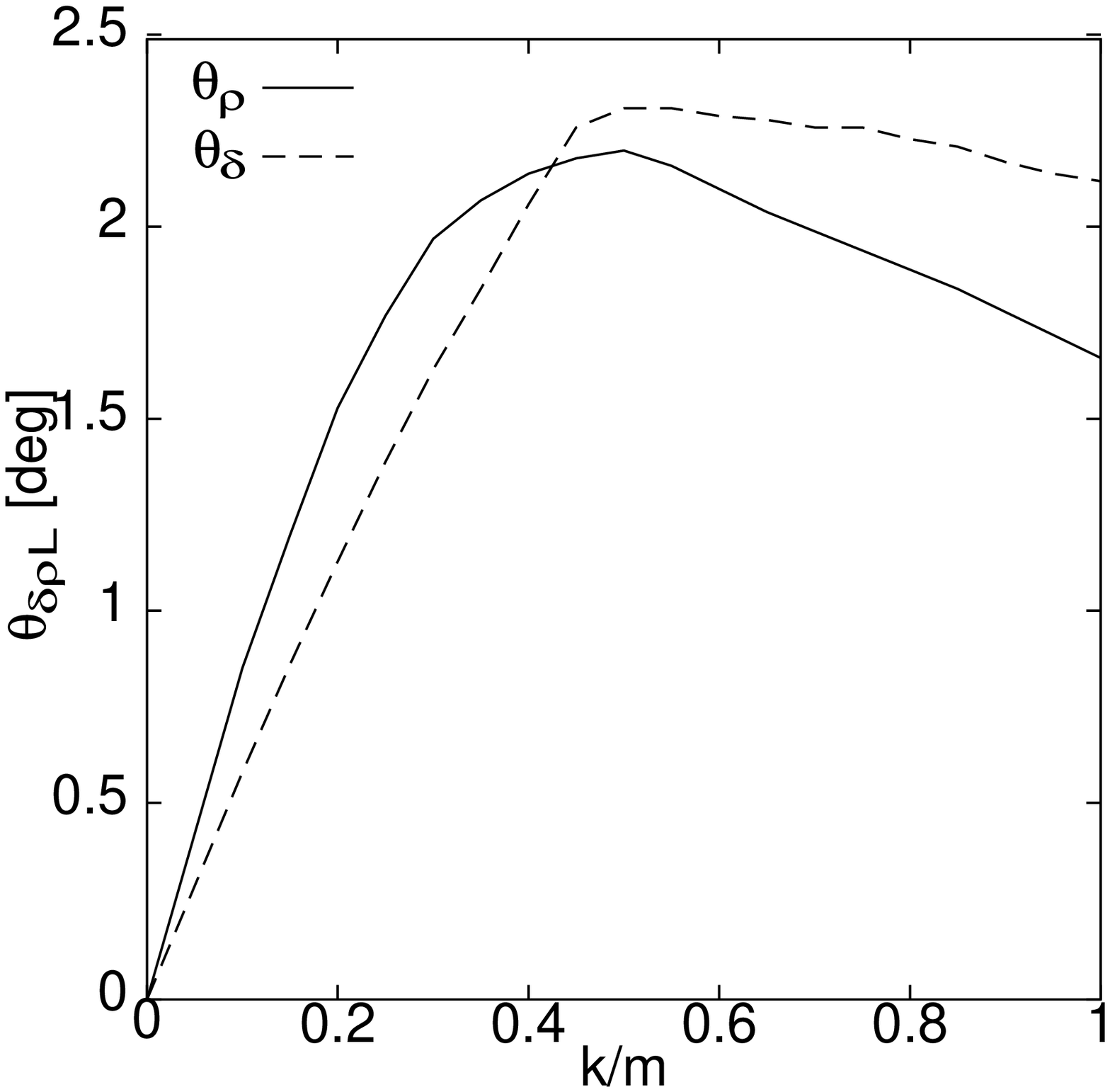,width=8.2cm}}
\parbox{0.1cm}{\phantom{a}}
\parbox{8.2cm}{\epsfig{file=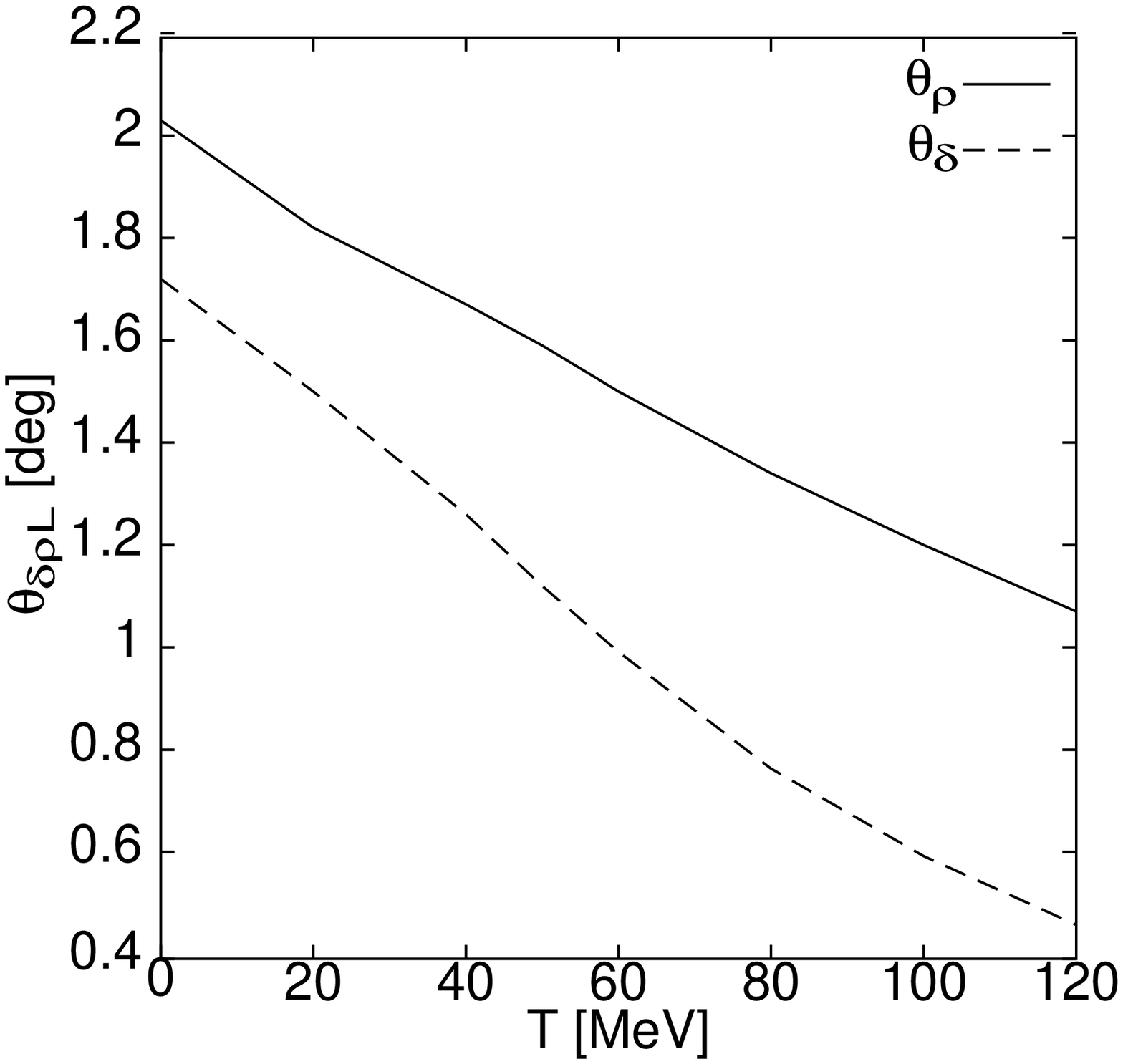,width=8.2cm}}
}
\vskip 0.1cm
\mbox{%
\parbox{16cm}{{\bf Fig. 5} Mixing angle beween the $\delta$ and 
$\rho$ in the longitudinal mode represented as a function of momentum
(left panel) at $n_B =3\ n_{\rm sat}$ and $T=0$, and as a function of 
temperature (right panel) at $n_B=3\ n_{\rm sat}$ and $k=300$ MeV. 
Both figures were obtained with renormalization scheme B}
}
\end{figure}

\begin{figure}[htb]
\mbox{%
\parbox{8.2cm}{\epsfig{file=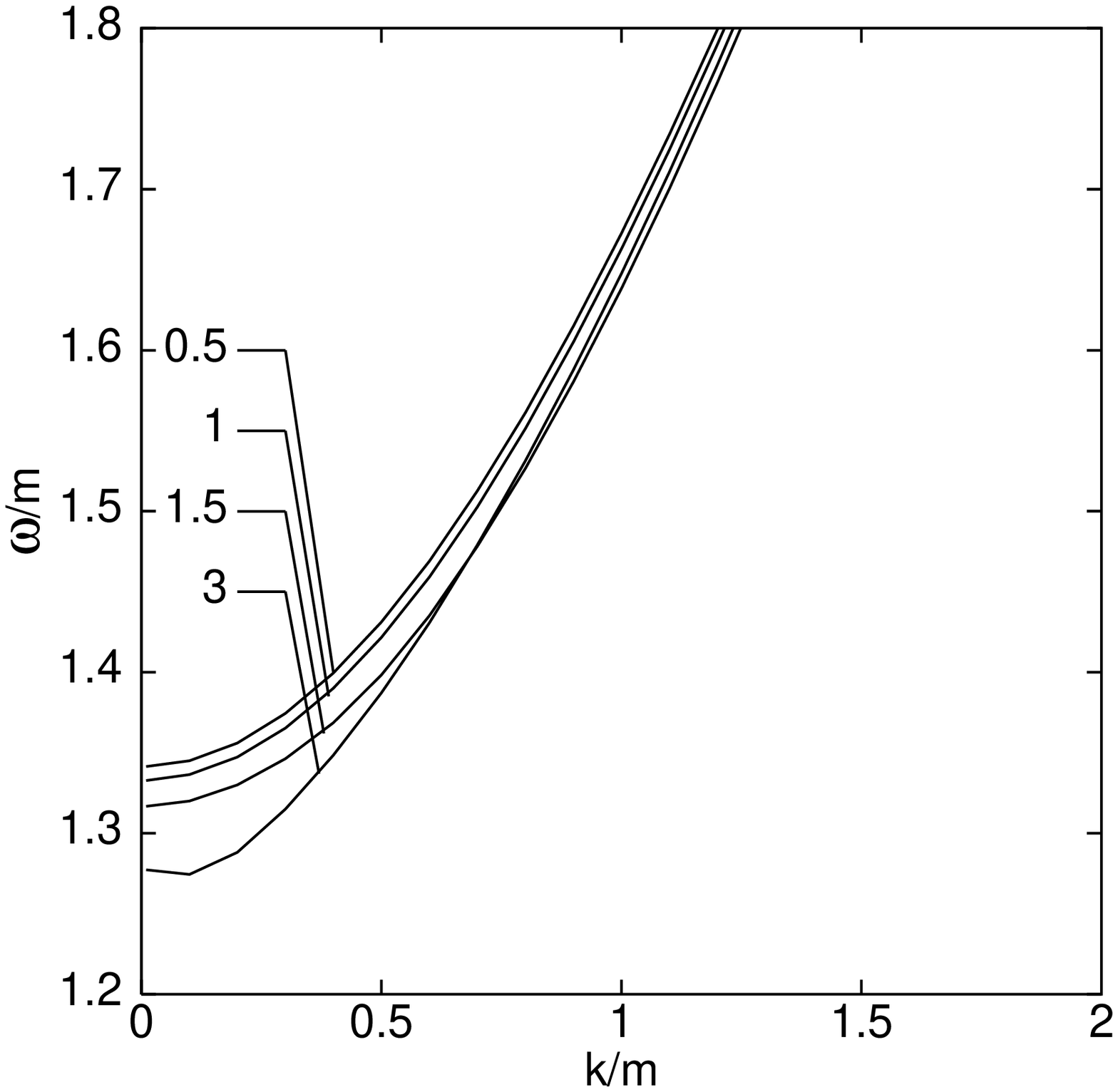,width=8.2cm}}
\parbox{0.2cm}{\phantom{a}}
\parbox{8.2cm}{\epsfig{file=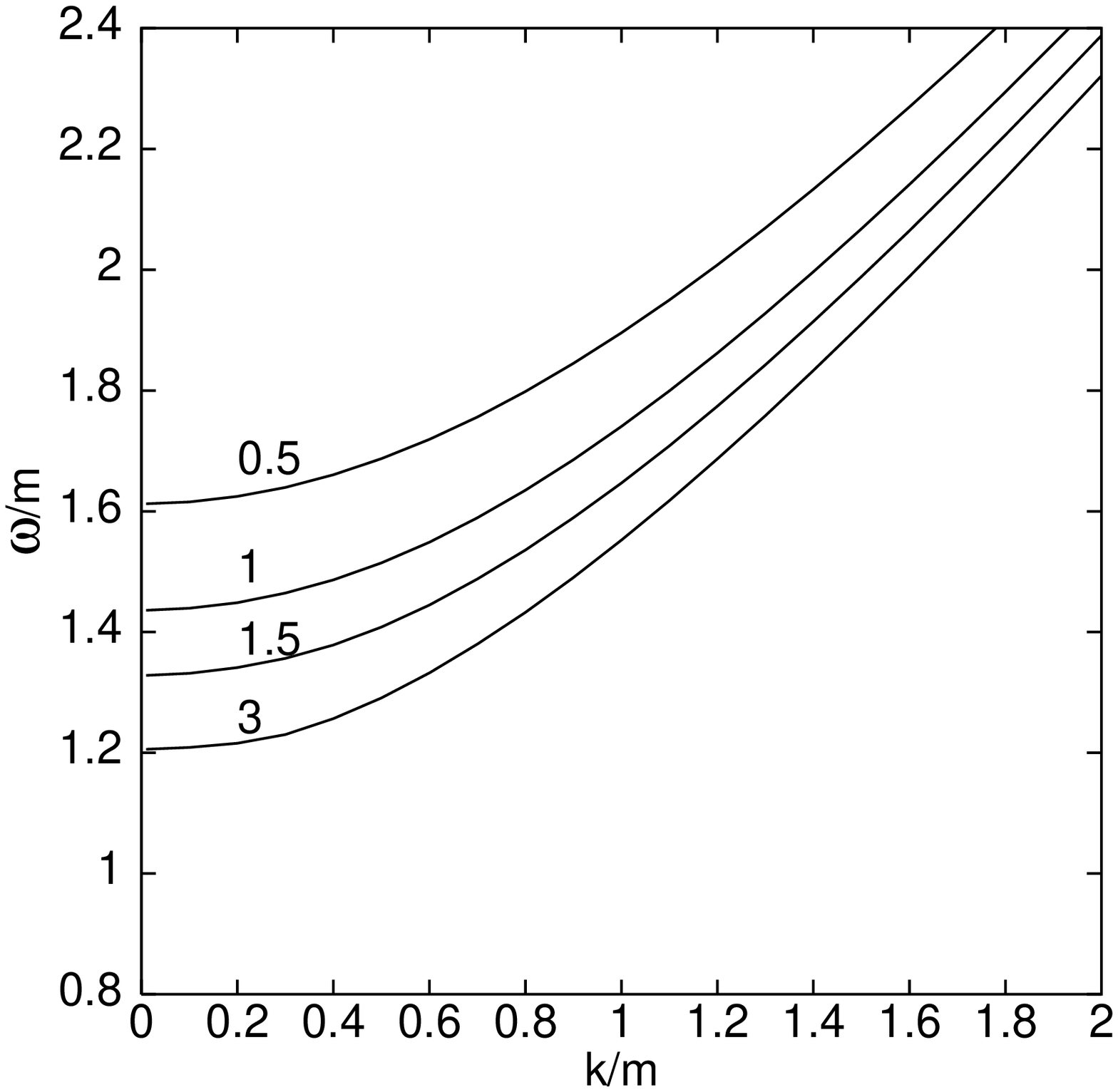,width=8.2cm}}
}
\vskip 0.1cm
\mbox{%
\parbox{16cm}{{\bf Fig. 6} Dispersion relation of the $a_1$ longitudinal mode,
as calculated with renormalization scheme A (left panel) or B (right panel). The curves are labelled by the value of the
density in saturation units $n_B/n_{\rm sat}$.}
}
\end{figure}

\begin{figure}[htb]
\mbox{%
\parbox{8.2cm}{\epsfig{file=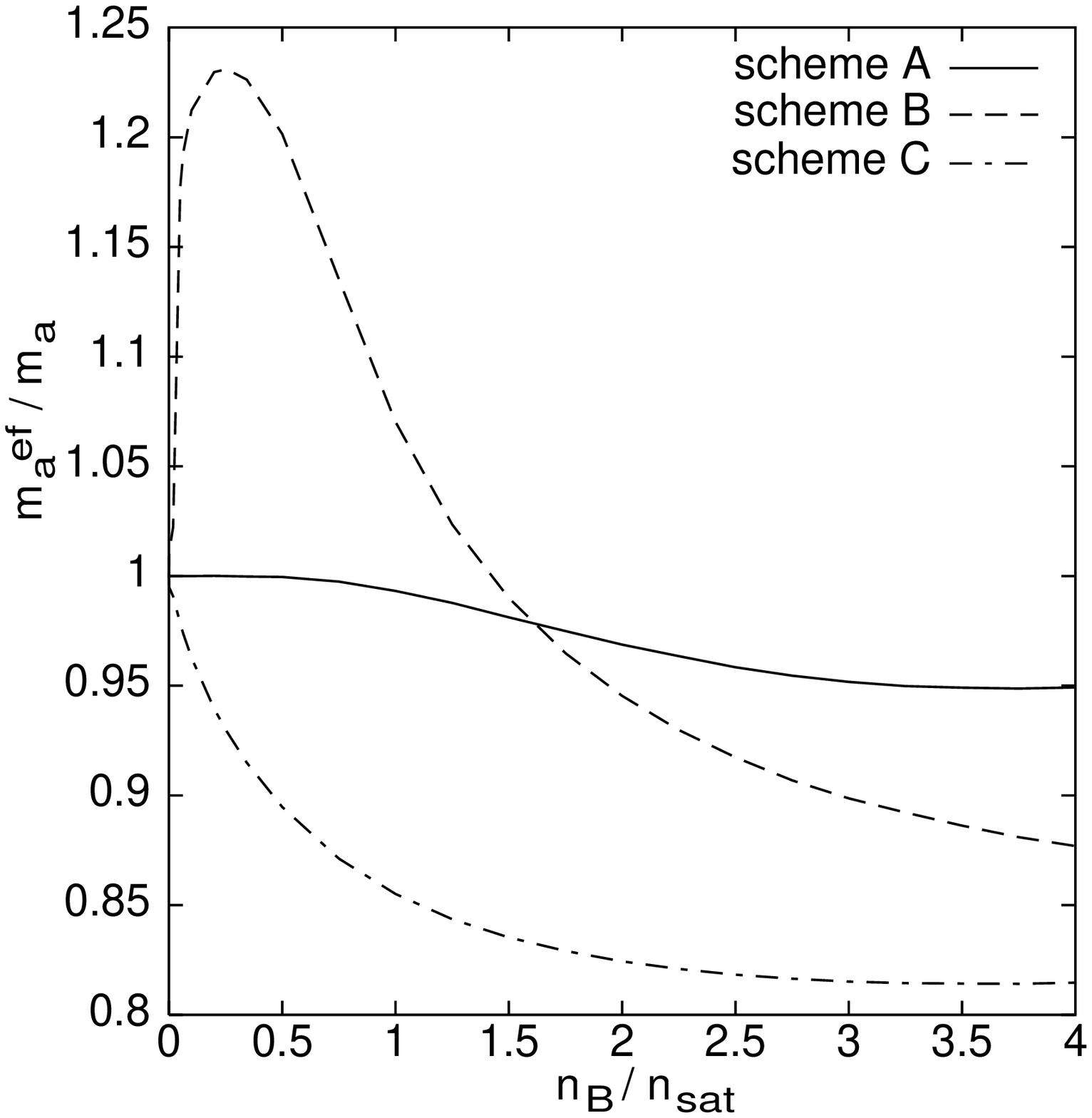,width=8.2cm}}
\parbox{0.1cm}{\phantom{a}}
\parbox{8.2cm}{\epsfig{file=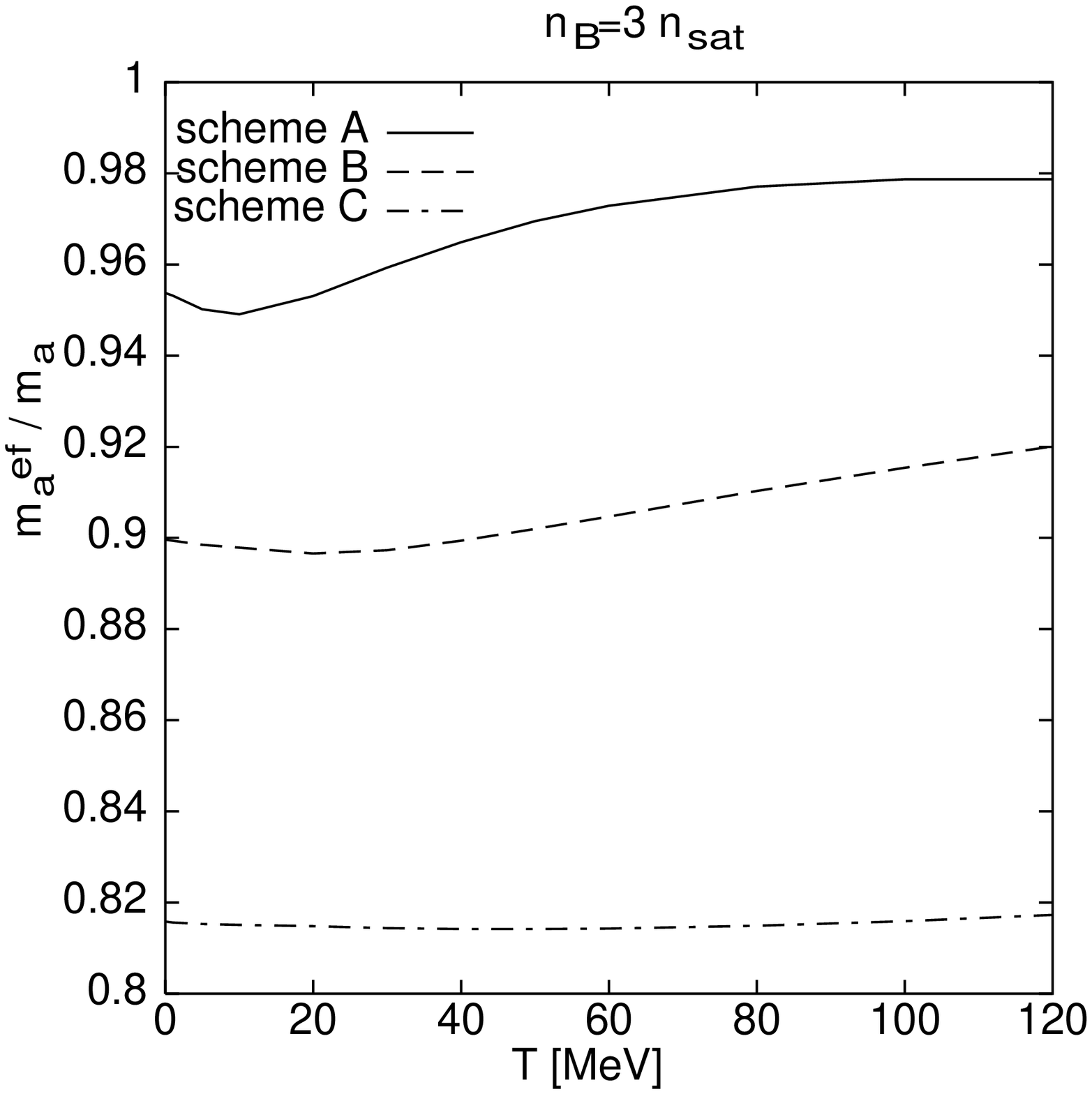,width=8.2cm}}
}
\vskip 0.1cm
\mbox{%
\parbox{16cm}{{\bf Fig. 7} Effective mass of the $a_1$ meson,
as calculated with renormalization schemes A, B or C, plotted
as a function of density at $T=0$ (left panel), and as a function of
temperature T at $n_B=3\ n_{\rm sat}$ (right panel) }
}
\end{figure}

\begin{figure}[htb]
\mbox{%
\parbox{8.2cm}{\epsfig{file=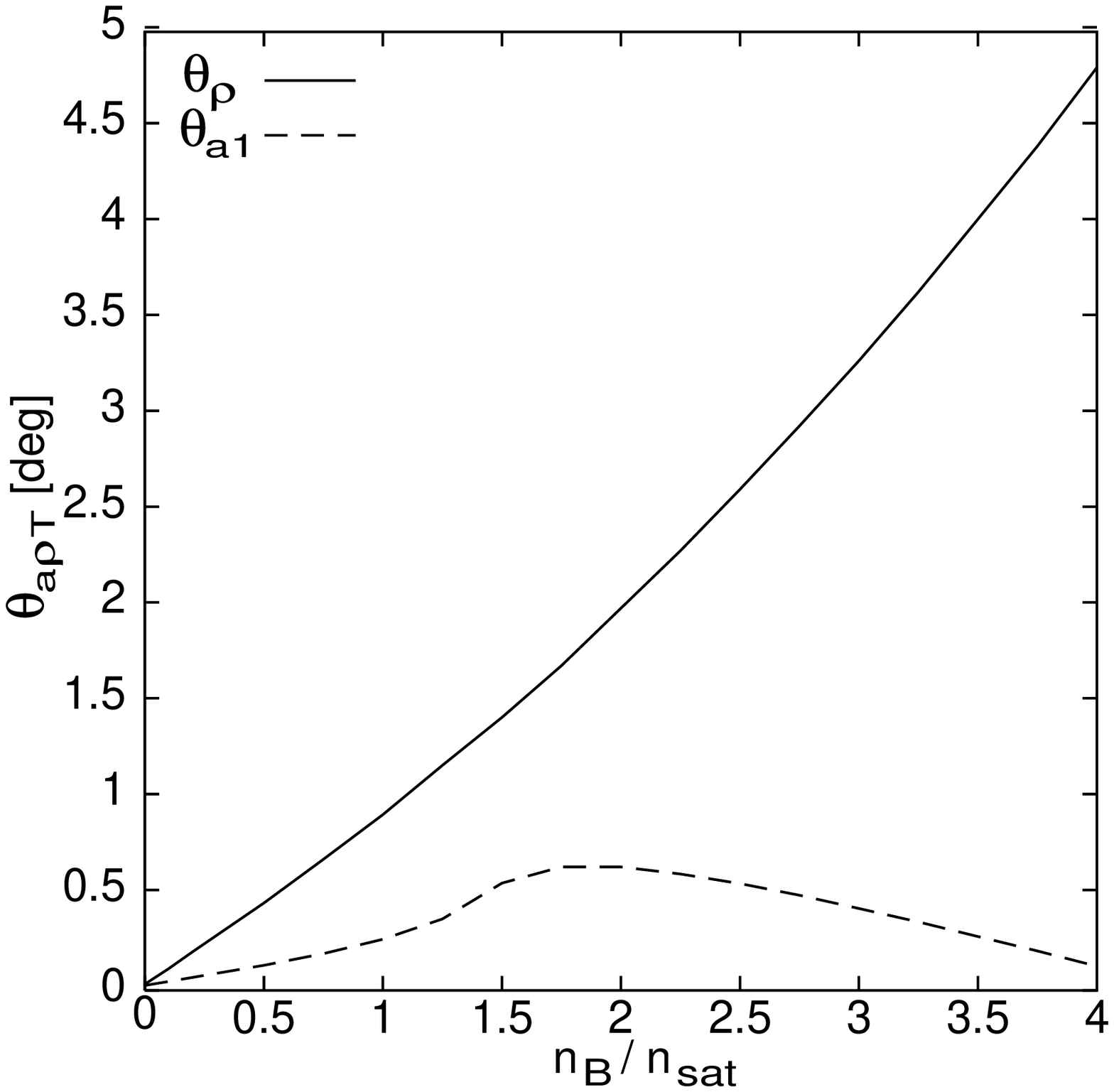,width=8.2cm}}
\parbox{0.1cm}{\phantom{a}}
\parbox{8.2cm}{\epsfig{file=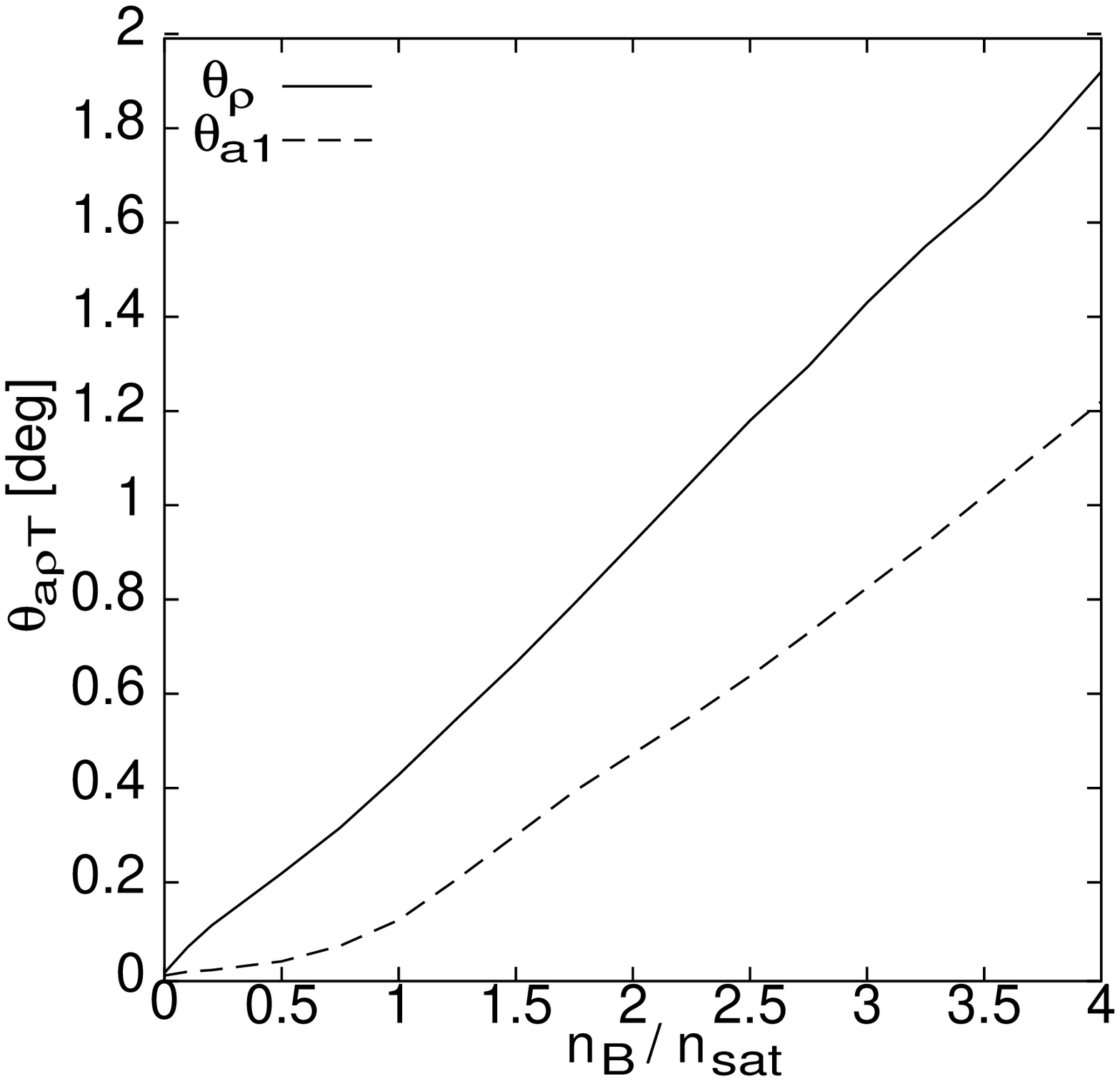,width=8.2cm}}
}
\vskip 0.1cm
\mbox{%
\parbox{16cm}{{\bf Fig. 8} Mixing angle in the $a_1$-$\rho$ 
transverse mode at $k$=300 MeV and $T=0$, as a function of density. 
On the left panel with renormalization scheme A; on the right, with 
scheme B.}
}
\end{figure}

\begin{figure}[htb]
\mbox{%
\parbox{8.2cm}{\epsfig{file=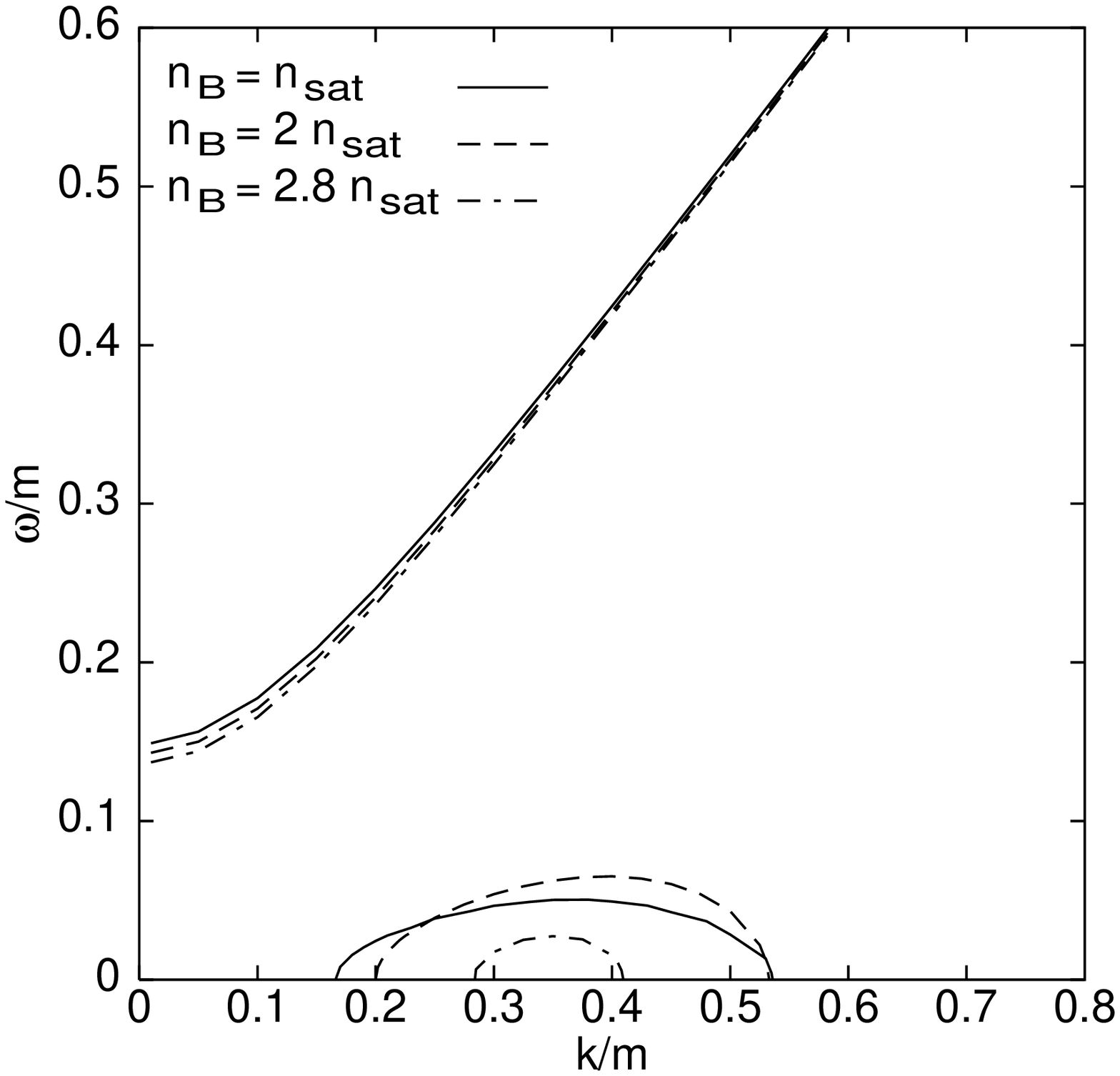,width=8.2cm}}
\parbox{0.1cm}{\phantom{a}}
\parbox{8.2cm}{\epsfig{file=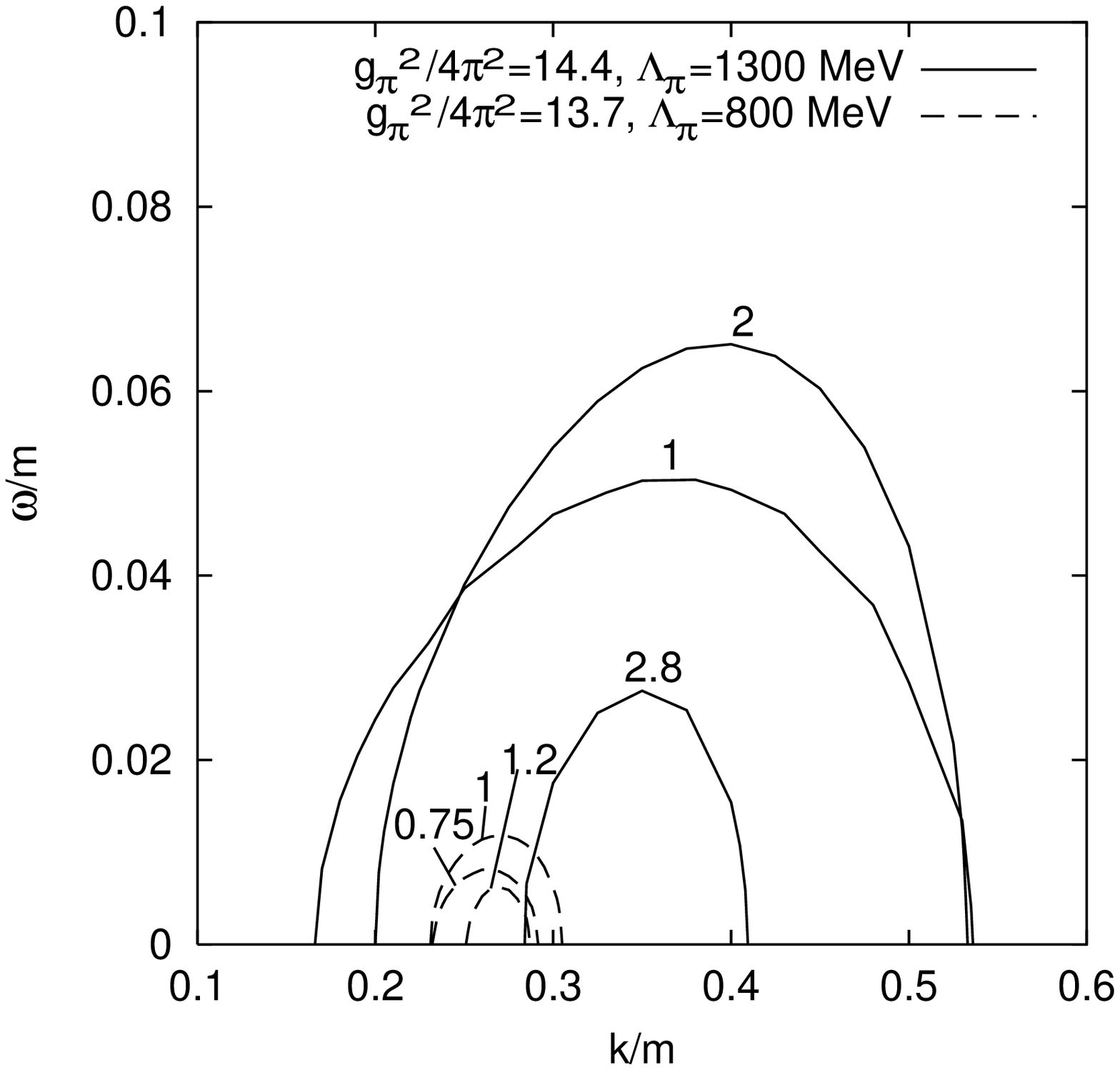,width=8.2cm}}
}
\vskip 0.1cm
\mbox{%
\parbox{8.2cm}{\epsfig{file=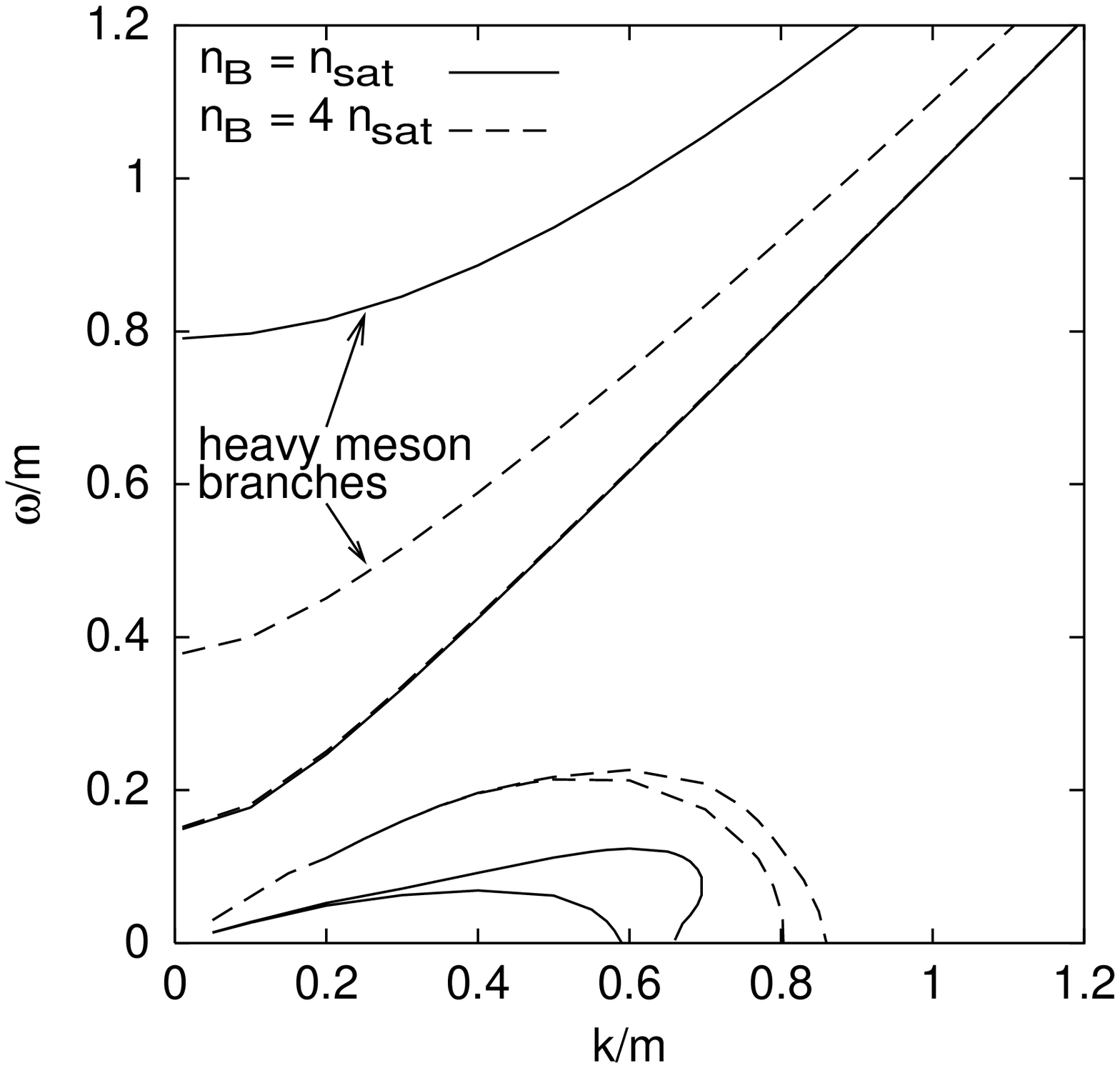,width=8.2cm}}
\parbox{0.1cm}{\phantom{a}}
\parbox{8.2cm}{{\bf Fig. 9} Dispersion relation for the pion with pure
pseudovector coupling, with and without
  mixing with the $a_1$. The curves are labelled by the value of the
density in saturation units $n_B/n_{\rm sat}$ and were obtained with
renormalization scheme A. }
}
\end{figure}

\begin{figure}[htb]
\mbox{%
\parbox{8.2cm}{\epsfig{file=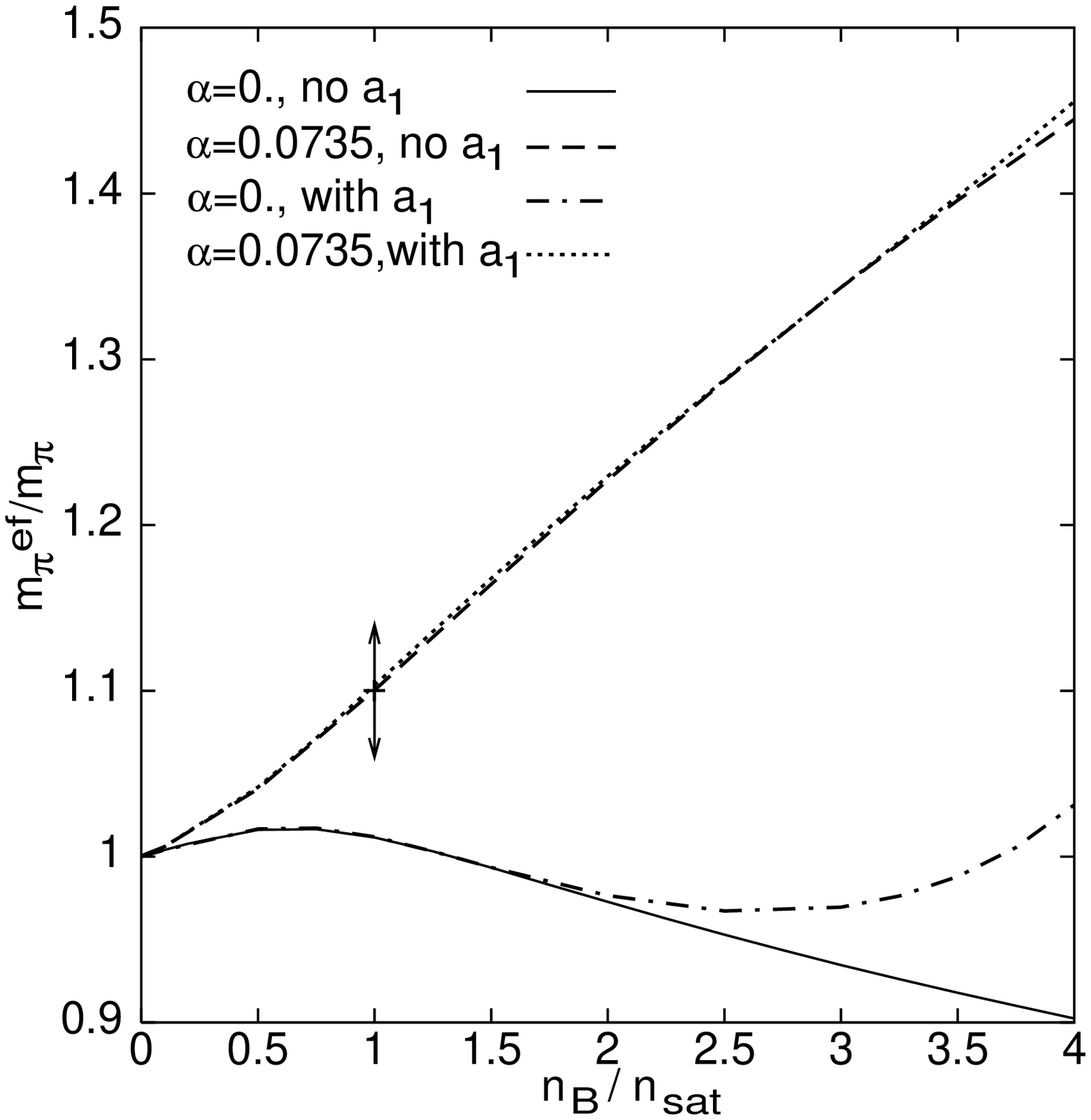,width=8.2cm}}
\parbox{0.1cm}{\phantom{a}}
\parbox{8.2cm}{\epsfig{file=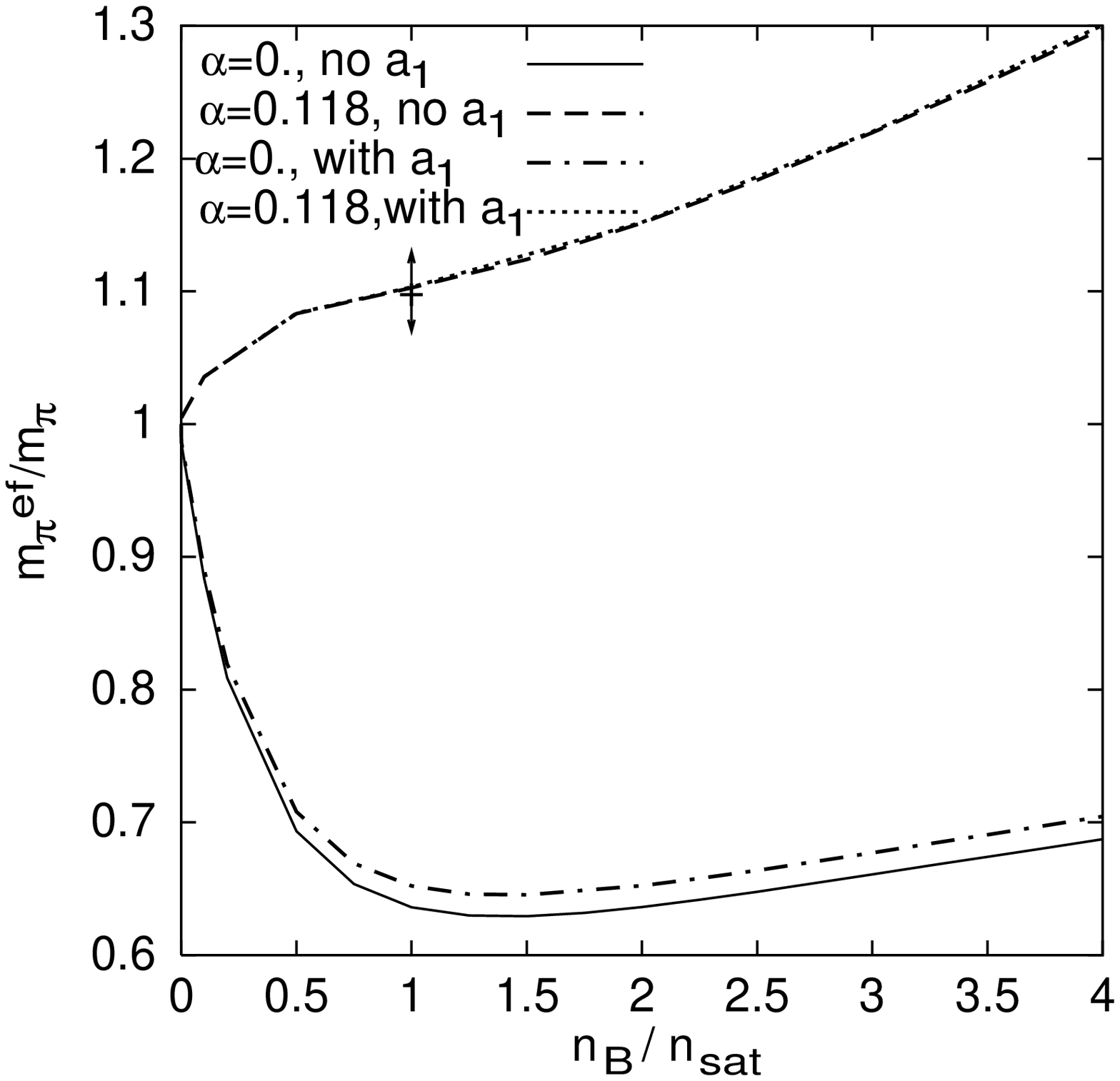,width=8.2cm}}
}
\vskip 0.1cm
\mbox{%
\parbox{16cm}{{\bf Fig. 10} Effective mass of the pion, with 
and without mixing with the $a_1$ and PS admixture, obtained with
renormalization scheme A (left) or B (right). The experimental
determination of the effective pion mass at $n_B=n_{\rm sat}$ 
is indicated on the figures.}
}
\end{figure}


\begin{thebibliography}{99}

\bibitem{DA85}  J. Diaz Alonso,  Ann. Phys. {\bf 160} 
(1985), 1.

\bibitem{DP91} J. Diaz Alonso and A. P\'erez Canyellas, Nucl. Phys. 
{\bf A526} (1991) 623.

\bibitem{DPS89}  J. Diaz Alonso, A. P\'{e}rez and H. Sivak, 
Nucl. Phys. {\bf A505} (1989) 695.

\bibitem{GDP94}  E. Gallego, J. Diaz Alonso and A. P\'{e}rez,  Nucl.
Phys. {\bf A578} (1994) 542.

\bibitem{MGP01} L. Mornas, E. Gallego and A. P\'erez, submitted to
Nucl. Phys. {\bf A}

\bibitem{BonnPot} R. Machleidt, K. Holinde and Ch. Elster, Phys. Rep.
{\bf 149} (1987) 1. 

\bibitem{W67-WZ67-GG69} 
S. Weinberg, Phys. Rev. Lett. {\bf 18} (1967) 188; \\
J. Wess and B. Zumino, Phys. Rev. {\bf 163} (1967) 1727; \\
S. Gasiorowicz and D.A. Geffen, Rev. Mod. Phys. {\bf 41} (1969) 531.

\bibitem{LS89} I. Lovas and K. Sailer, Phys. Lett. {\bf 220} 
(1989) 229.

\bibitem{SW92} B.D. Serot and J.D. Walecka, Acta Phys. Polon.
{\bf B23} (1992) 655.      

\bibitem{P98-99} G. Pr\'ezeau, Phys. Rev. {\bf C58} (1998) 1853,
{\it ibid} {\bf C59} (1999) 2301.

\bibitem{KR94} P. Ko and S. Rudaz, Phys. Rev. {\bf D50} (1994) 6877. 

\bibitem{FRS94} V.N. Fomenko, P. Ring and L.N. Savushkin, 
Nucl. Phys. {\bf A579} (1994) 438.

\bibitem{BMNQFS96}  P. Bernardos, S. Marcos, R. Niembro, M.L. Quelle, 
V.N. Fomenko and L.N. Savushkin, J. Phys. {\bf G22} (1996) 361.

\bibitem{BFMNLQS01} P. Bernardos, V.N. Fomenko, S. Marcos, 
R. Niembro, M. Lopez-Quelle and L.N. Savushkin, 
J. Phys. {\bf G27} (2001) 147.

\bibitem{SR97} V.G.J. Stoks and Th.A. Rijken, Nucl. Phys. {\bf A613} 
(1997) 311.

\bibitem{dileptonsa1} L. Xiong, E.V. Shuryak and G.E. Brown, Phys.
Rev. {\bf D46} (1992) 3798. \\
C. Song, C.M. Ko and C. Gale, Phys. Rev. 
{\bf D50}  (1994) R1827. \\
J.K. Kim, P.W. Ko, K.Y. Lee and S. Rudaz, Phys. Rev. {\bf D53} 
(1996) 4787. \\
S. Gao and C. Gale,  Phys. Rev. {\bf C57} (1998) 254. \\
S. Gao, C. Gale, C. Ernst, H. St\"ocker and W. Greiner, 
Nucl. Phys. {\bf A661} (1999) 518. 


\bibitem{DBS84} J.W. Durso, G.E. Brown and M. Saarela, 
Nucl. Phys. {\bf A430} (1984) 653.  

\bibitem{JHS96}  G. Janssen, K. Holinde and J. Speth, Phys. Rev.
{\bf C54} (1996) 2218. 

\bibitem{M89}  R. Machleidt. ''The Meson Theory of Nuclear Forces and
Nuclear Structure.'' Adv. in Nucl. Phys. Vol.{\bf 19}; 
J.W. Negele and E. Vogt Edts. (Plenum, New York, 1989.)

\bibitem{S99} R. Shyam, Phys. Rev. {\bf C60} (1999) 055213.

\bibitem{ADMSM96} A. Engel, A. K. Dutt-Mazumder, R. Shyam and U. Mosel,
Nucl.Phys. {\bf A603} (1996) 387.

\bibitem{HP94} C.J.  Horowitz and J. Piekarewicz, Phys. Rev. {\bf C50} 
(1994) 2540.

\bibitem{KPH95} H. Kim, J. Piekarewicz and C.J. Horowitz,
Phys. Rev. {\bf C51} (1995) 2739.

\bibitem{RPLP99} S. Reddy, M. Prakash, J.M. Lattimer and J. Pons,
Phys. Rev. {\bf C59} (1999) 2888.

\bibitem{YT00} S. Yamada and H. Toki, Phys. Rev. {\bf C61} (2000) 
015803.

\bibitem{MP01} L. Mornas and A. P\'erez, eprint nucl-th/0106058         submitted to Eur. Phys. J. {\bf A}
              
\bibitem{Angeles} M$^{\rm a}$ A. P\'erez Garc{\'\i}a, Ph.D Thesis, Oviedo 
University, Spain (2001)

\bibitem{DM98a} J. Diaz Alonso and L. Mornas,  Nucl. Phys. 
{\bf A629}  (1998) 679.

\bibitem{DM98b} J. Diaz Alonso and L. Mornas, Phys. Lett {\bf B437} 
(1998) 12.

\bibitem{MK95}  K. Morawetz and D. Kremp Z. Phys. {\bf A352} (1995) 
265.

\bibitem{Wehrberger} K. Wehrberger, Phys. Rep. {\bf 225} (1993) 273.

\bibitem{DF90} J.F. Dawson and R.J. Furnstahl, Phys. Rev. {\bf C42}
(1990) 2009.

\bibitem{Piek01} J. Piekarewicz, eprint nucl-th/0103016

\bibitem{ringvangiai01} P. Ring, Zhong-yu Ma, Nguyen Van Giai, 
D. Vretenar, A. Wandelt and Li-gang Cao, eprint nucl-th/0106061

\bibitem{M01b} L. Mornas, submitted to J. Phys. G

\bibitem{SW86-Se97}  B.D. Serot and J.D. Walecka, 
``{\it The relativistic nuclear many-body problem}'' Adv. Nucl. 
Phys. {\bf 16} (1986) 1. \\
B.D. Serot and J.D. Walecka, Int. J. Mod. Phys. {\bf E6} (1997) 515.

\bibitem{C77} S.A. Chin, Ann. Phys. (N.Y.) {\bf 108} (1977) 301.

\bibitem{KS88} H. Kurasawa and T. Suzuki,  Nucl. Phys. {\bf A490} 
(1988) 571.

\bibitem{LH89} K. Lim and C.J. Horowitz, Nucl. Phys. {\bf A501} 
(1989) 729. 

\bibitem{CPS92} L.S. Celenza, A. Pantziris and C.M. Shakin, Phys. Rev. 
{\bf C52} (1992) 205.

\bibitem{JP94}  H.C. Jean and J. Piekarewicz, Phys. Rev. {\bf C49}
(1994) 1981.

\bibitem{SH94}  H. Shiomi and T. Hatsuda, Phys. Lett. {\bf B334}
(1994) 281.

\bibitem{CL95} J.C. Caillon and J. Labarsouque, Phys. Lett. {\bf B352} 
(1995) 193.

\bibitem{TDMG00} O. Teodorescu, A. K. Dutt-Mazumder and C. Gale,
Phys. Rev. {\bf C61} (2000) 051901.      

\bibitem{KKS98} S. Kubis, M. Kutschera and S. Stachniewicz,
Acta Phys. Polon. {\bf B29} (1998) 809; \\
S. Kubis and M. Kutschera Phys. Lett. {\bf B399} (1997) 191.

\bibitem{dJL98} F. de Jong and H. Lenske, Phys. Rev. {\bf C57} (1998) 3099.

\bibitem{SST97} H. Shen, Y. Sugahara and H. Toki, Phys. Rev. {\bf C55}
(1997) 1211.

\bibitem{BR91} G. E. Brown and M. Rho,  Phys. Rev. Lett. {\bf 66}
(1991) 2720.

\bibitem{RCW97} R. Rapp, G. Chanfray and J. Wambach, 
Nucl. Phys. {\bf A617} (1997) 472.

\bibitem{KRBR00} Y. Kim, R. Rapp, G.E. Brown and M. Rho, Phys. Rev.
{\bf C62} (2000) 015202.

\bibitem{GL94} W.R. Gibbs and B. Loiseau, Phys. Rev. {\bf C50} 
(1994) 2742.

\bibitem{DAM01} J. Diaz-Alonso and L. Mornas, in preparation

\bibitem{MSTV90} A.B. Migdal, E.E. Saperstein, M.A. Troitsky and 
D.N. Voskresensky, Phys. Rep. {\bf 192} (1990) 179.

\bibitem{SDEM94} A. Sch\"afer, H.C. D\"onges, A. Engel and  U. Mosel,
Nucl. Phys. {\bf A575} (1994) 429.

\bibitem{BS69} R. Bryan, B.L. Scott, Phys. Rev. {\bf 177} (1969) 1435.

\bibitem{Dickoff} W.H. Dickhoff, A. Faessler, J. Meyer-Ter-Vehn and 
H. M\"uther, Nucl. Phys. {\bf A368} (1981) 445 \\
W.H. Dickhoff, A. Faessler, H. M\"uther and Shi Shu Wu, 
Nucl. Phys. {\bf A405} (1983) 534 

\bibitem{SST99} T. Suzuki, H. Sakai and T. Tatsumi, Phys. Lett. {\bf B455}
(1999) 25.

\bibitem{Oset} E. Oset, W. Weise and H. Toki, Phys. Rep. {\bf 83}
(1982) 281-380.

\bibitem{ATG00} C. Arias Tobe\~na and  M.F. Gari, eprint
 hep-ph/0008304   

\bibitem{Furnstahl} R.J. Furnstahl and B.D. Serot, 
 Comments Nucl. Part. Phys. {\bf 2} (2000) A23 \\
R.J. Furnstahl, B.D. Serot and H.-B. Tang, Nucl. Phys. {\bf A618}
(1997) 446.

\bibitem{LP01} M. Lehmann, G. Pr\'ezeau, eprint hep-ph/0102161

\bibitem{DM97} A.K. Dutt-Mazumder, Phys. Lett. {\bf B399} (1997) 196.

\bibitem{HKL93} T. Hatsuda, Y. Koike and S.J. Lee, Nucl. Phys. 
{\bf B394} (1993) 221.

\bibitem{P95} R.D. Pisarski, Phys. Rev. {\bf D52} (1995) R3773.


\bibitem{DEI90} M. Dey, V.L. Eletsky and B.L. Ioffe, 
 Phys. Lett. {\bf B252} (1990) 620.

\bibitem{Zhu} Shi-Lin Zhu, Phys. Rev. {\bf C59} (1999) 435.

\bibitem{DawPiekcond} J.F. Dawson and J. Piekarewicz, Phys. Rev. 
{\bf C43} (1991) 2631.

\bibitem{mefpi} T. Yamazaki {\it et al.}, Z. Phys. {\bf A355} (1996) 219. \\
T. Waas, R. Brockmann  and W. Weise, Phys. Lett. {\bf B405} (1997) 215. \\
E. Friedman and A. Gal, Phys.Lett. {\bf B432} (1998) 235.

\bibitem{DMP01} J. Diaz Alonso, L. Mornas and M.A. P\'erez-Garc{\'\i}a,
in preparation 

\bibitem{NFSLQMP98} 
R. Niembro, V.N. Fomenko, L.N. Savushkin, M. L\'opez-Quelle,
S. Marcos and P. Bernardos, J. Phys. {\bf G24} (1998) 1945.

\bibitem{GLMBG94} P.F.A. Goudsmit, H.J. Leisi, E. Matsinos, B.L. Birbrair
and  A.B. Gridnev, Nucl. Phys. {\bf A575} (1994) 673.

\bibitem{Gross} F. Gross, K.M. Maung, L.W. Townsend and S.J. Wallace,
Phys. Rev. {\bf C40} (1989) R10 \\
F. Gross, J.W. van Orden and K. Holinde, Phys. Rev. {\bf C41} (1990) R1909

\bibitem{BLLQMN00} P. Bernardos, R.J. Lombard, M. L\'opez-Quelle,
S. Marcos and R. Niembro, Phys. Rev. {\bf C62} (2000) 024314.


\end{thebibliography}
\end{document}